\documentclass[numberedappendix,apj,twocolumn]{emulateapj}
\usepackage{graphicx}
\usepackage{subfigure}
\usepackage{amssymb,amsmath}
\usepackage{natbib}
\usepackage{epsfig}

\begin{document}

\title{Optical-faint, Far-infrared-bright {\textit{\textbf Herschel}} Sources
in the CANDELS Fields: Ultra-Luminous Infrared Galaxies at \boldmath$z>1$ and
the Effect of Source Blending
}
\author{Haojing Yan \altaffilmark{1}, 
Mauro Stefanon \altaffilmark{1}, 
Zhiyuan Ma \altaffilmark{1},
S. P. Willner \altaffilmark{2},
Rachel Somerville \altaffilmark{3},
Matthew L. N. Ashby \altaffilmark{2},
Romeel Dav{\'e} \altaffilmark{4},
Pablo G. P\'erez-Gonz\'alez \altaffilmark{5},
Antonio Cava \altaffilmark{6},
Tommy Wiklind \altaffilmark{7},
Dale Kocevski \altaffilmark{8},
Marc Rafelski \altaffilmark{9},
Jeyhan Kartaltepe \altaffilmark{10,12},
Asantha Cooray \altaffilmark{11}
}

\altaffiltext{1}{Department of Physics \& Astronomy, University of Missouri,
Columbia, MO 65211, USA}
\altaffiltext{2}{Harvard-Smithsonian Center for Astrophysics, 60 Garden Street, Cambridge, MA 02138, USA}
\altaffiltext{3}{Department of Physics and Astronomy, Rutgers University, 136 Frelinghuysen Rd., Piscataway, NJ 08854, USA}
\altaffiltext{4}{University of the Western Cape, 7535 Bellville, Cape Town, South Africa}
\altaffiltext{5}{Departamento de Astrof\'{\i}sica, Facultad de CC. F\'{\i}sicas, Universidad Complutense de Madrid, E-28040 Madrid, Spain}
\altaffiltext{6}{Observatoire de Gen{\`e}ve, Universit{\'e} de Gen{\`e}ve, 51 Ch. des Maillettes, 1290 Versoix, Switzerland}
\altaffiltext{7}{Joint ALMA Observatory, Alonso de Cordova 3107, Vitacura, Santiago, Chile}
\altaffiltext{8}{Department of Physics and Astronomy, University of Kentucky, Lexington, KY 40506, USA}
\altaffiltext{9}{Infrared Processing and Analysis Center, California Institute of Technology, Pasadena, CA 91125, USA}
\altaffiltext{10}{National Optical Astronomy Observatory, 950 North Cherry Avenue, Tucson, AZ 85719, USA}
\altaffiltext{11}{Department of Physics and Astronomy, University of California, Irvine, California 92697, USA}
\altaffiltext{12}{Hubble Fellow.}

\begin{abstract}

  The {\it Herschel} very wide-field surveys have charted hundreds of square
degrees in multiple far-IR (FIR) bands. While the Sloan Digital Sky Survey
(SDSS) is currently the best resource for optical counterpart identifications
over such wide areas, it does not detect a large number of {\it Herschel} FIR
sources and leaves their nature undetermined. As a test case, we studied seven
``SDSS-invisible'', very bright 250~$\mu m$ sources ($S_{250} > 55$~mJy) in the
Cosmic Assembly Near-infrared Deep Extragalactic Legacy Survey (CANDELS) fields
where we have a rich multi-wavelength data set. We took a new approach to
decompose the FIR sources, using the near-IR or the optical images directly for
position priors. This is an improvement over the previous decomposition efforts
where the priors are from mid-IR data that still suffer from the source blending
problem in the first place. We found that in most cases the single
{\it Herschel} sources are made of multiple components that are not necessarily
at the same redshifts. Our decomposition succeeded in identifying and
extracting their major contributors. We show that these are all ULIRGs at 
$z\sim 1$--2 whose high $L_{IR}$ is mainly due to dust-obscured star formation.
Most of them would not be selected as sub-mm galaxies. They all have
complicated morphologies indicative of merger or violent instability, and their
stellar populations are heterogeneous in terms of stellar masses, ages and
formation histories. Their current ULIRG phases are of various degrees of
importance in their stellar mass assembly. Our practice provides a promising
starting point to develop an automatic routine to reliably study bright
{\it Herschel} sources.

\end{abstract}


\section{Introduction}

   In its more than four years of operation, the
{\it Herschel} Space Observatory (Pilbratt et al. 2010), the largest FIR/sub-mm
space telescope ever flown, produced a wealth of data awaiting exploration. It
carried two imaging spectrometers, namely, the Photodetector Array Camera and 
Spectrometer (PACS, Poglitsch et al. 2010) observing in 100 (or 70) and 
160~$\mu$m, and the Spectral and Photometric Imaging REceiver (SPIRE, Griffin 
et al. 2010) observing in 250, 350 and 500~$\mu$m. Together they sampled the
peak of heated dust emission from $z = 0$ to 6 and possibly beyond, and 
offered the best capability to date in the {\it direct} measurement of the 
total infrared (IR) luminosities for a large number of galaxies at $z>1$.
Two of the largest {\it Herschel} extragalactic surveys, the Herschel 
Astrophysical Terahertz Large Area Survey (H-ATLAS; Eales et al. 2010) and the
Herschel Multi-tiered Extragalactic Survey (HerMES; Oliver et al. 2012), have
mapped the FIR/sub-mm universe in unprecedented detail. H-ATLAS has surveyed
$\sim 570$~deg$^2$ over six areas at a uniform depth, while HerMES has observed
a total of $\sim 380$~deg$^2$ in several levels of depth and spatial coverage
combinations (``L1'' to ``L7'', from the deep-and-narrow to wide-and-shallow).

   The true power of these {\it Herschel} data can only be achieved when they
are combined with observations at other wavelengths, most importantly in 
optical to near-IR (NIR) as this is the traditional regime where the stellar
population of galaxies is best studied. The Sloan Digital Sky Survey (SDSS;
York et al. 2000) is the most natural choice when identifying optical
counterparts of the FIR sources over such large areas. However, its limited
depth does not allow us to take the full advantage of these already sensitive
FIR data. For example, Smith et al. (2011) cross-matched the 6,876 sources in
the $\sim 16$~deg$^2$ H-ATLAS
Science Demonstration Phase field to the SDSS, and only 2,422 (35.2\%) of them
have reliable counterparts. While some of identification failures are caused by
the ambiguity in assigning the counterpart due to the large {\it Herschel} beam
sizes, most of these FIR sources that are not matched in the SDSS
are genuinely faint in the optical. This probably should not be surprising
because FIR sources could be very dusty. Nevertheless, it is still
interesting that some of the brightest {\it Herschel}\, sources, whose flux 
densities are a few tens of mJy, are not visible in the SDSS images at all. The
nominal 5~$\sigma$ limits of the SDSS is 22.3, 23.3, 23.1, 22.3, 20.8 ~mag in
$u$, $g$, $r$, $i$, and $z$, respectively (York et al. 2000). Using
a 2~$\sigma$ limit, a conservative estimate of
the FIR-to-optical flux density ratio of such SDSS-invisible FIR sources is
$S_{FIR}/S_{opt}\gtrsim 10^4$. 

   Naturally, one would speculate that these optical-faint {\it Herschel}
sources are Ultra-Luminous InfraRed Galaxies (ULIRGs; see e.g. Lonsdale, 
Farrah \& Smith 2006 for a review) at high redshifts. If they are at 
$z\gtrsim 1$, their absence from the SDSS images can be easily explained by 
their large luminosity distances. If they are ULIRGs, their FIR brightness
could also be understood. In this sense, the closest analogs to such objects
are sub-mm galaxies (SMGs), which are usually selected at $\sim 850$~$\mu$m and
are found to be ULIRGs at $z\sim 2$--3. On average, a typical SMG would have 
850~$\mu$m flux density $S_{850}\sim 5.7$~mJy and optical brightness 
$R\sim 24.6$~mag (see e.g.  Chapman et al. 2005). The SMGs at the faint-end of
the optical brightness distribution are likely at $z\sim 4$--5 or even higher
redshifts (Wang et al. 2007; Capak et al. 2008; Schinnerer et al. 2008; Coppin
et al. 2009; Daddi et al. 2009; Younger et al. 2009). The most extreme example
is the historical HDF850.1 (Hughes et al. 1998), which has no detectable 
counterpart in even the deepest optical/NIR images (Cowie et al. 2009 and
the references therein) and now has been confirmed to be at $z=5.183$ based on
its CO lines (Walter et al. 2012). It has also been suggested that extremely
dusty galaxies like HDF850.1 could play a major role in the star formation 
history in the early universe (Cowie et al. 2009). If this is true, the very
wide field {\it Herschel}\, surveys should be able to reveal a large number of
such objects at very high redshifts. In fact, recently a bright FIR galaxy
discovered in the HerMES, which again has no detectable optical counterpart 
($z>25.9$~mag) and is only weakly visible in NIR, set a new, record-high
redshift of $z = 6.337$ for ULIRG (Riechers et al. 2013), approaching the end
of the cosmic H I reionization epoch.

   It is thus important to investigate in detail the nature of such 
SDSS-invisible, bright {\it Herschel}\, sources. Obviously, the minimum
requirement to move forward is to acquire optical data that are much deeper
than the SDSS. In this paper, we present our study of seven such sources that
happen to be covered by the Cosmic Assembly Near-infrared Deep Extragalactic
Legacy Survey (CANDELS; PIs: Faber \& Ferguson; Grogin
et al. 2011; Koekemoer et al. 2011) and therefore have a rich multiwavelength
data set. Our main scientific goal is to understand whether these sources are
indeed high-z ULIRGs and the stellar populations of their host galaxies.
By targeting some of the brightest {\it Herschel} sources, our study will also
help address one of the most severe problems at the FIR bright-end, namely,
the discrepancy between the observed bright FIR source number counts and 
various model predictions (see e.g., Clements et al. 2010, and the references
therein; Niemi et al. 2012).
While the vast majority of other {\it Herschel}\, sources do not have
comparable ancillary data as used in this current work, our study here will
serve as an useful guide to future investigations. 

   The most severe technical obstacle that we need to overcome is the
long-standing source confusion (i.e., blending) problem in the FIR/sub-mm
regime. The beam sizes (measured as Full Width at Half Maximum; FWHM) of PACS
are $\sim 6$--7\arcsec\, and $\sim 11$-–14\arcsec\, at 70/100~$\mu$m and 
160~$\mu$m, respectively, depending on the scanning speeds. Similarly, the beam
sizes of SPIRE are $\sim 18$\arcsec, 25\arcsec\, and 36\arcsec\, at 250, 350 and
500~$\mu$m, respectively. Therefore, source confusion can still be severe in
the {\it Herschel} data, which causes ambiguity in assigning the correct
counterparts. It also raises the possibility that the very high FIR flux
densities of the seemingly single {\it Herschel}\, sources might be caused by
the blending of multiple objects, each being less luminous, within a single
beam. SMGs are known to suffer from exactly the same problem because of the
coarse angular resolutions of the sub-mm imagers used for their discovery at
single-dishes.  In fact, using the accurate positions determined by the sub-mm 
interferometry at the Submillimeter Array (SMA; Ho et al. 2004), it has been
unambiguously shown that some of the brightest SMGs indeed are made of multiple
objects that may or may not be physically related (Wang et al. 2011; Barger 
et al. 2012). Recently, a large, high-resolution 870~$\mu$m interferometry
survey of 126 SMGs with the Atacama Large Millimeter/submillimeter Array (ALMA)
has shown that $>35$\% of the SMGs originated from single-dish observations
could consist of multiple objects (Hodge et al. 2013; Karim et al. 2013).

    Even with the unprecedented sensitivity of the ALMA, however, it 
is still impractical to pin down the locations of the large number
of {\it Herschel}\, sources and to resolve their multiplicities through sub-mm
interferometry. We therefore employed an alternative, less expensive,
empirical approach in this work, using deep optical/NIR data to decompose a 
given {\it Herschel}\, source and to identify its {\it major} counterpart(s) in
the process. This is different from the statistical ``likelihood ratio'' method
(LR; Sutherland \& Saunders 1992), which would assign a counterpart based on
the probabilities of all candidates but would not apportion the flux in
case of multiplicity. It is also different from the de-blending approach where
mid-IR data of better resolution (albeit still being coarse) would be used as
the position priors (e.g., Roseboom et al. 2010; Elbaz et al. 2010; Magnelli
et al. 2013). We do not
take this latter approach because it could be that a prior mid-IR source is 
already a blend of multiple objects that are not necessarily associated. In
this paper, we show that in most cases our method can successfully extract the
{\it major} contributors to the FIR sources, which is sufficient if we are
mainly interested in the ULIRG population with the current {\it Herschel}\, 
very-wide-field data. While it is still in its rudimentary stage, this method
has the potential of being fully automated, and could be critical in the
{\it Herschel} fields where the ``ladders'' in the mid-IR are not available and
can no longer be obtained due to the lack of instruments.

    The paper is organized as follows. The sample and the relevant data are
presented in \S 2, followed by an outline of our analyzing methods in \S 3.
Due to the different data sets involved in different CANDELS fields, we
present the analysis of individual objects in \S 4, 5 and 6, respectively.
We present a discussion of our results in \S 7, and summarize in \S 8. 
We assume the
following cosmological parameters throughout: $\Omega_M=0.27$, 
$\Omega_\Lambda=0.73$ and $H_0=71$~km~s$^{-1}$~Mpc$^{-1}$. The quoted 
magnitudes are all in the AB system.

\section{Sample Description}

   The SDSS-invisible, bright {\it Herschel}\, sources in this work were
selected based on the first data release (DR1) of the HerMES team, which only
includes the SPIRE data. We used their band-merged ``xID'' catalogs, which
were constructed by fitting the point spread function (PSF) at the source
locations determined in the 250~$\mu$m images (Wang et al., in prep.).
These catalogs are cut at a bright flux
density level of 55~mJy for 250 and 350~$\mu$m, and 30~mJy for 500~$\mu$m.

    Among these released HerMES fields, six of them have SDSS coverage, 
three of which have overlap with the CANDELS fields. These three fields
are the ``L2\_GOODS-N'' ($\sim 0.59$~deg$^2$, 53 sources in the HerMES DR1
catalog), ``L3\_Groth-Strip'' ($\sim 1.15$~deg$^2$, 74 sources), and 
``L6\_XMM-LSS-SWIRE'' fields ($\sim 22.58$~deg$^2$, 2,320 sources),
which cover the CANDELS ``GOODS-N'', ``EGS'' and ``UDS'' fields, respectively
\footnote{``GOODS-N'' stands for the northern field of the Great Observatories 
Origins Deep Survey (Giavalisco et al. 2003), ``EGS'' stands for the extended
Groth Strip, and ``UDS''stands for the Ultra-Deep Survey component of the 
United Kingdom Infrared Telescope (UKIRT) Infrared Deep Sky Survey (UKIDSS).}.
The CANDELS {\it HST}\, data, taken by the Advanced Camera for Surveys
(ACS) and/or the IR channel of the Wide Field Camera 3 (WFC3), 
include 5 (in GOODS-N), 13 (in EGS) and 4 (in UDS) of these sources. We
examined them in the SDSS DR9 images and selected those that do not have any
optical detections in any of the five bands within 6\arcsec\ 
(approximately $\sim 3\times$ of the positional accuracy)
to the reported 250~$\mu$m source centroids. This resulted
in 2, 3 and 2 sources in GOODS-N, EGS and UDS, respectively, which form the
sample studied by this work. Table 1 lists the photometry of these seven
sources from the HerMES public catalogs, and Figure 1 shows their images in the
SPIRE 250~$\mu$m and the SDSS $i^\prime$ bands. In addition to the HerMES
three-band SPIRE data, the sources in L2\_GOODS-N and L3\_Groth-Strip also have
the PACS 100 and 160~$\mu$m data from the DR1 of the PACS Evolutionary Probe
program (PEP; Lutz et al. 2011). 

\begin{figure*}[tbp]
\centering 
\includegraphics[width=\textwidth]{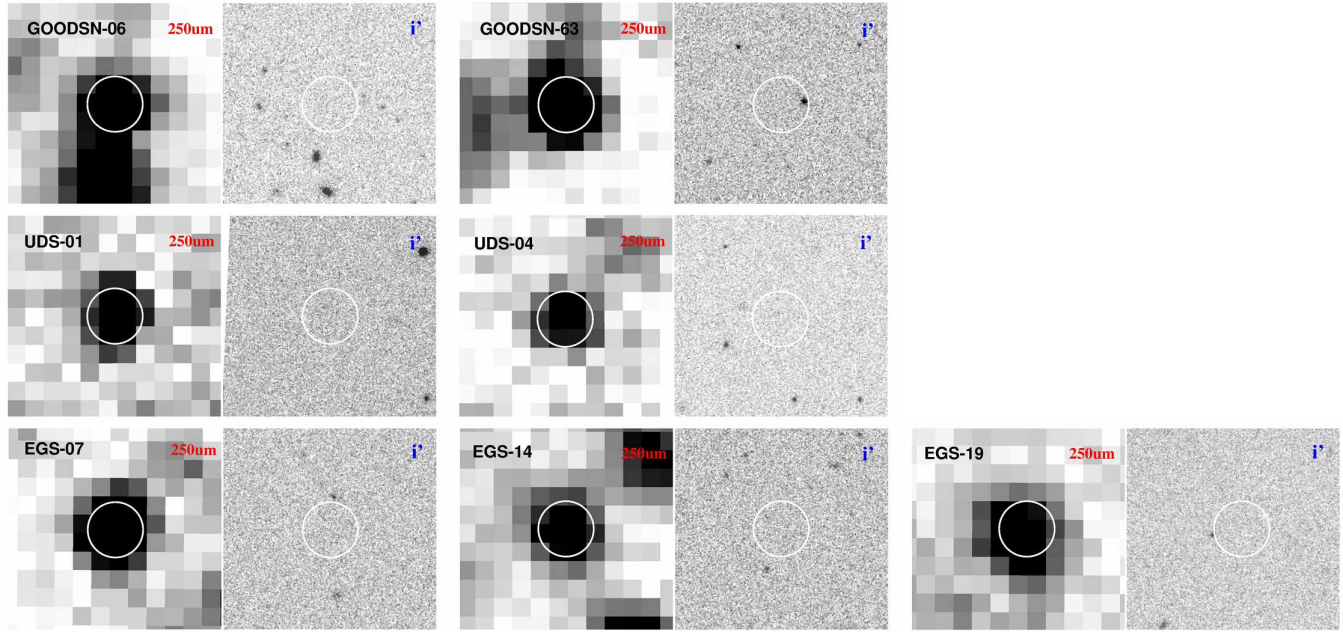}
\caption{Image cutouts of the seven SDSS-undetected, bright {\it Herschel} 
sources in our sample. Two of them are in the GOODS-N field (top), two are in 
the UDS field (middle), and three are in the EGS field (bottom). For each
source, the 250~$\mu$m image is shown in left and the SDSS $i^\prime$ image is
shown in right. All images are $1\farcm 2\times 1\farcm 2$ in size. North is up
and east is to left. The white circles, which are 9\arcsec\, in radius and
resemble the 250~$\mu$m beam size, center
on the 250~$\mu$m source centroids as reported by the HerMES DR1 catalogs.
} 
\end{figure*}

    The three CANDELS fields have a wide range of ancillary data, and those
used in the spectral energy distribution (SED) analysis and/or morphological
study are listed below:

   --- GOODS-N: {\it HST} ACS F435W (hereafter $B_{435}$), F606W ($V_{606}$),
F775W ($i_{775}$) and F850LP ($z_{850}$) from the GOODS program;
F098M ($Y_{098}$), F105W ($J_{105}$ and F160W ($H_{160}$) from the CANDELS
program; {\it Spitzer} InfraRed Array Camera (IRAC) 3.6, 4.5, 5.8, and 
8.0~$\mu$m, and Multiband Imaging Photometer for Spitzer (MIPS) 24~$\mu$m
from the GOODS program; MIPS 70~$\mu$m from the Far-Infrared Deep Extragalactic
Legacy survey (FIDEL; Dickinson et al., in prep.); ground-based deep $U$-band
image taken by the GOODS team at the KPNO 4m MOSAIC (Dickinson, priv. comm.).

  --- UDS: ACS $V_{606}$ and F814W ($I_{814})$, WFC3 IR $J_{105}$ and $H_{160}$
from the CANDELS program; IRAC 3.6 and 4.5~$\mu$m from the {\it Spitzer}
Extended Deep Survey (SEDS, PI. Fazio; Ashby et al. 2013); MIPS 24~$\mu$m from
the {\it Spitzer} UDS program (SpUDS, PI. Dunlop); ground-based $u^\ast$,
$g^\prime$, $r^\prime$, $i^\prime$, and $z^\prime$ data from the final data
release (``T0007'') of the CFHT Legacy Survey Wide component (CFHTLS-Wide);
the ground-based $J$, $H$, and $K_s$ data from the Ultra Deep Survey (UDS)
component of the UKIRT Deep Sky Survey (UKIDSS).

  --- EGS: ACS $V_{606}$ and $I_{814}$ from the All-wavelength Extended Groth
strip International Survey (AEGIS; Davis et al. 2007); WFC3 IR $J_{105}$ and
$H_{160}$ from the CANDELS program; ground-based $u^\ast$, $g^\prime$, 
$r^\prime$, $i^\prime$, and $z^\prime$ data from the final release of the
CFHTLS Deep component (CFHTLS-Deep); MIPS 24 and 70~$\mu$m from the FIDEL
program; IRAC 3.6 and 4.5~$\mu$m that incorporate the data from both the
SEDS program and the {\it Spitzer} Guaranteed Time Observing program 8
(Barmby et al. 2008).

   In addition, public X-ray and high-resolution radio images and/or catalogs
are also available in these fields, which improve the interpretation of
our sources. The GOODS-N has the {\it Chandra}\, 2~Ms data (Alexander et al.
2003) and the VLA 1.4~GHz data (Morrison et al. 2010), the UDS field has the
X-ray source catalog based on the XMM-Newton observations (Ueda et al. 2008)
and the VLA 1.4~GHz source catalog (Simpson et al. 2006), and the EGS field has
the Chandra 800~ks data from the AEGIS program (Nandra et al. in prep.; Laird
et al. 2009) and the VLA 1.4~GHz data (Ivison et al. 2007; Willner et al. 2012).

\section{Overview of Methods for Analysis}

   As the deep optical-to-NIR data reveal, there are always multiple candidate
counterparts for any 250~$\mu$m source in our sample. Our basic goal is to 
decompose the blended objects, to determine the major contributors to the FIR
emissions, and to reveal their nature by analyzing their optical-to-far-IR
SEDs. Here we briefly describe our methods.

\subsection{Source Decomposition and Flux Calibration}

   We used the GALFIT package developed by Peng et al. (2002) to do the
decomposition. While its wide usage by the community is mostly to study galaxy
morphologies, GALFIT has a straightforward capability of fitting the PSF
at multiple, fixed locations. This well suits our cases, because all the
potential components are effectively point sources at the angular resolutions
of the {\it Herschel} instruments. Although other PSF fitting softwares tailored
for crowded field photometry (such as ALLSTAR of DAOPHOT, Steston 1987), could
also be used in the decomposition of {\it Herschel} data (e.g., Rawle et al.
2012), we chose GALFIT because it produced the best results in our tests.
When the source S/N is sufficient, GALFIT
is capable of differentiating the offset of the input source position as small
as $\sim 1/10$ of a native pixel (Peng, priv.comm.). This corresponds to 
$\sim 2$\arcsec\, in the 250~$\mu$m band, and is comparable to the astrometric
accuracy (i.e., the accuracy of their centroids) of our bright 250~$\mu$m
sources. 

   As we are interested in constructing the FIR SEDs, we decomposed not only
the 250~$\mu$m but also the PACS 160 and 100~$\mu$m images when they are 
available. We also extended the same decomposition to the MIPS 70 and 24~$\mu$m
data, because the point-source assumption still largely holds at the MIPS
resolution. On the other hand, we did not attempt to decompose the SPIRE 350
and 500~$\mu$m data, because the beam sizes of these two bands are so large
that it is difficult to achieve reliable results using our current approach.
We generated the 250~$\mu$m PSF by using a symmetric 2D Gaussian of 
18.15\arcsec\ FWHM, which has already been proved to be a good fit to the point 
sources in the HerMES images (see e.g. Roseboom et al. 2010, Smith et al. 2011).
For the PACS images, we adopted the \emph{green} (for $100\mu$m) and \emph{red}
(for $160\mu$m) PSFs provided by the PEP team along with their DR1, which were
built by stacking a set of bright, isolated and point-like sources. The MIPS 
70 and 24~$\mu$m PSFs were created using the \texttt{daophot.psf} task in 
\texttt{IRAF}. For each field and band, a PSF was constructed from a set of
7 to 15 bright and isolated sources, which are all point sources as judged from
the WFC3 $H_{160}$ image. 

   For a given 250~$\mu$m source, we identified its potential components by 
searching for objects in the high-resolution near-IR or optical images within
$r=18$\arcsec\, of the 250~$\mu$m centroid. In other words, the diameter of the
searching area is twice the FWHM of the 250~$\mu$m beam. The positions of the
potential optical/near-IR components were then used as the priors for GALFIT to
extract the fluxes at these locations in various bands. This is different from
using the 24~$\mu$m source positions as the priors, and we did not
take the latter approach because, as mentioned in \S 1, it is not uncommon that
a single 24~$\mu$m source is actually a blended product of multiple objects
that are not necessarily related. In addition, using optical/near-IR position
priors has the advantage of being able to directly tie the FIR source to the
ultimate counterpart(s). Our first choice of the
detection image was the WFC3 $H_{160}$ image, and the second choice was the ACS
$I_{814}$ image if $H_{160}$ is not available. The third choice was the 
ground-based $i$-band image if it is more appropriate than the {\it HST} images
(for instance to avoid splitting one galaxy of complex morphology into several
sub-components at the {\it HST} resolution). Regardless of the exact choice,
the search always resulted in several tens of objects. Currently, it is not
desirable for our routine to deal with such a large number of objects because
there are still a large number of intermediate steps that require human 
intervention (see below). As only a fraction of these objects actually
contribute to the FIR emission, in this current work we chose to use the PACS
160 and 100~$\mu$m and the MIPS 24~$\mu$m images to narrow down the input list
for GALFIT. The general guideline was that only the objects that could have
non-negligible 24~$\mu$m emissions should be further considered. While we
could have developed some quantitative criteria for this process, we opted to
simply rely on visual inspection because this was only an interim step that we
are planning to replace in the near future. For this work, we were conservative
in eliminating objects from the input list, and included those that are at the
outskirts of the 24~$\mu$m source footprints.

    The above procedure narrowed the input lists to $\lesssim 10$ objects 
each. Similar to the philosophy of PSF fitting in crowded stellar fields
(e.g., Stetson 1987), one should seek to fit {\it simultaneously the exact
objects} that are contributing flux, because this is when the most accurate
result can be obtained. The implementation was rather difficult, however. In
our case, fitting all the objects from the input list simultaneously was often
not satisfactory. The symptoms were bad residual image after subtracting 
off the fitted objects, large errors associated with the decomposed fluxes, or
in some cases, the crashing of the decomposition process. Therefore, we took
two complementary approaches. 

    In the first approach, which we dub as the ``automatically iterative''
approach, all the potential components were still fitted simultaneously,
however those with their derived fluxes smaller than the associated flux
errors were deemed negligible and removed from the input list for the next
round. The simultaneous fit was repeated using the cleaned input list until
no negligible objects were left. This procedure usually converged after 2--3
iterations.  Among the surviving components, some could have fitted fluxes an 
order of magnitude smaller than the others. When this happened, a new fit was
performed using an input list that contained only the major components. If the
residual image was of the same quality (judged by $\chi^2$ and also visual 
inspection) as from the previous round, these major 
components were deemed as the only contributors and
those less important ones were ignored. Otherwise we kept the results from the
previous round. Finally, a ``sanity check'' was carried out on the surviving
objects by fitting them one at a time. Usually it was obvious that there
were residuals left at the locations of the other components, indicating that
more than one component contributes significantly to the source and hence the
simultaneous fit was necessary. However, in some cases the fit to only one 
object produced a clean residual image of the same quality as the one from the
simultaneous fit to multiple objects, and when this happened we flagged this
source as being a degenerate case and would offer alternative interpretations. 

    The other approach was highly interactive, which we call the 
``trial-and-error'' approach. This method had to be used when the automatically
iterative method failed at the first step: it either crashed or resulted in
unreasonable survivors with very large errors that could not be improved by
removing any of the survivors (i.e., the iteration failed). In this case, we
started from our best guess of the most likely counterpart, and fit for only
this object in the first round. We then checked the residual map to see if
there were any residuals left at the positions of any other objects in the
original input list. If yes, these objects were added to the fitting list, one
at a time, and the fit iterated until reaching the best result possible. The
iteration stopped when adding more objects either produced only negligible
contributors or started to produce unreasonable results. 

    We note that these two approaches can be integrated and further improved in
their implementation. In particular, it is possible to not only automate the 
trial-and-error approach but also use it to narrow down the input list without
the reference to additional data such as the MIPS 24~$\mu$m image. While it
is beyond the scope of this paper to fully develop this automatic routine,
we demonstrate its feasibility in APPENDIX.
We also note that we did not use the radio source positions as the priors. This
is because some objects that have non-negligible contributions to the FIR
emissions could fall below the sensitivity of the currently available
radio data.

    The PSF-fitting fluxes obtained in the procedures above should be corrected
for the finite PSF sizes, which could be achieved by comparing our PSF-fitting 
results to the curve-of-growth aperture photometry on a set of isolated point 
sources. In addition, the PACS images also suffer from light lost
caused by the application of the high-pass filtering in the reduction process,
which is not compensated in the currently released images and thus should also
be corrected. To simply the process, we did the following for this work.
For the PACS data, we derived the total correction factors for 100 and 
160~$\mu$m bands in each field by comparing our PSF-fitting results against
those in the PEP DR1 single-band catalogs. For the SPIRE data, we adopted the
photometry reported in the HerMES DR1 catalog as the total flux densities and 
proportionated among the multiple contributors according to our decomposition
results.

    To obtain realistic uncertainties in the {\it Herschel} bands, we
carried out extensive simulations.
For each source, we simulated it by adding the PSFs according to the final
decomposition result, put the simulated object in several hundreds of
random locations on the real image, ran the decomposition at these locations,
and obtained a distribution of the recovered fluxes. The dispersion of this
distribution was adopted as the instrumental errors. The errors due to the
confusion noise were then added in quadrature to the instrumental errors to 
obtain the final errors. For PACS 100 and 160~$\mu$m, we adopted the confusion
noise values of 0.15 and 0.68~mJy/beam from Magnelli et al. (2013), and
for the SPIRE data we inherited the values from the HerMES DR1 catalogs, which
are 5.8, 6.3 and 6.8~mJy/beam based on Nguyen et al. (2010). For our objects,
the confusion noise only contributes insignificantly to the total error in
the PACS bands, however it dominates the total error in the SPIRE bands. For
the sake of consistency, the errors in the MIPS 24 and 70~$\mu$m were obtained
in the same way but without the confusion term.

   As mentioned above, we did not apply the decomposition to the 350 and the 
500~$\mu$m data where the resolutions are much worse and the source $S/N$ is 
generally much lower than at the 250~$\mu$m. This means that generally these
two bands cannot be incorporated in our analysis. However, in few cases we
find that the vast majority of the 250~$\mu$m flux is from only one object
(others contribute a few per cent at most), or that the major contributors 
are likely at the same redshift and thus we can discuss their combined 
properties. For such sources, we directly use the 350 and the 500~$\mu$m
measurements from the HerMES DR1 catalogs in our study.

\subsection{SED Fitting}

   In order to understand the objects under question, their redshifts are
needed. While some have spectroscopic redshifts, most of our objects have to
rely on photometric redshift ($z_{ph}$) estimates. To derive $z_{ph}$, we used
the SEDs that extend from optical to the IRAC wavelengths (to 8.0~$\mu$m when
possible). The inclusion of the IRAC data was important but often
non-trivial. As many of our sources have input objects that are close to each
other, they are usually blended in the IRAC images and thus need to be
decomposed as well.  For most objects, the point-source assumption does not
hold anymore in IRAC. Therefore, we used the TFIT technique (Laidler et al.
2006) for this purpose, where the image templates were constructed from the 
$H_{160}$ images, and then were convolved by the IRAC PSFs to fit the IRAC
images.

    We used the Hyperz software (Bolzonella et al. 2000) and the stellar
population synthesis (SPS) models of Bruzual \& Charlot (2003; hereafter BC03)
to estimate $z_{ph}$. The latest implementation of Hyperz also includes the
tool to derive other physical quantities such as the stellar mass ($M^\ast$),
the age ($T$) etc. (``hyperz\_mass'', M. Bolzonella, priv. comm.) from the 
models. We adopted the BC03 models of solar metallicity and the initial mass
function (IMF) of Chabrier (2003), and used a series of exponentially declining
star formation histories (SFHs) with $\tau$ ranging from 1~Myr to 20~Gyr. The
simple stellar population (SSP) model was also included, which was treated as
$\tau=0$. The models were allowed to be reddened by dust following the
Calzetti's law (Calzetti 2001), with $A_V$ ranging from zero to 4~mag.

    The above SED fitting procedure does not consider the effect of emission
lines. To assess how this might impact the derived physical properties, we
also used the LePhare software (Arnouts \& Ilbert 2007) to analyze the same
SEDs. LePhare can calculate the strengths of the common emission lines based
on the star formation rate (SFR) of the underlying SPS models, which we chose
to be the same sets of models used in the Hyperz analysis. The results derived
by LePhare in most cases are very close to those derived using Hyperz, and thus
we only include the discussion of the Hyperz results in this paper.

    A key quantity in revealing the nature of our sources is the total IR 
luminosity, $L_{IR}$, traditionally calculated over restframe 8 to 1000~$\mu$m.
To derive $L_{IR}$, we used our own software tools to fit the 
mid-to-far-IR SED to the starburst models of Siebenmorgen \& Kr{\"u}gel
(2007; hereafter SK07) at either the spectroscopic redshift,
when available, or the adopted $z_{ph}$ from above. We adopted their
``9kpc'' and ``15kpc'' models, which SK07 produces for the starbursts at high
redshifts. When possible, we also derive the dust temperature ($T_d$) and mass
($M_d$) by using the code of Casey (2012), which fit the FIR SED to a modified
blackbody spectrum combined with a power law extending from the mid-IR. As we
only have limited passbands, we fixed the slope of the mid-IR power law to
$\alpha=2.0$ and the emissivity of the blackbody spectrum to $\beta=1.5$
(Chapman et al. 2005; Pope et al. 2006; Casey et al. 2009). This approach
usually does not produce a good fit at 24~$\mu$m, presumably due to the fixed
$\alpha$. Therefore, we ignored the 24~$\mu$m data points when deriving $T_d$
and $M_d$. The code derives two different dust temperatures,
one being the best-fit temperature of the graybody ($T_d^{fit}$) and the other
being the temperature according to Wien's displacement law that corresponds to
the peak of the emission as determined by the best-fit graybody ($T_d^W$).
We adopted $T_d^{fit}$. Finally, we also obtained the gas mass, $M_{gas}$, by
applying the nominal Milky Way gas-to-dust-mass ratio of 140 (e.g. Draine 
et al. 2007), with the caveat that this ratio could strongly depend on the
metallicities (e.g., Draine et al. 2007; Galametz et al. 2011;
Leroy et al. 2011).

   Another important quantity is the SFR. SED fitting in the optical-to-NIR
regime provides the SFR intrinsic to the SFH of the best-fit BC03 model
(hereafter $SFR_{fit}$), which naturally takes into account the effect of
dust extinction. However, there could still be star formation processes
completely hidden by dust, which would not be counted by $SFR_{fit}$ and could
only be estimated through the measurement of $L_{IR}$. A common practice is to
exercise the conversion given by Kennicutt (1998), which is
$SFR_{IR}=1.0\times 10^{-10}L_{IR}$ after adjusting for a Chabrier IMF (see
e.g., Riechers et al. 2013)
\footnote{Using the conversion for a Salpeter IMF, the SFR will be a factor of
1.7 higher.}. This conversion assumes solar metallicity and is valid for
a star-bursting galaxy that has a constant SFR over 10--100~Myr and whose
dust re-radiates all of the bolometric luminosity. The total SFR would then be
some combination of $SFR_{fit}$ and $SFR_{IR}$. However, this should not be
a straightforward sum of the two. The reason is that our measured $L_{IR}$ 
includes not only the IR emission from the region completely blocked by dust
(hereafter $L_{IR}^{blk}$) but also the contribution from the exposed region 
where a fraction of its light is extincted by dust and is re-radiated in the
FIR (hereafter $L_{IR}^{ext}$), i.e., $L_{IR}=L_{IR}^{blk}+L_{IR}^{ext}$.
In other words, $SFR_{IR}$ derived by applying the Kennicutt's conversion
directly to $L_{IR}$ would have already included part of the contribution from
$SFR_{fit}$. To deal with this problem, we calculated $L_{IR}^{ext}$ by
integrating the difference between the reddened and the de-reddened spectra
from the best-fit BC03 model and assuming that this amount of light is
completely re-radiated in the FIR. We then obtained
$L_{IR}^{blk}=L_{IR}-L_{IR}^{ext}$, and calculated 
$SFR_{IR}^{blk}=1.0\times 10^{-10}L_{IR}^{blk}$, where $SFR_{IR}^{blk}$ is
the ``net'' SFR in the region completely blocked by dust.

   From the derived SFR and $M^\ast$, one can calculate the specific SFR as
$SSFR=SFR_{tot}/M^\ast$. The caveat here is that the stellar mass derived by
SED fitting as mentioned above is only for the relatively exposed region and
does not include the completely obscured region. Nevertheless, we can still use
a related but more appropriate quantity in this context, namely the stellar
mass doubling time ($T_{db}$), which is defined as the time interval necessary
to further assemble the same amount of stellar mass of the exposed region
should the galaxy keeps its SFR constant into the future. Specifically, we can
obtain $T_{db}^{tot}=M^\ast/SFR_{tot}$ and $T_{db}^{blk}=M^\ast/SFR_{IR}^{blk}$.
Comparing the latter quantity to the age ($T$) of the existing stellar
population in the exposed region is particularly interesting, as this is a 
measure of the importance of the completely dust-blocked region in the
future evolution of the galaxy.

   Finally, the measurement of $L_{IR}$ enables us to examine the well-known
FIR-radio relation (see Condon 1992 for review) for the sources that have radio
data. We used the conventional formalism in Helou et al. (1985), which takes
the form of
\begin{displaymath}
q_{IR}=\log[(S_{IR}/3.75\times 10^{12}W m^{-2} Hz^{-1})/(S^0_{1.4GHz}/W m^{-2}Hz^{-1})],
\end{displaymath}
where $S_{IR}$ is the integrated IR flux while $S^0_{1.4GHz}$ is the radio flux
density at the restframe 1.4~GHz. For the k-correction in radio, we assume
$S_\nu \propto \nu^{-0.8}$. The original usage of $S_{IR}$ (the quantity
``FIR'' in Helou et al. 1985) is defined between the restframe 42.5 and
122.5~$\mu$m. In order to take the full advantage of the {\it Herschel}
spectral coverage, we opted to adopt $S_{IR}$ as in Ivison et al. (2010), where
this quantity is defined from 8 to 1000~$\mu$m in the restframe. In practice,
we calculated $S_{IR}$ using the best-fit SK07 model. For reference, the mean 
value that Ivison et al. (2010) obtain using {\it Herschel} sources in the 
GOODS-N is $q_{IR}=2.40\pm 0.24$.

\section{Sources in the GOODS-N}

\subsection {GOODSN06 (J123634.3+621241)}

  This source is in the SMG sample of Wang et al. (2004; their GOODS 850-19),
however it was detected at $<4 \sigma$ level
($S_{850}=3.26\pm 0.85$~mJy) and was not included in the SMA observations of 
Barger et al. (2012).

\subsubsection {Morphologies and Potential Components}

   Fig. 2 shows this source in various bands from 250~$\mu$m to $H_{160}$.
Within 18\arcsec\, radius, there are 74 objects in $H_{160}$ that have 
$S/N\geq 5$ (measured in the MAG\_AUTO aperture). By inspecting the FIR and 
the 24~$\mu$m images, it is obvious that only six of these objects could
possibly contribute to GOODSN06. In order of the distances between their 
centroids to the 250~$\mu$m position, these potential components are marked 
alphabetically from ``{\tt A}'' to ``{\tt F}''. 
Objects {\tt A}, {\tt B} and {\tt C}, which are $1\farcs 38$, $2\farcs 50$, 
and $3\farcs 15$, away from the 250~$\mu$m position, respectively, are 
completely blended in the 24~$\mu$m image as one single source. This is the 
brightest 24~$\mu$m source within 18\arcsec\, ($S_{24}=446\pm 5$~$\mu$Jy), and
also dominates the flux in 70~$\mu$m. The centroid of this single 24~$\mu$m
source also coincident with the centroid of the emission in 100, 160 and
250~$\mu$m. In the 8.0~$\mu$m image, 
{\tt A} and {\tt C} are separated and both are well detected, and hence it is
reasonable to assume that they both contribute to the 24~$\mu$m flux. {\tt B}
is still blended with {\tt A} in the 8.0~$\mu$m image, and we cannot rule it
out as a contributor to the 24~$\mu$m flux. Under our assumption, this means 
that these three objects could all be contributors to the FIR emission. 

   As it turns out, {\tt A} and {\tt C}
have spectroscopic redshifts of 1.224 and 1.225, respectively
(Barger et al. 2008). Their separation is $2\farcs 64$, which corresponds to
22.1~kpc at these redshifts, and is well within the scale of galaxy groups.
This suggests that they could indeed be associated.
The high resolution {\it HST} images reveal that both {\tt A} and {\tt C} have
complicated morphologies and that {\tt B} could be part of {\tt A}. This is
show in Fig. 3. While it is difficult to be separated
photometrically, {\tt A} is actually made of three sub-components, which we
label as ``{\tt A-1}'', ``{\tt A-2}'' and ``{\tt A-3}'' (Fig. 3, left). The ACS
images show that {\tt A-1} has a compact core and a one-side, curved tail. 
{\tt A-2} is invisible in the ACS optical images, but it is the most 
prominent of the three in the WFC3 IR, and also has a compact core. {\tt A-3}
is similar to {\tt A-1} in the WFC3 IR but extends to the opposite direction
and is much fainter. The one-side, curved tail of {\tt A-3} is invisible in the
ACS optical. The core of {\tt A-2} has nearly the same angular separation from
those of {\tt A-1} and {\tt A-3}, which is about 0\arcsec.38, or 3.2~kpc at 
$z=1.22$. Therefore, {\tt A} is likely a merging system. In fact, 
{\tt B} could well be part of this system, as it seems to connect with the tail
of {\tt A-3}. {\tt C} looks smooth and regular in the WFC3 images, however in
the higher resolution ACS images it shows two nuclei, which is also indicative
of merging process (Fig. 3, right). All this suggests that it will be 
reasonable to consider the {\tt A/B/C} complex as a single system when treating
the FIR emission.

     While it is only marginally detected in 24~$\mu$m, component {\tt D}
seems to have non-negligible fluxes in both 70 and 100~$\mu$m. This object is
invisible in $B_{435}$ and $V_{505}$, and is marginally detected in $i_{775}$,
but starts to be very prominent in $z_{850}$ and in the WFC3 IR and the IRAC
bands. Its morphology in $z_{850}$ is rather irregular and shows two 
sub-components in $Y_{105}$ and $J_{125}$. However in $H_{160}$ it looks like
a normal disc galaxy.  While all this suggests that it could be a dusty galaxy
and thus could have non-negligible FIR emission, its position is $8\farcs 88$
offset from the the 250~$\mu$m centroid, and thus is less
likely a major contributor to the 250~$\mu$m flux.  

     Finally, {\tt E} and {\tt F} cannot be associated with any of the above 
because they have spectroscopic redshifts of 0.562 and 0.9617, respectively
(Barger et al. 2008). While both of them are prominent sources in 24~$\mu$m, 
only {\tt E} is well visible in 160~$\mu$m and thus could be a significant
contributor to the FIR emission.

\subsubsection{Optical-to-near-IR SED Analysis}

     The optical-to-near-IR SED analysis for objects {\tt A} to {\tt D}
follows the procedure outlined in \S 3.2. The SED was constructed based on the
photometry in the ACS, the WFC3 IR and the IRAC bands. The ACS and
the WFC3 photometry were obtained by running the SExtractor software 
(Bertin \& Arnouts 1996) in dual-image mode on the set of images that are 
registered and PSF-matched to the $H_{160}$-band. The $H_{160}$ image was used
as the detection image, and the colors were measured in the {\tt MAG\_ISO}
apertures. The $H_{160}$ {\tt MAG\_AUTO} magnitudes are then adopted as the
reference to convert the colors to the magnitudes that go into the SED. The
IRAC magnitudes are the TFIT results using the $H_{160}$ image as the 
morphological template and the PSF derived from the point sources in the
field.

    The results are summarized in Fig. 4, which include $z_{ph}$ and the
physical properties of the underlying stellar populations such as the stellar
mass ($M^*$), the extinction $A_V$, the age ($T$), the characteristic star
forming time scale ($\tau$), and the dust-corrected star formation rate (SFR).
This figure also shows $P(z)$, which is the probability density function of
$z_{ph}$, for all these objects. The available $z_{spec}$ agree with our 
$z_{ph}$ reasonably well. The largest discrepancy happens in {\tt C}, which has 
$\Delta z=0.185$, or $\Delta z/(1+z)=0.08$. This discrepancy is explained by
the secondary $P(z)$ peak at $z_{ph}\sim 1.2$. Object {\tt B}, a very close 
neighbor to {\tt A}, has $z_{ph}=0.93$, which may seem to suggest that it is
not related to {\tt A}. However, its $P(z)$ is rather flat over all redshifts
and does not have any distinct peak, and therefore its $z_{ph}$ cannot be
used as a strong evidence to argue for or against its relation with {\tt A}.
From these results and the morphologies, it is reasonable to believe that
{\tt B} is only a less important satellite to {\tt A}.
Object {\tt D} has
$z_{ph}=1.34\pm 0.04$, and is different from $z_{spec}$ of the {\tt A/C}
complex by $\Delta z/(1+z)=0.05$. This is within the accuracy of our $z_{ph}$
technique, suggesting that {\tt D} could be in the same group.  Nevertheless,
in the following analysis we still treat it separately.
    
\begin{figure*}[tbp]
\centering
\includegraphics[width=\textwidth]{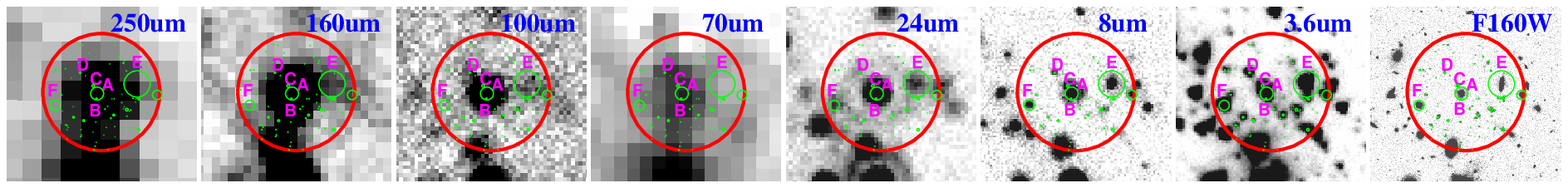} \\
\includegraphics[width=\textwidth]{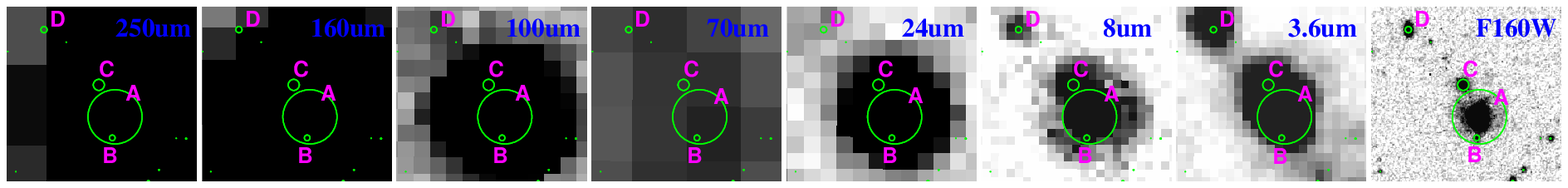}
\caption{Vicinity of GOODSN06 (top) and zoomed-in view of its possible
contributors (bottom) in SPIRE 250~$\mu$m, PACS 160 and 100~$\mu$m, MIPS 70 and
24~$\mu$m, IRAC 8 and 3.6~$\mu$m, and WFC3 $H_{160}$. North is up and east is
to the left. The red circles have 36$\arcsec$ diameter, twice the beam FWHM
at 250~$\mu$m. The objects identified in $H_{160}$ within this radius are 
marked by the green circles whose sizes are proportional to their $H_{160}$
brightnesses. The possible contributors to the 250~$\mu$m emission are labeled
by the letters in purple.
}
\end{figure*}

\begin{figure*}[tbp]
\centering
\subfigure{
  \includegraphics[width=0.5\textwidth]{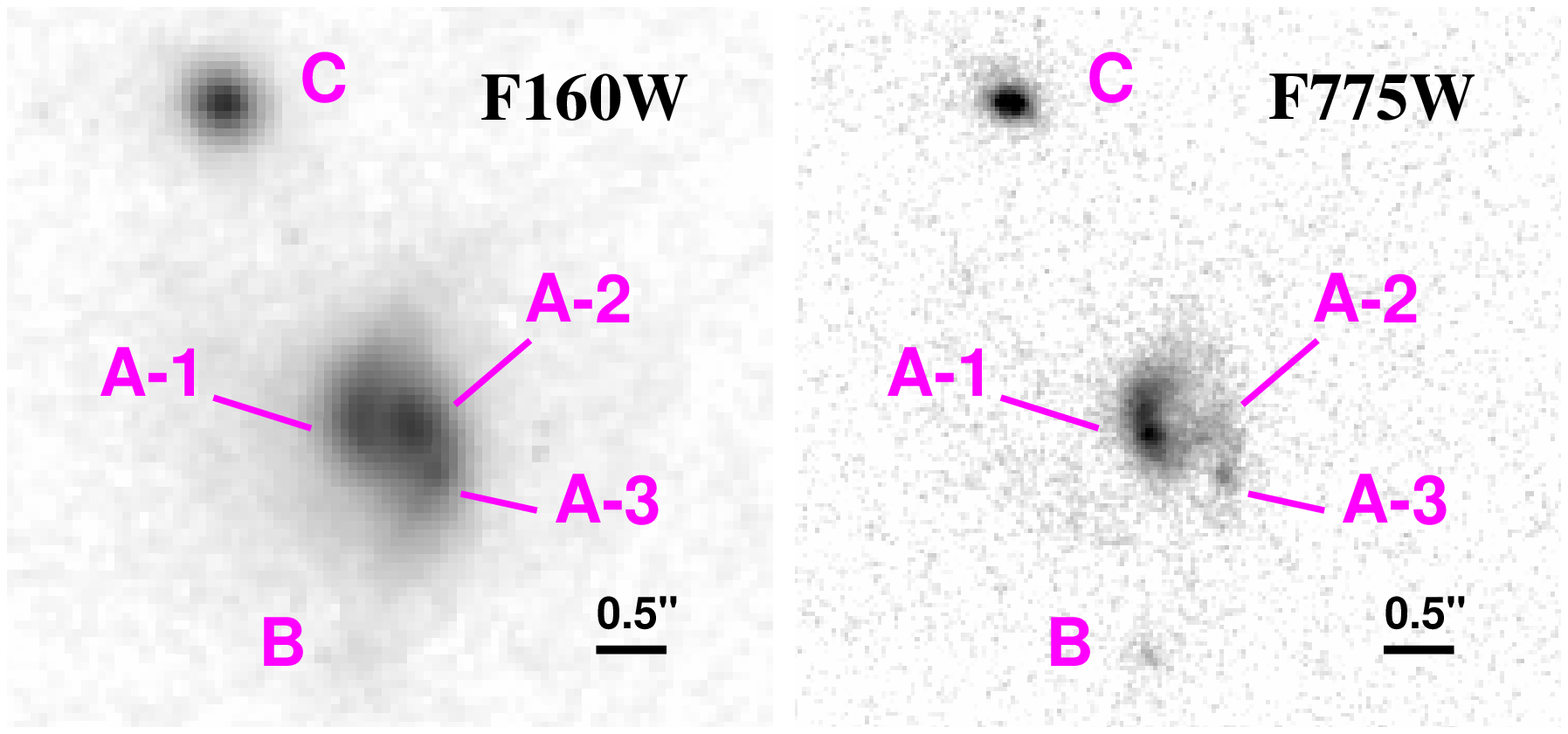} 
  \includegraphics[width=0.5\textwidth]{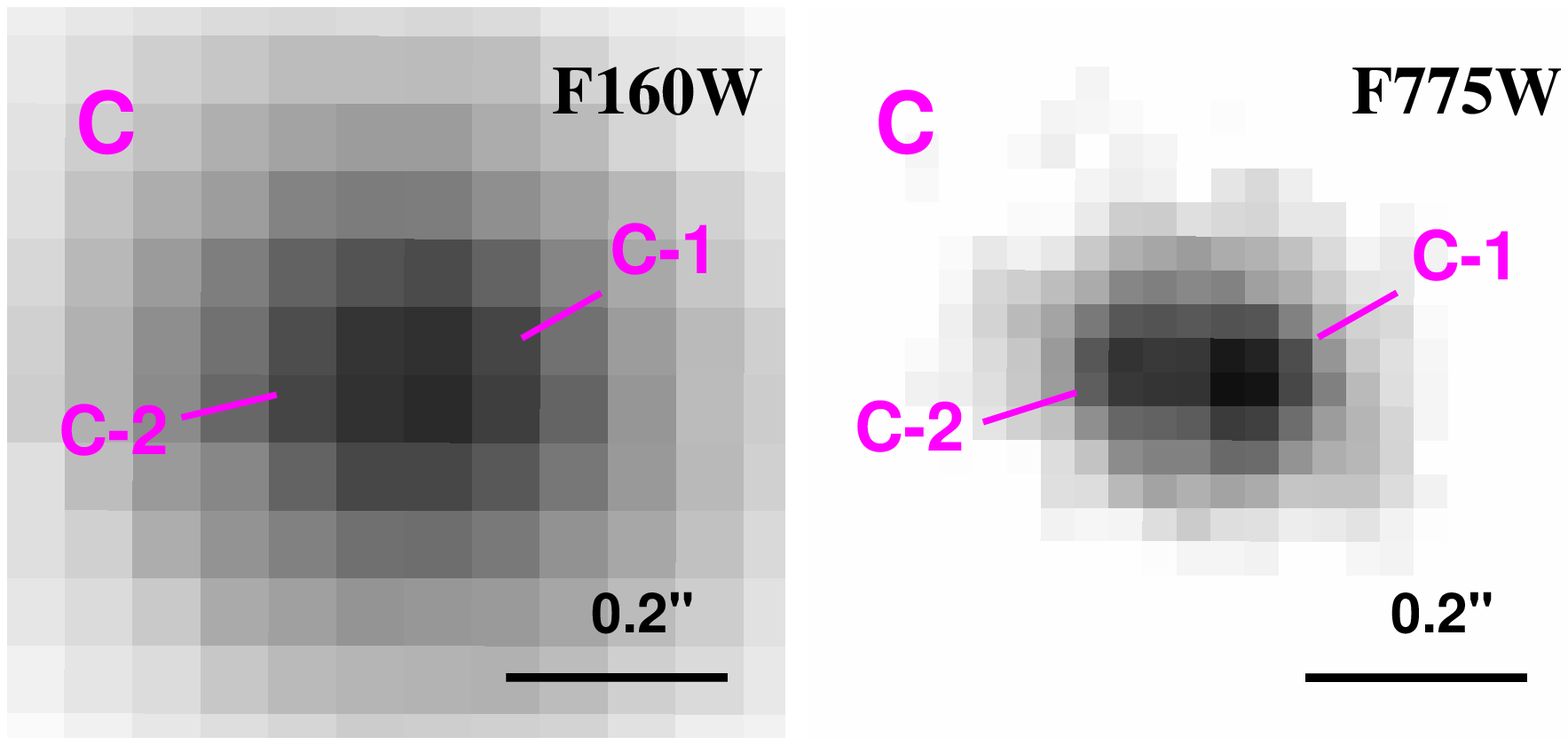}
}
\caption{Detailed morphologies of {\tt A} (left) and {\tt C} (right) in 
$H_{160}$ and $i_{775}$. The sub-components are labeled numerically. {\tt A-2}
is prominent in NIR but is very weak in optical, indicating that it is obscured
by dust. {\tt C} shows double nuclei in the higher resolution $i_{775}$
image.
}
\end{figure*}

\begin{figure*}[btp]
\centering
\includegraphics[width=\textwidth]{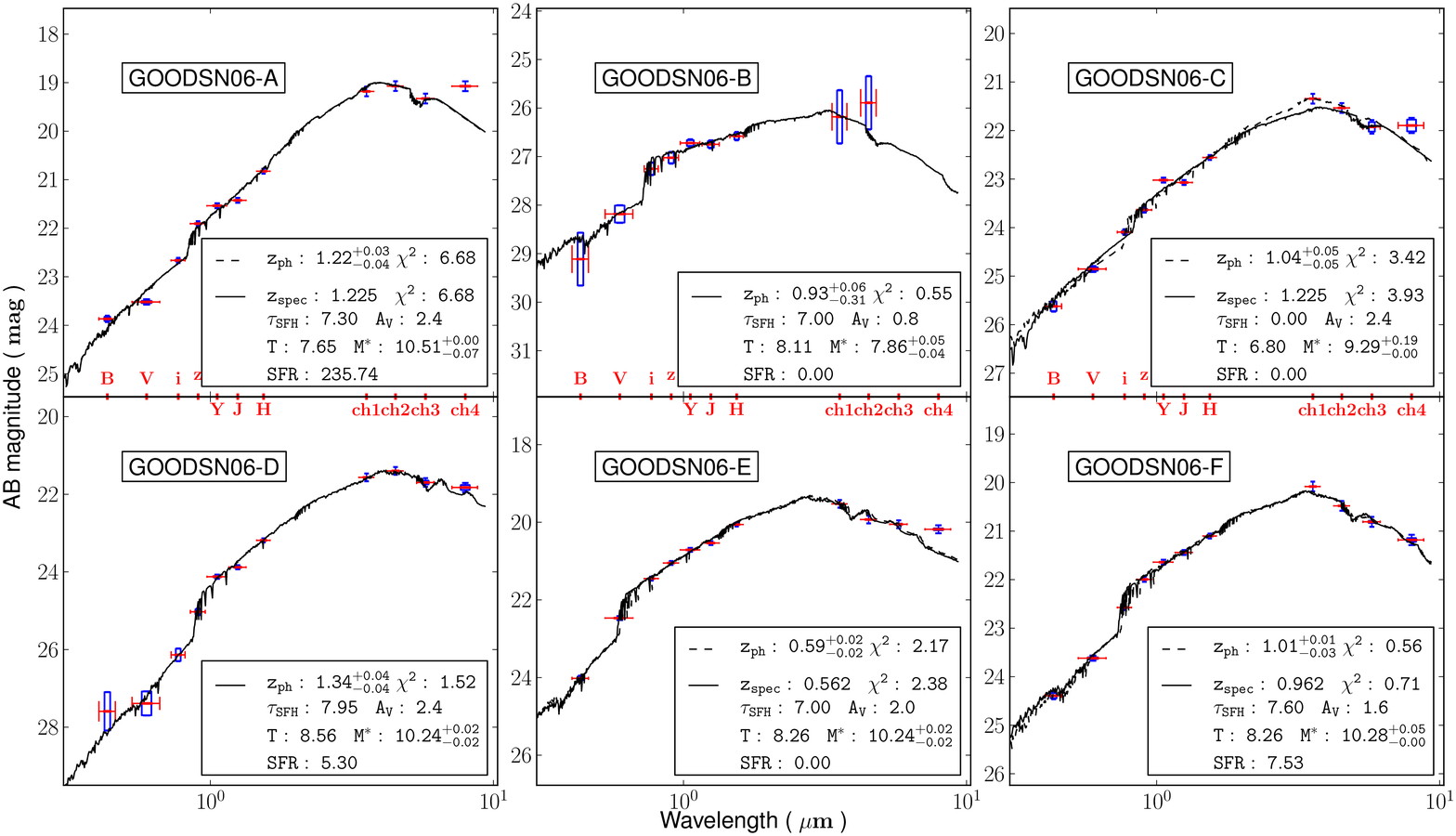} 
\includegraphics[width=0.5\textwidth]{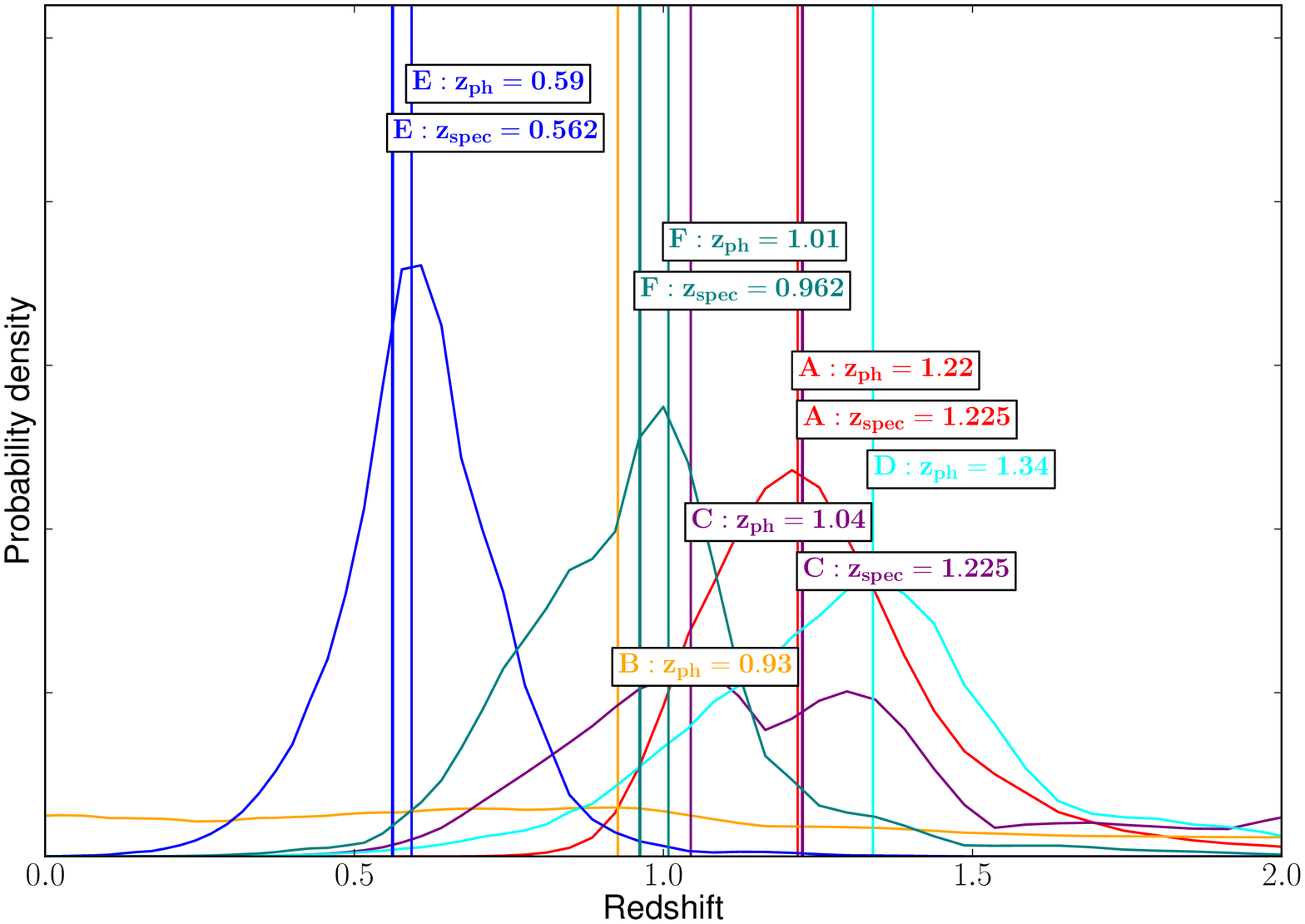}
\caption{Optical-to-NIR SED analysis of the possible contributors to GOODSN06.
The top panel summarizes the SED fitting results of these objects individually.
The data points are shown together with the best-fit BC03 models. For the 
objects that have $z_{spec}$, the solid curves are the best-fit model at this
redshift, while the dashed curves are the best-fit model when $z$ is a free
parameter. The most relevant physical properties inferred from the models
are summarized in the boxed region: reduced $\chi^2$, $\tau$ (in Gyr), $A_V$,
$T$ ($\log T$ in year), $M\ast$ ($\log M^\ast$ in $M_\odot$), and $SFR_{fit}$
(in $M_\odot/yr$).
The bottom panel shows the probability distribution functions of the 
photometric redshifts ($P(z)$) of these objects, which are distinguished by
different colors. The best-fit $z_{ph}$ and the $z_{spec}$ (when available)
values are shown in boxes, and are also marked by the vertical lines. 
}
\end{figure*}

\subsubsection{Decomposition in Mid-to-Far-IR}

   The decomposition of GOODSN06 was done in the SPIRE 250~$\mu$m, the PACS
160 and 100~$\mu$m, and the MIPS 70 and 24~$\mu$m. This source has two 
neighbors to the south (``{\tt S1}'' and ``{\tt S2}''), which, while being
outside of $r=18$\arcsec\, , could still contaminate the target in the 250, 160
and 70~$\mu$m bands where the beam sizes are large. While in other cases we
only fit the objects within $r=18$\arcsec\, , for this source we must consider
these two contaminators that are further out. We adopted their positions in 
$H_{160}$, as there is no ambiguity in their identifications, and fit them 
together with all the components of GOODSN06 when decomposing in
these three bands. They are well separated from the target in 100 and 
24~$\mu$m, and hence did not enter the decomposition process in these two bands.

   For illustration, Fig. 5 shows the decomposition in 250~$\mu$m. The
simultaneous fit to all the five objects within $r=18$\arcsec\, ({\tt A} to
{\tt E}) resulted in {\tt B} and {\tt C} as the only two survivors. However,
GALFIT reported very large flux errors for these two objects, and their
extracted fluxes would have very low GALFIT fitting S/N ($<3$ and $<0.5$, 
respectively). Removing either of
the two did not improve the result. Therefore we had to use the
trial-and-error approach. Judging from the mid-IR images
(\S 4.1.1, in particular the 24~$\mu$m image) and from the 
optical-to-NIR SED analysis (\S 4.1.2), it is reasonable to conclude that
{\tt A}, {\tt B} and {\tt C} are associated, with {\tt A} being by far the
most dominant. The forced fit for {\tt A} only, however, showed a severe
oversubtraction and yet a clear residual at the location of {\tt E}. The
forced fit at {\tt A} and {\tt E} significantly improved, however the
extraction at {\tt A} still had a large error. When we included the two
southern neighbors ({\tt S1} and {\tt S2})
and fit them together with {\tt A} and {\tt E}, the error
of {\tt A} was greatly reduced. If we left out {\tt E}, the residual
persistently showed up at this location, indicating that the contamination 
from {\tt S1} and {\tt S2} is not the reason and that the inclusion of {\tt E}
is necessary. The fit with {\tt A}, {\tt E}, {\tt S1} and {\tt S2} seems to
be the best among all possibilities, and forcing the fit to additional
objects would create completely non-physical results. Therefore, we adopted
the {\tt A+E+S1+S2} scheme, which concludes that {\tt A} contributes 82\% of
the total flux to GOODSN06 and {\tt E} contributes the other 18\%.

\begin{figure*}[btp]
\centering
\includegraphics[width=\textwidth]{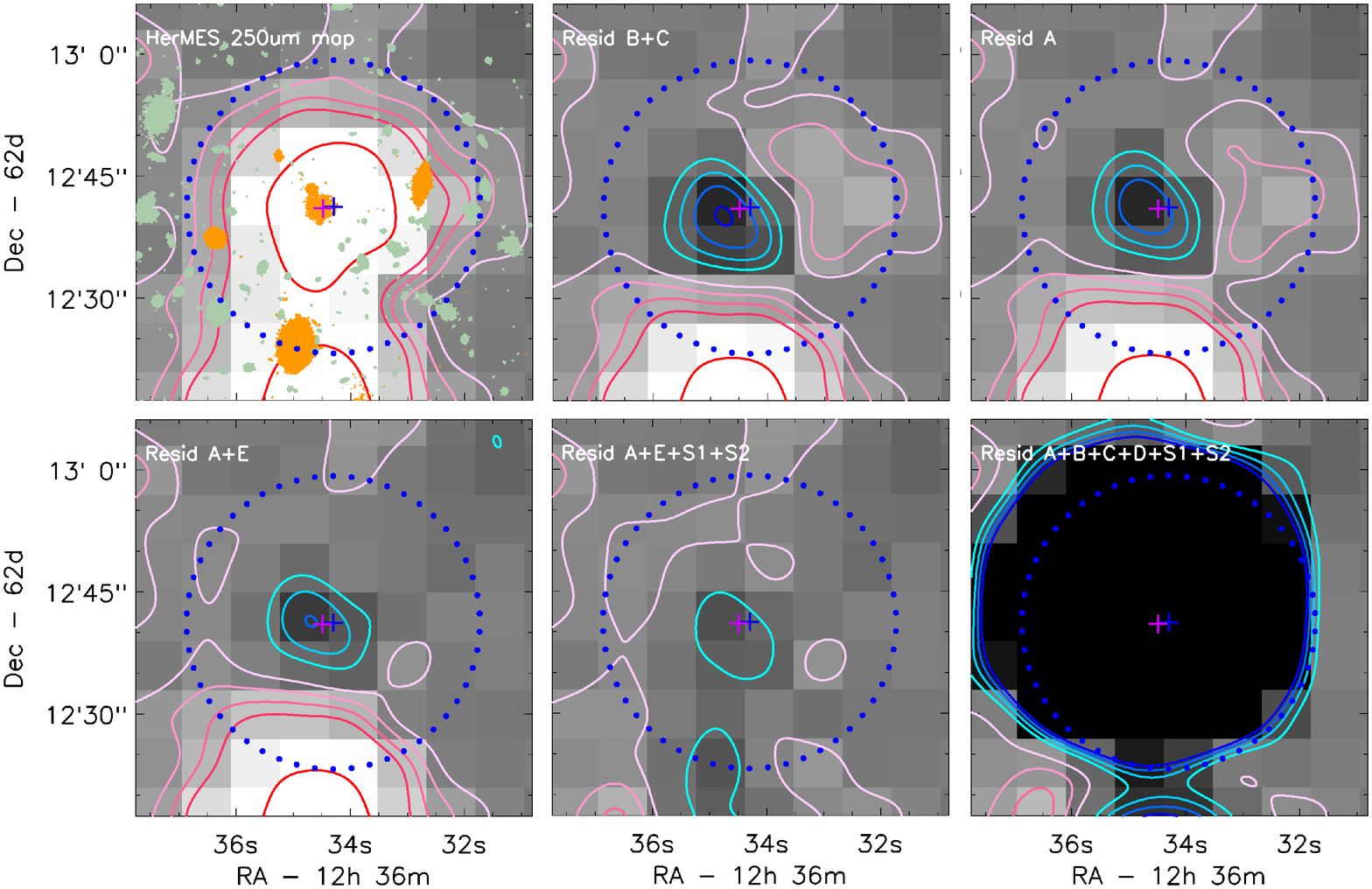}
\caption{Demonstration of the decomposition in 250~$\mu$m for GOODSN06. The
first panel shows the original 250~$\mu$m image, while the others show the
residual maps of the different decomposition schemes where different input 
sources are considered (labeled on top). The ``{\tt B+C}'' panel is for the
automatically interactive fit that fails to provide a satisfactory result, while
the rest are for different trial-and-error runs where the input objects are
interactively decided. The best result is produced by the ``{\tt A+E+S1+S2}''
scheme (see text). All images are displayed in positive, i.e., using white for
the positive values and black for the negative values. The contours coded in
``warm'' and ``cold'' colors indicate the different positive (i.e., residuals)
and negative (i.e., over-subtraction) levels, respectively: the positive level increases from light pink to dark red,
and the negative level increases from light green to deep blue. The contour
starts at the level of $\pm$ 0.4~mJy/pix, and increases or decreases at a step
size of 0.4~mJy/pix. In the first panel that shows the original 250~$\mu$m
image, the $H_{160}$-band objects are superposed as the gray-green segmentation
maps, among which the candidate counterparts are further highlighted in orange.
In all panels, the blue, dotted circle is 18\arcsec\, in radius. The blue and
the magenta crosses mark the 250~$\mu$m source centroid and the VLA 1.4~GHz
source position, respectively.
}
\end{figure*}
   
   The automatically iterative decomposition procedure was successful in other
bands. In 160~$\mu$m, it converged on {\tt A+E}, and introducing {\tt S1}
further reduced the errors and improved the residuals. In 100~$\mu$m, it
converged on {\tt A+D+E} and did not need to include the southern neighbors. 
In 70~$\mu$m, it converged on {\tt A+E+F}, and including both {\tt S1} and
{\tt S2} further improved the results. In 24~$\mu$m, it converged on
{\tt A+E+F}. In all these bands, {\tt A} is the most dominant 
object as well.

    The flux densities of the major component {\tt A} based on the above
decomposition results are summarized in Table 2.
The VLA 1.4~GHz data from Morrison et al. (2010), which have the positional
accuracy of $0\farcs 02$--$0\farcs 03$,
reveal a source with $S_{1.4GHz}=0.201\pm 0.010$~mJy at $RA=12^h36^m34^s.49$,
$DEC=62^o12^{'}41^{''}.0$, and is right at the location of {\tt A-2}. {\tt A} is
also a moderate X-ray source in Alexander et al. (2003), which has rest-frame
0.5--2~keV luminosity of $L_{0.5-2}=8.1\times 10^{41}$~erg/s 
(Barger et al. 2008), and hence most likely is powered by stars rather than
AGN (see e.g., Georgakakis et al. 2007).

\subsubsection{Total IR Emission and Stellar Populations}

\begin{figure*}[btp]
\centering
\subfigure{
  \includegraphics[width=0.5\textwidth]{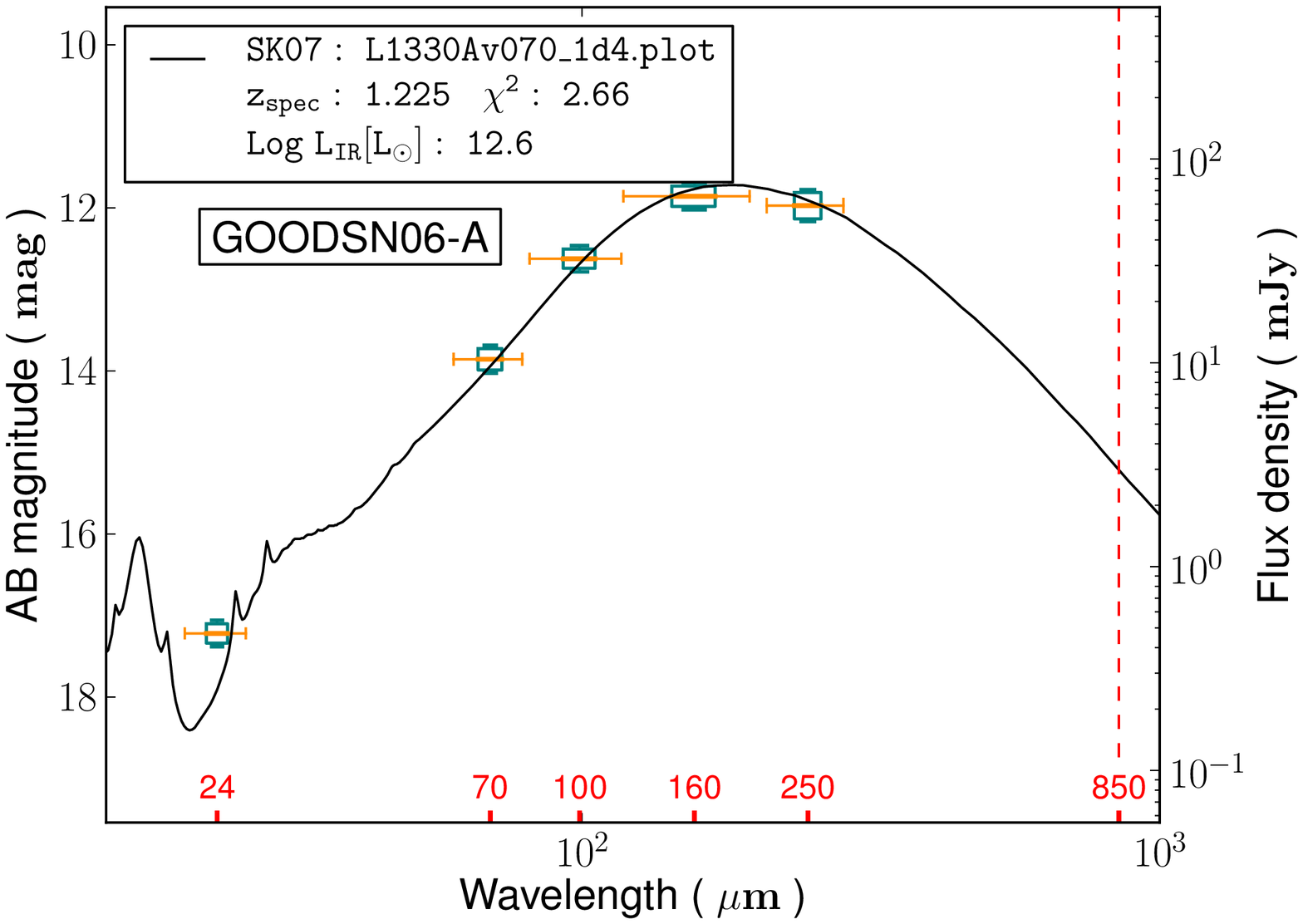}
  \includegraphics[width=0.5\textwidth]{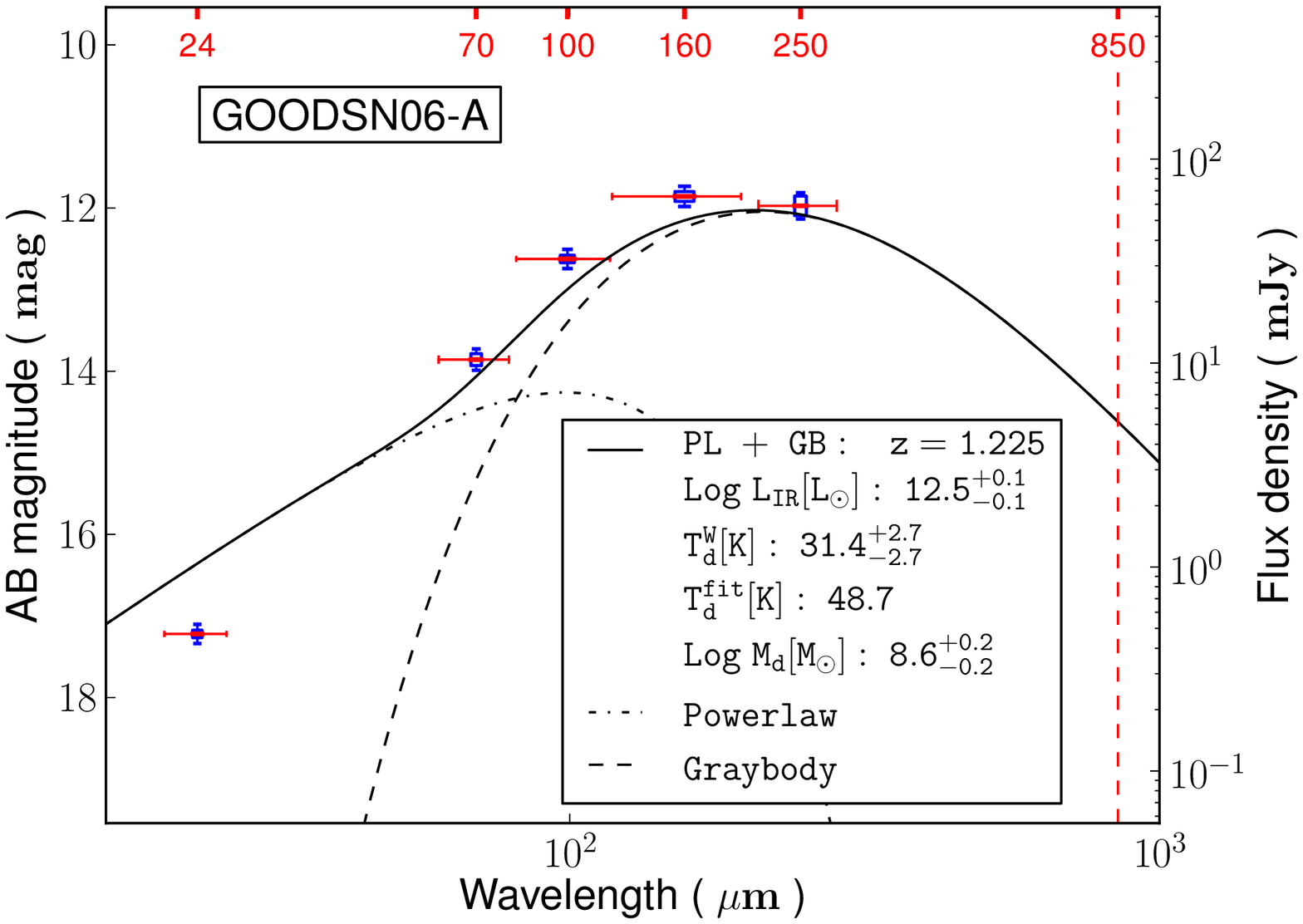}
}
\caption{Mid-to-FIR SED fitting of GOODSN06-{\tt A} at $z_{spec}=1.225$.
The relevant parameters derived from the fits are summarized in the boxed 
regions. The left panel shows the fit to the SK07 models to derive $L_{IR}$,
where the green boxes are the data points and the superposed curve is the 
best-fit model. The right panel shows the fit to derive the dust temperature
and the dust mass, where the analytic models are made of a power law and a 
graybody (see \S 3.2). The data points are shown in blue. The best-fit model is
shown as the black, solid curve, while its power-law and graybody components
are shown as the dot-dashed and the dashed curves, respectively. In both 
panels, the most relevant best-fit parameters are summarized in the boxes. The
red vertical dashed lines are at 850~$\mu$m, which is the wavelength where
SMGs are selected.
}
\end{figure*}

   Fig. 6 shows our fit to the mid-to-far-IR SED, which results in 
$L_{IR}=4.0\times 10^{12} L_\odot$ at $z_{spec}=1.225$. This is in the ULIRG
regime. Based on the optical-to-near-IR SED fitting in \S 4.1.2., we integrated
the difference between the reddened, best-fit model template spectrum of {\tt A}
and its de-reddened version (see \S 3.2), and obtained
$L_{IR}^{ext}=2.3\times 10^{12} L_\odot$. The ``net'' IR luminosity from the
region completely blocked by dust is therefore
$L_{IR}^{blk}=1.7\times 10^{12} L_\odot$, which implies 
$SFR_{IR}^{blk}=170$~$M_\odot/yr$. The best-fit BC03 model for {\tt A} in 
\S 4.1.2 gives $SFR_{fit}=236$~$M_\odot/yr$. Thus $SFR_{tot}=406$~$M_\odot/yr$.
While the {\tt A/B/C} complex is very likely an interacting system, it is 
interesting to notice that the ULIRG phenomenon only happens to {\tt A}, and
that {\tt B} and {\tt C} are not triggered. Furthermore, it seems that the
current star formation activities all concentrate on
{\tt A}. The best-fit model for {\tt C} is a SSP, which formally has no 
on-going star formation. It is plausible that {\tt A} itself is an on-going
merger (judging from its complex morphology) and hence is undergoing an
intense starburst, while its companion {\tt C} and the smaller satellite
{\tt B} are still on the way falling to it.

   The stellar mass of this system is also dominated by {\tt A}, for which we
obtained $3.2 \times 10^{10}$~$M_\odot$ (\S 4.1.2). The stellar mass of {\tt C}
is one order of magnitude less at $1.9 \times 10^9$~$M_\odot$ and is 
negligible. Therefore, this implies $SSFR=12.7$~Gyr$^{-1}$. Using the stellar
mass doubling time scales as discussed in \S 3.2., we obtained
$T_{db}^{tot}=79$~Myr and $T_{db}^{blk}=188$~Myr, respectively.
Interestingly, the optical-to-near-IR SED fitting for {\tt A} in \S 4.1.2 gives
an age of $T=45$~Myr. Its SFH has $\tau=20$~Myr, which is comparable to this
best-fit age. In other words, the SFH of the exposed region is a rather short
burst, and this every young galaxy has been in star-bursting mode ever since
its birth, although it is unclear whether it could be an ULIRG in the earlier
time. While the star formation in the completely dust-blocked region is
playing a less important role as the compared to that in the exposed region
($T_{db}^{blk}$ is more than a factor of four longer than $T$), it will still
be significant should the gas reservoir suffice (see below).

   The fit for the dust temperature and the dust mass is shown Fig. 6.
$T_d^{fit}=48.7$~K is on the high side of the usual dust temperatures of the 
SMGs, which are around $\sim 30$--40~K (e.g., Chapman et al. 2005, 2010; 
Pope et al. 2006; Magnelli et al. 2010, 2012; Swinbank et al. 2013).
In fact, as mentioned earlier, GOODSN06 is only a marginal
SMG and has $S_{850}=3.26\pm 0.85$~mJy. This is a demonstration that the
{\it Herschel}\, bands are less biased against ULIRGs of high dust temperature.
From the best-fit SK07 model, one would
predict $S_{850}=2.29$~mJy, in agreement with the actual measurement to 
1.14~$\sigma$. The dust mass we obtained is $M_d=3.7\times 10^8 M_\odot$. If
assuming the nominal gas-to-dust ratio of $\sim 140$, the gas reservoir is 
$5.2\times 10^{10} M_\odot$, sufficient for the whole system to continue for
another $\sim 130$~Myr or the dust-blocked region for $\sim 300$~Myr.

    Finally, we examine the FIR-radio relation using the formalisms summarized
in \S 3.2, for which we obtained $q_IR=2.44$, which is in good agreement with
the mean value of $2.40\pm 0.24$ of Ivison et al. (2010).

\subsection{GOODSN63 (GOODSN-J123730.9+621259)}

  This source has a multi-wavelength dataset nearly as rich as GOODSN06 does
except that it lacks 70~$\mu$m image. While being a single source in 250~$\mu$m,
it breaks up into two sources (but still blended) in 160 and 100~$\mu$m.
According to the PEP DR1 catalog, these two PACS components have
$S_{160}=22.8\pm 1.4$ and $9.5\pm 1.2$~mJy, and $S_{100}=4.9\pm 0.4$ and
$3.6\pm 0.4$~mJy, respectively. This source is a previously discovered SMG
(Wang et al. 2004, their GOODS 850-6), whose position has been precisely
determined by the SMA observations to be
$RA=12^{h}37^{m}30.80^{s}$, $DEC=62{^o}12^{'}59.00^{''}$
(J2000), and has $S_{850}=14.9\pm 0.9$~mJy (Barger et al. 2012).

\subsubsection{Morphologies and Potential Components}

  Fig. 7 shows the images of GOODSN63. Within 18\arcsec\, radius, there are 63
objects in $H_{160}$ with $S/N\geq 5$. The inspection shows that only four of
them could be significant contributors in the FIR. These four objects,
marked alphabetically as ``{\tt A}'' to {\tt D}'', are $1\farcs 68$, 
$3\farcs 40$, $6\farcs 86$, and $10\farcs 27$ away from the 250~$\mu$m 
position, respectively. In 24~$\mu$m, {\tt A} and {\tt B} are blended as a
single source, while {\tt C} and {\tt D} are blended as another.
These are the two brightest 24~$\mu$m sources within $r=18$\arcsec, and their
positions are consistent with the two local maxima in 160 and 100~$\mu$m. The
brightest object in $H_{160}$ (marked by the largest green circle in Fig. 7)
has spectroscopic redshift of $z=0.5121$ (Barger et al. 2008) and is not
visible in 24~$\mu$m at all, and hence it is not likely to have any significant
FIR emission.

   Objects {\tt A} and {\tt B} show complex morphologies. The WFC3 IR 
images reveal that {\tt A} consists of two subcomponents, {\tt A-1} and
{\tt A-2}, which are only $0\farcs 30$ apart (Fig. 8, left). {\tt A-2} is 
completely invisible in all the ACS bands. In the IR images, {\tt B} shows a 
dominant core that is irregular in shape, and also has irregular, satellite
components to its north-western side.
In the ACS images, its core is less distinct, and the satellite components are 
invisible. All this suggests that both {\tt A} and {\tt B} are likely very
dusty systems, and that both could be going through violent merging
process. While the small separation between them ($1\farcs 80$) suggests
that they could be an interacting pair, our SED analysis below shows that they
actually are not associated.

   Objects {\tt C} and {\tt D} are $3\farcs 42$ apart. They have spectroscopic
redshifts of 0.9370 and 1.3585, respectively (Barger et al. 2008), and 
therefore are not associated. The morphologies of these two objects are shown
in Fig. 8 (right) in details. While it is a rather smooth disc galaxy in the
restframe optical, {\tt C} shows knotty structures and dust lanes in restframe
UV, suggestive of active star formation. On the other hand, {\tt D} appears as
a smooth spheroid in restframe UV to optical. All this suggests that {\tt C} is
the major contributor to the 24$\mu$m flux, which is consistent with the 
24~$\mu$m morphology as the 24~$\mu$m centroid is closer to the position of 
{\tt C} than to {\tt D}.

\begin{figure*}[tbp]
\centering
\includegraphics[width=\textwidth] {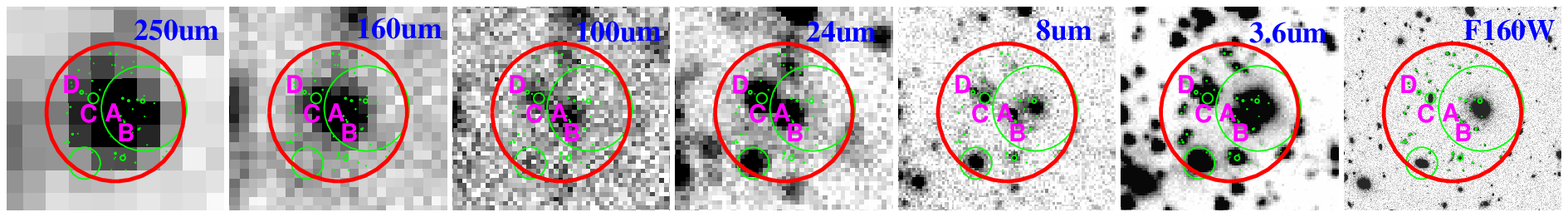}
\includegraphics[width=\textwidth] {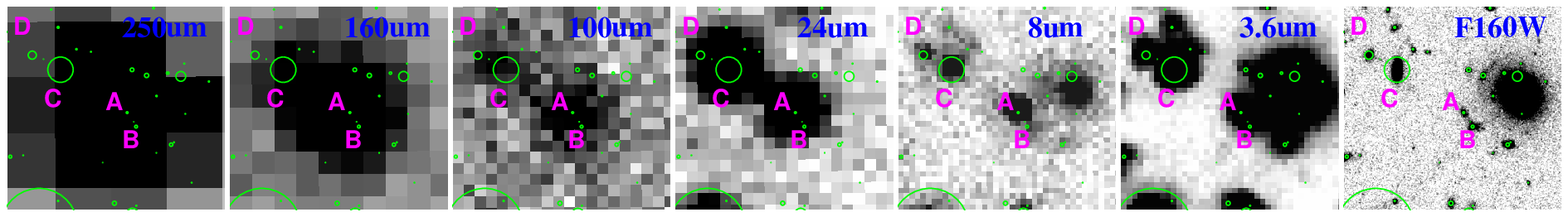}
\caption{FIR to near-IR images of GOODSN63. The organization and the legends
are the same as in Fig. 2. This source does not have MIPS 70~$\mu$m data. The
possible contributors to the 250 μm emission are labeled by the letters in
purple.}
\end{figure*}

\begin{figure*}[tbp]
\centering
\subfigure{
   \includegraphics[width=0.5\textwidth]{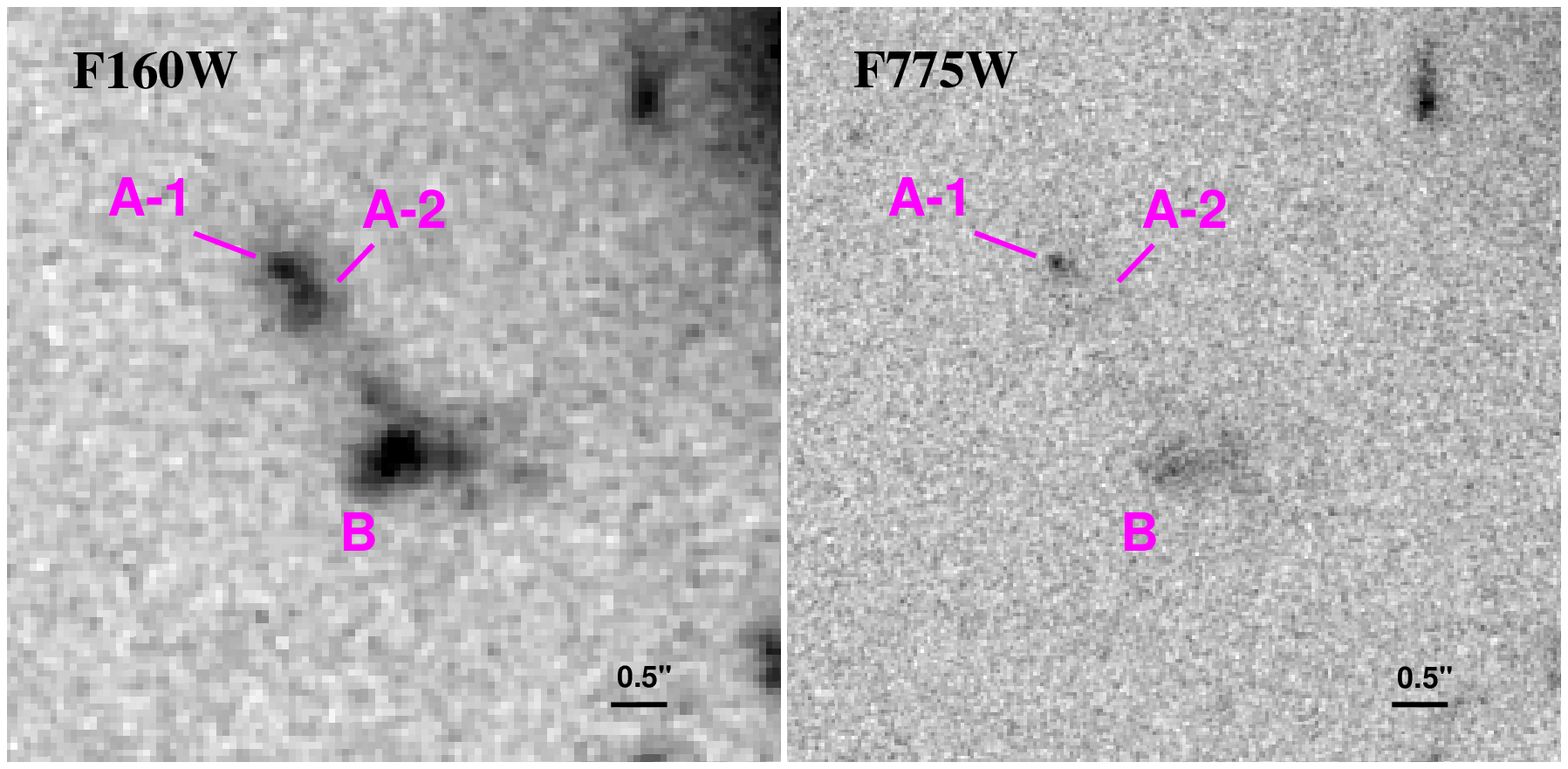}
   \includegraphics[width=0.5\textwidth]{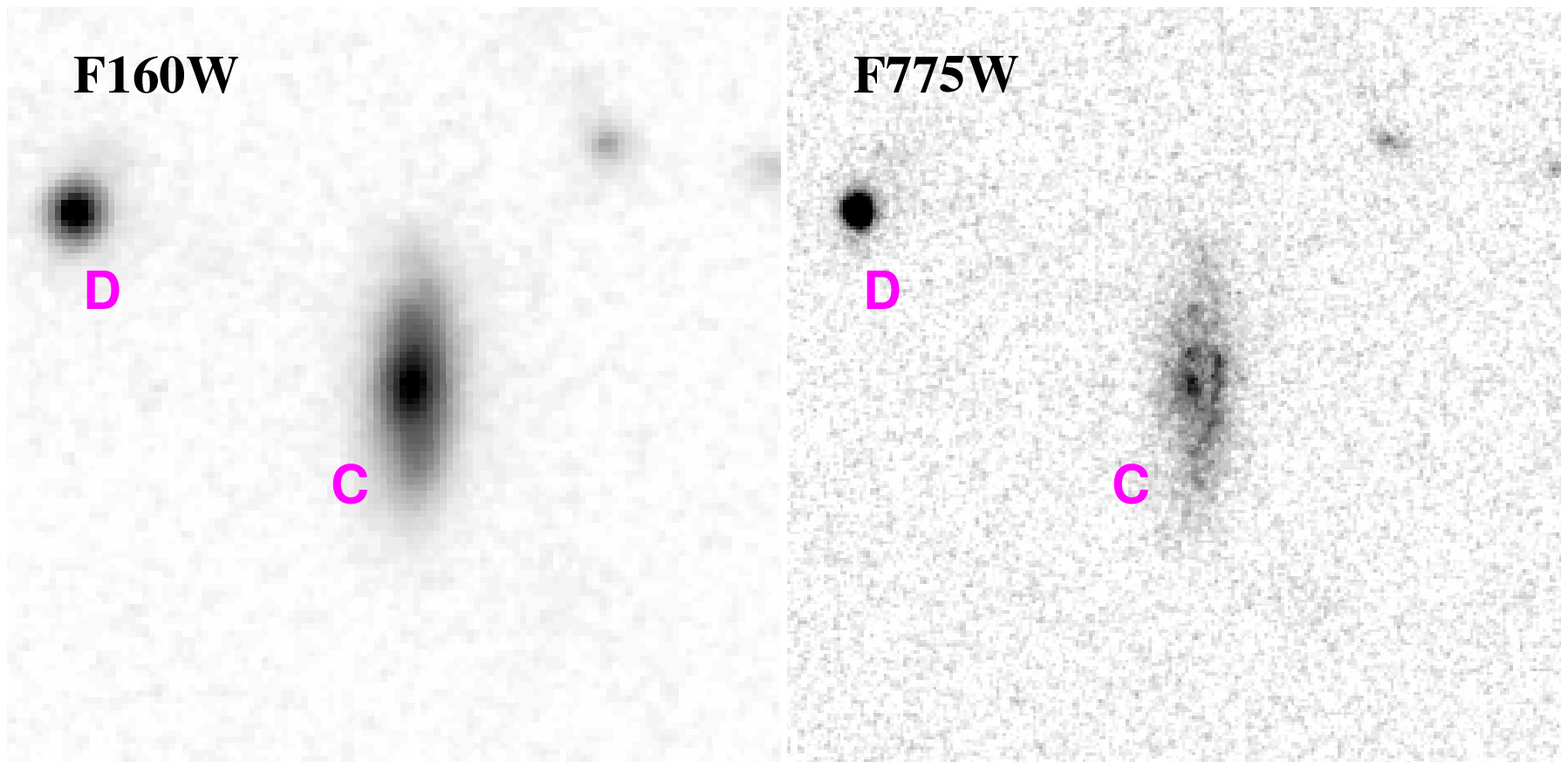}
}
\caption{NIR and optical morphologies of the possible contributors to GOODSN63,
shown in $H_{160}$ and $i_{775}$. The left panel is for {\tt A} and {\tt B},
and the right panel is for {\tt C} and {\tt D}. {\tt A} clearly breaks up into
two sub-components in $H_{160}$, labeled as ``{\tt A-1}'' and ``{\tt A-2}'',
the latter of which is only barely visible in optical.
}
\end{figure*}

\subsubsection{Optical-to-near-IR SED Analysis}

  Fig. 9 summarizes the results of the optical-to-near-IR SED analysis for
the four potential components. The SEDs used here are based on the same 
photometry as in \S 4.1.2 except for {\tt A}, where it is augmented by the
$U$-band image. 

  The fitting of {\tt C} gives $z_{ph}=0.93$, in excellent agreement with its
$z_{spec}=0.937$. {\tt D} has $z_{ph}=1.22$ as opposed to 
$z_{spec}=1.359$, nevertheless they are still in reasonable agreement
($\Delta_z/(1+z)=0.06$). {\tt A} and {\tt B} do not have spectroscopic
redshifts. {\tt B} is not detected in the $U$-band image, and would be selected
as an LBG at $z\approx 3$. In fact, the fit to its SED, either with or without
including the $U$-band constraint, gives $z_{ph}=3.04$.

\begin{figure*}[tbp]
\centering
\includegraphics[width=\textwidth] {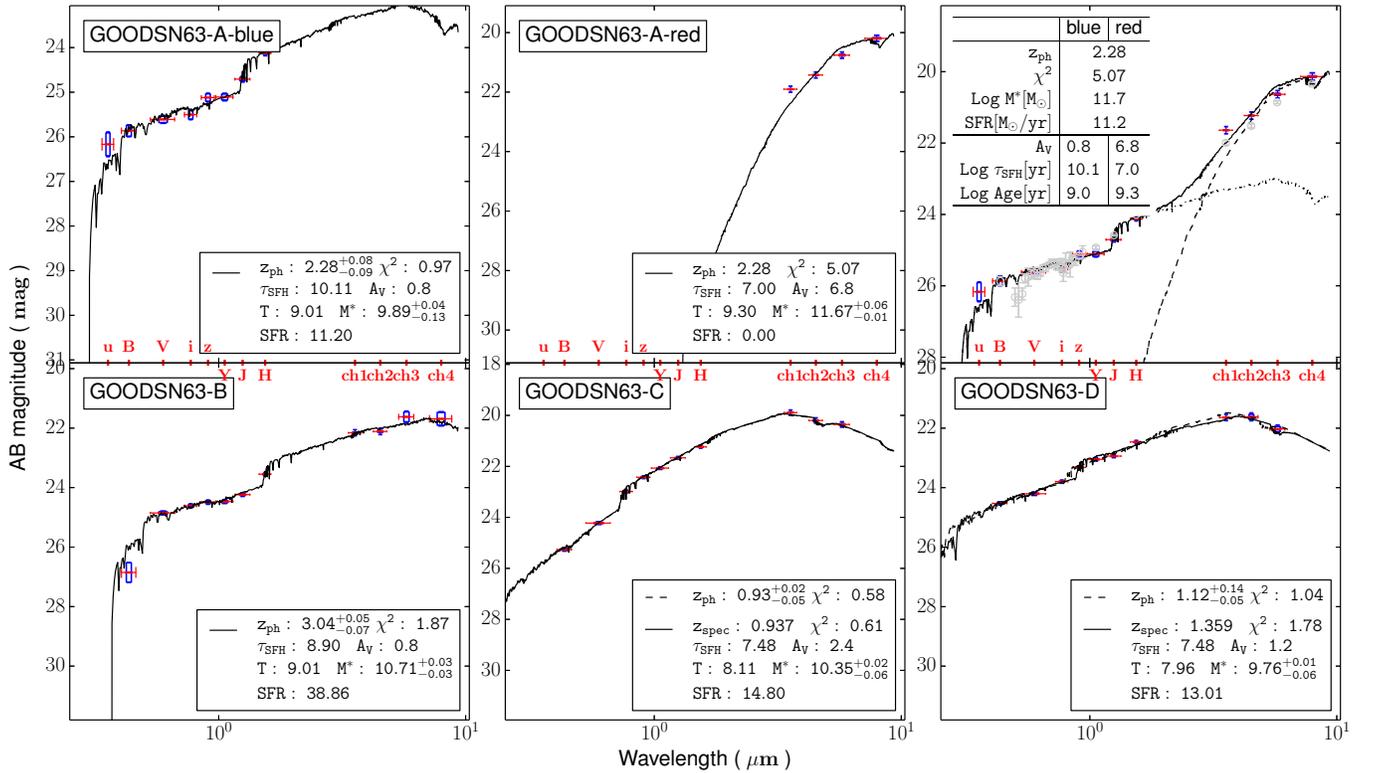}
\caption{Optical-to-NIR SED analysis of the possible contributors to GOODSN63.
Legends are the same as in Fig. 4. The three panels on the top are all for
object {\tt A}: from left to right, these are the fit to its blue
sub-component, its red sub-component, and the sum of the two. The grey data
points in the upper-right panel are from the Rainbow database. Note that
the ``blue'' and the ``red'' refer to the SED sub-components but not the
morphological sub-components {\tt A-1} and {\tt A-2}. See text for details.
}
\end{figure*}

   {\tt A} deserves some more detailed discussion. The single-component fit as
used for other objects fails to explain the SED of {\tt A}. We have also tested
a different set of SFHs where the SFR increases with time (Papovich et al.
2011), however {\tt A} still cannot be fitted by a single component.
Briefly, the strong IRAC fluxes and the large break between WFC3 IR and IRAC
tend to assign a high redshift of $z_{ph}\gtrsim 3$ to the object, which would
then give an implausibly high stellar mass of $> 10^{12} M_\odot$. On the other
hand, the optical fluxes (especially the detection in the $U$-band) are only
consistent with significantly lower redshifts. 
Therefore, we attempted to fit this object with two components.
We took a two-step
approach. We first fit the blue part of the SED up to $H_{160}$, using the
whole set of BC03 models as usual. We obtained the best-fit
at $z_{ph}=2.28$. The best-fit model is of moderate stellar mass 
($7.8 \times 10^9 M_\odot$) and age (1.0~Gyr), and has a very extended SFH 
($\tau=13$~Gyr) with the current $SFR_{fit}=11.2$~$M_\odot/yr$.
This best-fit template, which extends to the IRAC wavelengths,
was then subtracted from the original SED to produce the SED for the red part.
This red part remains very strong in the IRAC bands, as the blue component only
contributes a small amount at these wavelengths. It has nearly zero flux from 
$U$ to $J_{125}$, but still has non-zero residual flux in $H_{160}$. We then
fitted this red SED, from $H_{160}$ to 8.0~$\mu$m, at the fixed redshift of 2.28
from the blue component. We also relaxed the constraint on the dust extinction
and allowed $A_V$ to vary from 0 to 10~mag.
We indeed found a reasonable best-fit model, which is a highly extincted
($A_V=6.8$~mag), high stellar mass ($4.7 \times 10^{11} M_\odot$), old
(age 2.0~Gyr) burst ($\tau=10$~Myr). The result of this two-step fit is 
summarized again in Fig. 9 with more details. While this is probably not the
unique solution, it offers a plausible explanation to the SED of {\tt A}. 
In addition, we also examined if the red part could be explained by a 
power-law AGN in the form of $f_\nu \propto \nu^{-\alpha}$. The commonly
adopted slope is $\alpha=1.0$--1.3 in the restframe near-IR (i.e., the IRAC
bands in the observer's frame for this object), and changes to a flatter value
of $\alpha=0.7$--0.8 when extending into the restframe optical regime (see e.g.,
Elvis et al. 1994; Richards et al. 2006; Assef et al. 2010). Such power-laws,
however, cannot provide any reasonable fit to the data. We found that the IRAC
data points, with or without the extension to the MIPS 24~$\mu$m, could only
be explained by a very steep slope of $\alpha=1.8$--1.9. Adopting this slope
in the restframe optical and ignoring the possible flattening of the SED,
we would get still get $H_{160}\sim 23.6$~mag, which is much higher with the
requirement of $H_{160}=28.8$~mag based on the SED of the red part. In fact,
this is also significantly brighter than the observed total of $H_{160}$=24.12~
mag. While we cannot definitely rule out the possibility that there might be
an embedded AGN in GOODSN63-A, there is no compelling evidence that such an AGN
component exists. Based on our analysis, we believe that the optical-to-near-IR
emission of GOODSN63-A is dominated by stellar populations.

   Interestingly, Barger et al. (2012) derive $z_{ph}=2.7$ for GOODSN63 based
on the radio-to-sub-mm flux ratio. To further check on the redshift of this
GOODSN63-{\tt A}, we used the ``Rainbow'' database in the GOODS-N, which
incorporates the 24 narrow-band images (0.50--0.95~$\mu$m) from the Survey
of High-z Absorption Red and Dead Sources program (SHARDS;
P\'erez-Gonz\'alez et al. 2013) and the GOODS \& CANDELS ACS/WFC/IRAC data as
used above (see the grey data points in the upper right panel of Fig. 9).
This database gives $z_{ph}=2.4$ for this object, which agrees
with $z_{ph}=2.28$ quite well considering the different photometry in
these two approaches. We adopt $z_{ph}=2.28$ for the analysis below.

\subsubsection{Decomposition in Mid-to-Far-IR}

   The automatically iterative decomposition of GOODSN63 was successful in
all bands from 250~$\mu$m to 24~$\mu$m.
In 24~$\mu$m, {\tt B} has zero flux. We have tried to decompose by
subtracting {\tt A} first, and find that the residual image has no flux left at
the position of {\tt B}, consistent with the simultaneous fit. This probably is
not surprising, because {\tt B}, while being brighter than {\tt A} from
$V_{606}$ to $H_{160}$, steadily becomes much fainter than {\tt A} when moving
to the redder and redder IRAC channels ($>1.5\times$ fainter in 8.0~$\mu$m).
The simultaneous fit in other bands also gives zero flux for {\tt B}. Similar
to {\tt B}, {\tt D} also has zero flux in 100, 160 and 250~$\mu$m. Therefore,
the two distinct sources in the 100 and 160~$\mu$m described at the beginning
of \S 4.2 should corresponds to {\tt A} and {\tt C}, respectively.

\begin{figure*}[btp]
\centering
\includegraphics[width=\textwidth]{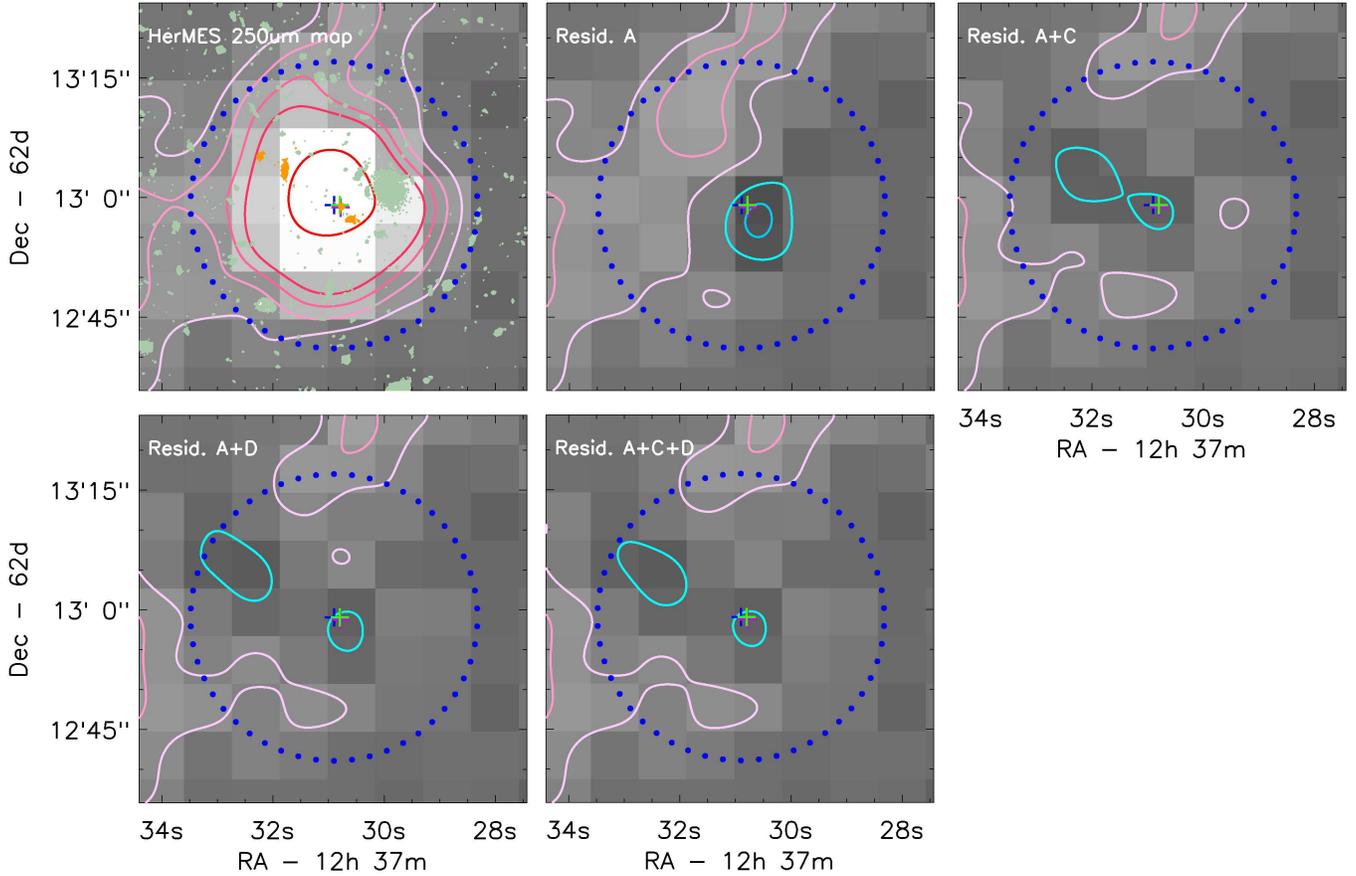}
\caption{Demonstration of the decomposition in 250~$\mu$m for GOODSN63.
The first panel shows the original 250~$\mu$m image, while the others show the
residual maps of the different decomposition schemes where different input
sources are considered (labeled on top). Legends are the same as in Fig. 5. In
the addition to the blue and the magenta crosses, the bright green cross 
indicates the position of the SMA detection.  The ``{\tt A}'' panel is for
the automatically iterative decomposition, which results in {\tt A} as the
only survivor and does not leave obvious residuals at the locations of other
possible contributors. While the trial-and-error fits
with the addition of {\tt C} and/or {\tt D} would produce equally acceptable
results (other panels), the reasonable result from the automatically iterative
approach always takes the precedence per our decomposition rule and hence
{\tt A} is deemed to be the sole contributor to 250~$\mu$m emission.
}
\end{figure*}

   The automatically iterative decomposition shows that
{\tt C} also has zero flux in 250~$\mu$m, leaving {\tt A} as the sole object 
responsible for the emission in this band. The decomposition in this band is 
shown in Fig. 10.  The decomposition settled on {\tt A} with a very reasonable
error estimate, and left no detectable fluxes at the positions of any other
sources. We also tried the trial-and-error approach by adding
{\tt C} and/or {\tt D} (although {\tt D} probably should not be involved in
the first place given its zero fluxes in 100 and 160~$\mu$m), which produced
equally acceptable results. However, our decomposition rule is that the
automatically iterative approach always takes the precedence as long as the 
results are reasonable, and therefore we adopted {\tt A} as the only contributor
to the 250~$\mu$m emission. This decision is also supported by the fact that
the flux ratio of {\tt A} to {\tt C} gradually increases with the wavelengths, 
increasing from 0.77:1 at 24~$\mu$m to 3.47:1 at 160~$\mu$m, consistent with
the picture that {\tt A} is much more dominant in the FIR.  This also raises
the point that one should not blindly take the brightest 24~$\mu$m source
within the {\it Herschel}\, beam as the counterpart.

   The above results mean that it is possible to include the data in 350 and
500~$\mu$m for further analysis, as 
there should not be any other neighbors that could contaminate the light from
{\tt A} in these two bands. The final mid-to-far-IR flux densities of 
GOODSN63-{\tt A} that we adopt are summarized in Table 2.

\begin{figure*}[btp]
\centering
\subfigure{
  \includegraphics[width=0.5\textwidth]{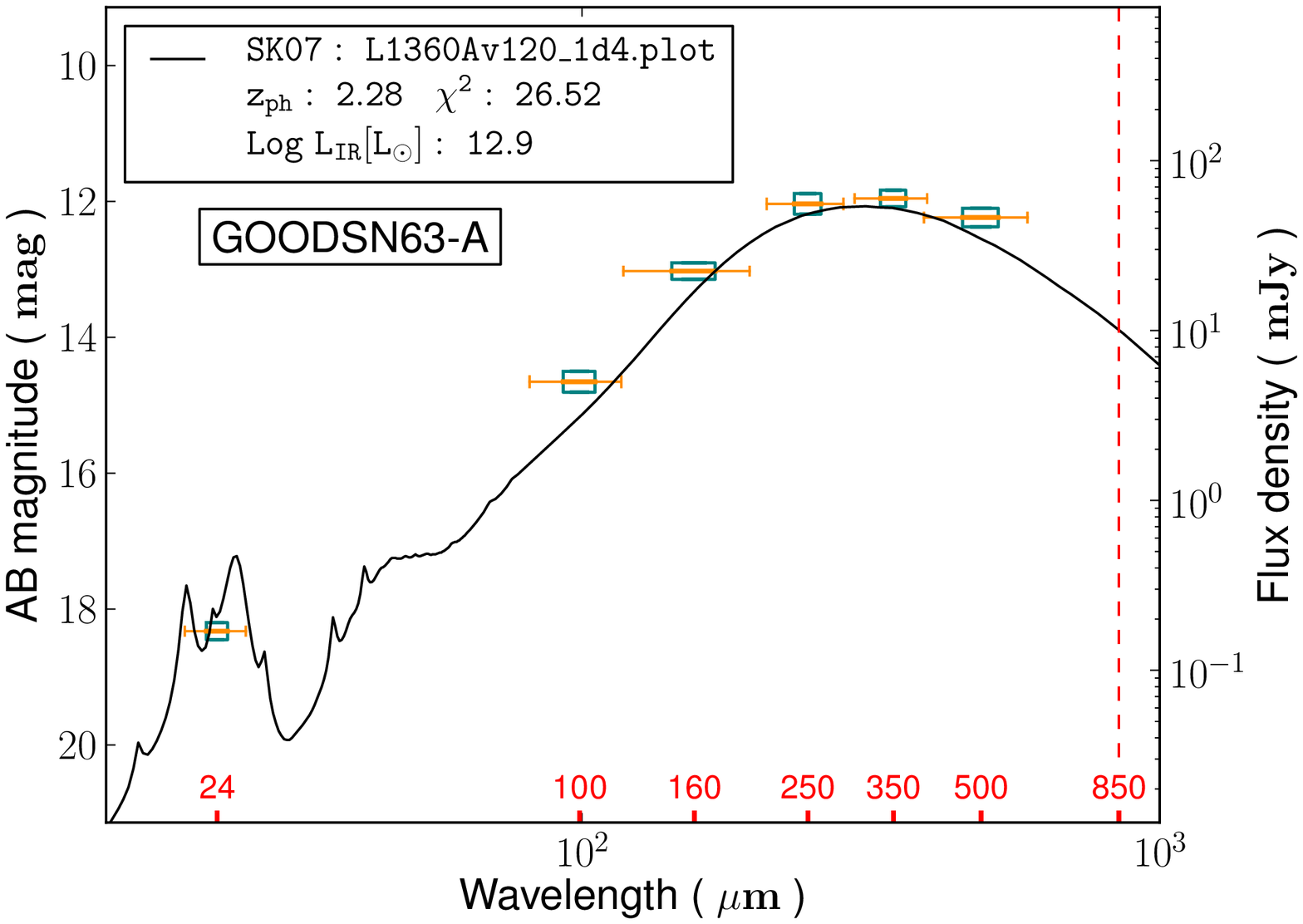}
  \includegraphics[width=0.5\textwidth]{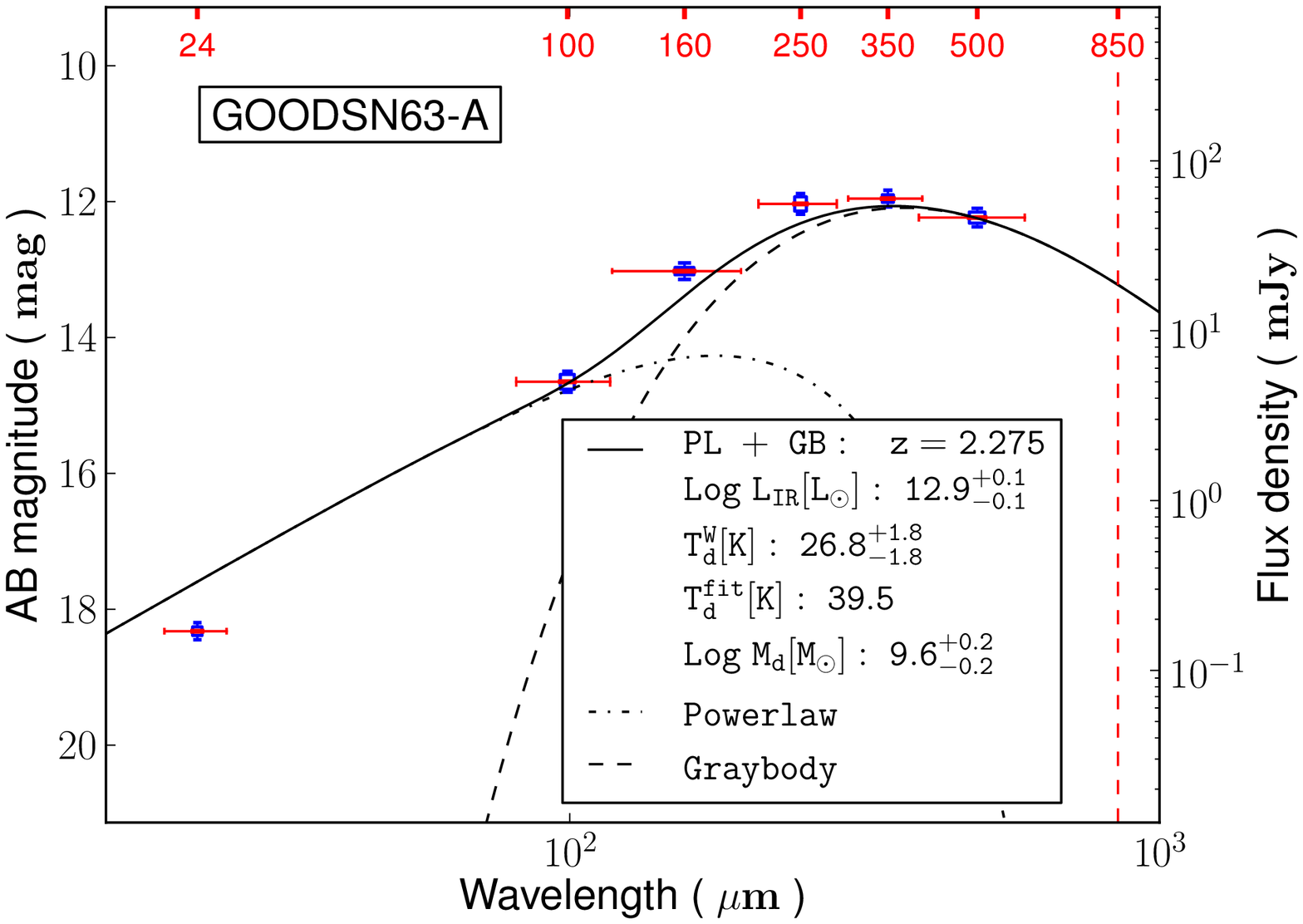}
}
\caption{Mid-to-FIR SED fitting for GOODSN63-{\tt A} at $z_{ph}=2.28$. The SED
incorporates the SPIRE 350 and 500~$\mu$m photometry directly from the HerMES
catalog because {\tt A} is the sole contributor at 250~$\mu$m and longer 
wavelengths and hence no decomposition in these two bands are necessary. 
The left panel shows the fit to the SK07 models, and the right panel shows the
fit using the ``power-law + graybody'' models. Legends are the same as in 
Fig. 6.
}
\end{figure*}

\subsubsection{Total IR Emission and Stellar Populations}
   
   The {\tt A}-only decomposition result above is supported by the SMA 
observation of this source (see \S 4.2; Barger et al. 2012), which reveals a
single source whose position is on {\tt A-1}. This is also supported by the
VLA 1.4~GHz data, where there is a source with $S_{1.4GHz}=0.123$~mJy 
coinciding with {\tt A}. The radio position is at $RA=12^h37^m30^s.78$, 
$DEC=62^o12^{'}58^{''}.7$, which is right in between {\tt A-1} and {\tt A-2}
(Morrison et al. 2010). There is no X-ray detection of GOODSN63 in the 2Ms
{\it Chandra} data (Alexander et al. 2003), which have the limit of
$2.5\times 10^{-17}$~$erg/s/cm^2$ in the most sensitive 0.5--2~keV band. At
$z=2.28$, this corresponds to an upper limit of $1.0\times 10^{42}$~$erg/s$ in
the restframe 1.6--6.6~keV, which is at the border line of AGN. Therefore, we
believe that the FIR emission of GOODSN63 is most likely powered by stars.

    We derived $L_{IR}$ using the mid-to-far-IR SED from 24 to 500~$\mu$m,
fixing the redshift at $z_{ph}=2.28$. The result is summarized in the left
panel of Fig. 11. The best-fit SK07 model is heavily dust-extincted, with 
$A_V=12.0$~mag. We obtained $L_{IR}=7.9\times 10^{12} L_\odot$, and therefore
this object is an ULIRG. The contribution from the dust re-processed light in
the exposed region only amounts to $L_{IR}^{ext}=8.7\times 10^{10} L_\odot$,
and thus $L_{IR}^{blk}=7.8\times 10^{12} L_\odot$ and
$SFR_{IR}^{blk}=781 M_\odot/yr$. In contrast, as
described in \S 4.2.2, it has $SFR_{fit}=11.2 M_\odot/yr$, which is contributed
solely by the blue component of {\tt A}. The stellar mass of this system
is dominated by the red subcomponent of {\tt A}, which is 
$4.7\times 10^{11} M_\odot$. Therefore, we get $SSFR=1.69 Gyr^{-1}$ (or
$T_{db}^{tot}=593$~Myr), and $T_{db}^{blk}=602$~Myr.
The latter doubling time is vastly
different from that inferred for GOODSN06, and is much longer than the typical
lifetime of 10--100~Myr of an ULIRG. Nevertheless, it is still much shorter
$T=2.0$~Gyr of the exposed stellar population.
The dust temperature and the dust mass of this system, 
$T_d=39.4$~K and $M_d=3.6\times 10^9 M_\odot$ are typical of
SMGs. The best-fit SK07 model predicts $S_{850}=9.81$~mJy, consistent with the
actual observation.
The dust mass implies the gas mass of
$M_{gas}=5.6\times 10^{11} M_\odot$, sufficient to fuel the
ULIRG for the next $\sim 700$~Myr.

     The red subcomponent of {\tt A} is a short burst
($\tau=10$~Myr) that is 2~Gyr old (\S 4.2.2), which implies that the majority
of the existing stellar mass of this system was formed in an intense
star-bursting phase at $z>5$, and that the current ULIRG phase is an episode
unrelated to the main build-up process of the existing system.

    We also examined the FIR-radio relation for this system, and obtained
$q_{IR}=2.29$. This is lower but still consistent with the mean value of
$2.40\pm 0.24$ of Ivison et al. (2010).

\section{Sources in the UDS}

   The two sources in the UDS have much shallower SPIRE data than those in the
GOODS-N, because they are in a HerMES ``L6'' field. Unfortunately the PACS data
are not yet released. 

\subsection{UDS01 (UDS-J021806.0-051247)}

   This source does not have data in the ACS, and we relied on the WFC3 IR data
for their morphologies. The optical data from the CFHTLS-Wide were used in
their optical-to-NIR SED analysis.

\subsubsection{Morphologies and Potential Components}

   Fig. 12 shows the images of UDS01.
Within 18\arcsec\, radius, of the 250~$\mu$m position, 
there are 35 objects in $H_{160}$ with $S/N\geq 5$. By comparing the 250~$\mu$m
and the 24~$\mu$m images, we found that only nine of these objects could be
significant contributors to the FIR flux. These objects are labeled from
``{\tt A}'' to ``{\tt H}'' according to their proximity to the 250~$\mu$m
centroid, which lies in between {\tt B} and {\tt C}. {\tt B} has spectroscopic
redshift of $z=1.042$ (Bradshaw et al. 2013; McLure et al. 2013).

   At 24~$\mu$m, these nine sources are already severely blended. 
Nevertheless, we can still roughly distinguish two major sets based on their
positions. The morphological details of these two sets in $H_{160}$, hereafter
the ``southern'' and the ``northern'' sets, are shown in Fig. 13.
The southern set includes {\tt A} and {\tt B}, which are $3\farcs 9$
apart, and are $2\farcs 2$ and $2\farcs 5$ from the 250~$\mu$m centroid, 
respectively. {\tt A} is a small, compact object. {\tt B} is
by far the more dominant of the two in both IRAC and MIPS bands. In $H_{160}$,
it  shows a dominant core and an extended, irregular halo that is suggestive of
a post-merger. The ``northern'' set includes {\tt C}, {\tt D}, {\tt E}, 
{\tt F}, and {\tt G}. The dominant member in this set, as seen from
IRAC to MIPS, is {\tt D}, while {\tt C} is the next in line. These two
objects are $2\farcs 8$ and $3\farcs 4$ from the 250~$\mu$m centroid, 
respectively, and they are $2\farcs 5$ apart from each other. {\tt C} has
a spheroidal-like core, while {\tt D} is more extended and has an irregular
halo suggestive of a post-merger similar to {\tt B}. {\tt E} seems to be
the companion of {\tt C}, while {\tt G} could be associated with
{\tt D}. {\tt F}, on the other hand, already diminishes to invisible in 
8~$\mu$m, and therefore is likely negligible at the longer wavelengths. 
Finally, {\tt I} and {\tt H}, which are small and compact and only $1\farcs 6$
apart in $H_{160}$, form a distinct, but the least important set.

\begin{figure*}[tbp]
\centering
\includegraphics[width=\textwidth] {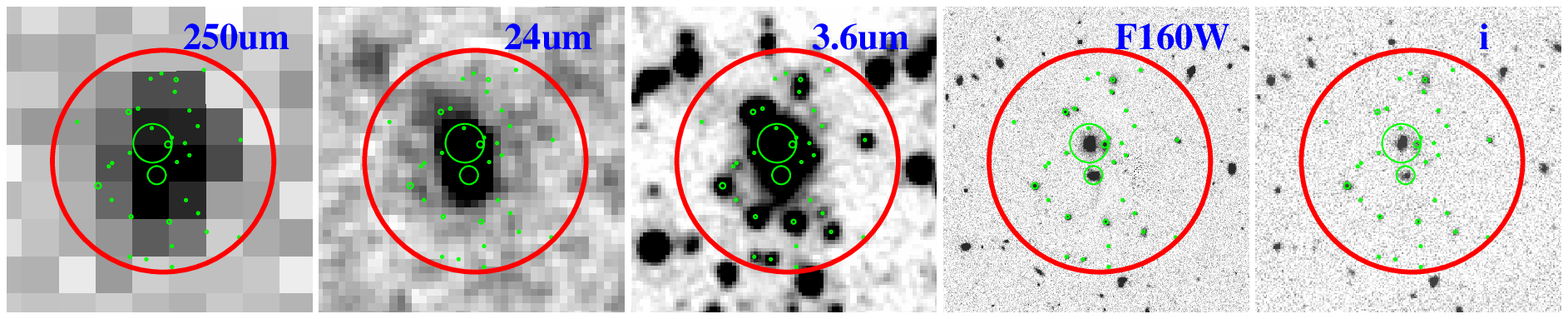}
\includegraphics[width=\textwidth] {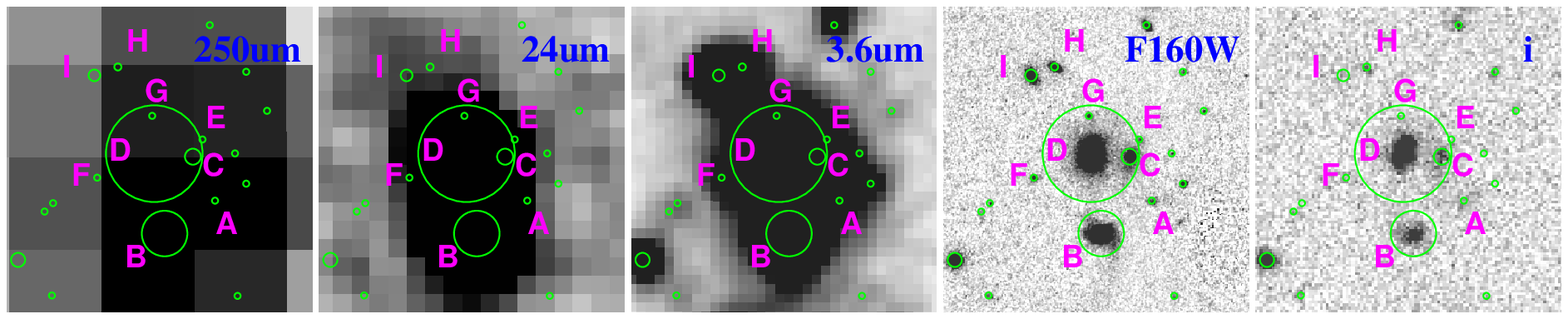}
\caption{FIR to optical images of UDS01. The organization and the legends are
the same as in Fig. 2. This source lacks 160, 100 and 70~$\mu$m data. The
$i$-band image is from the CFHTLS-Wide program.}
\end{figure*}

\begin{figure*}[tbp]
\centering
\subfigure{
  \includegraphics[width=0.5\textwidth] {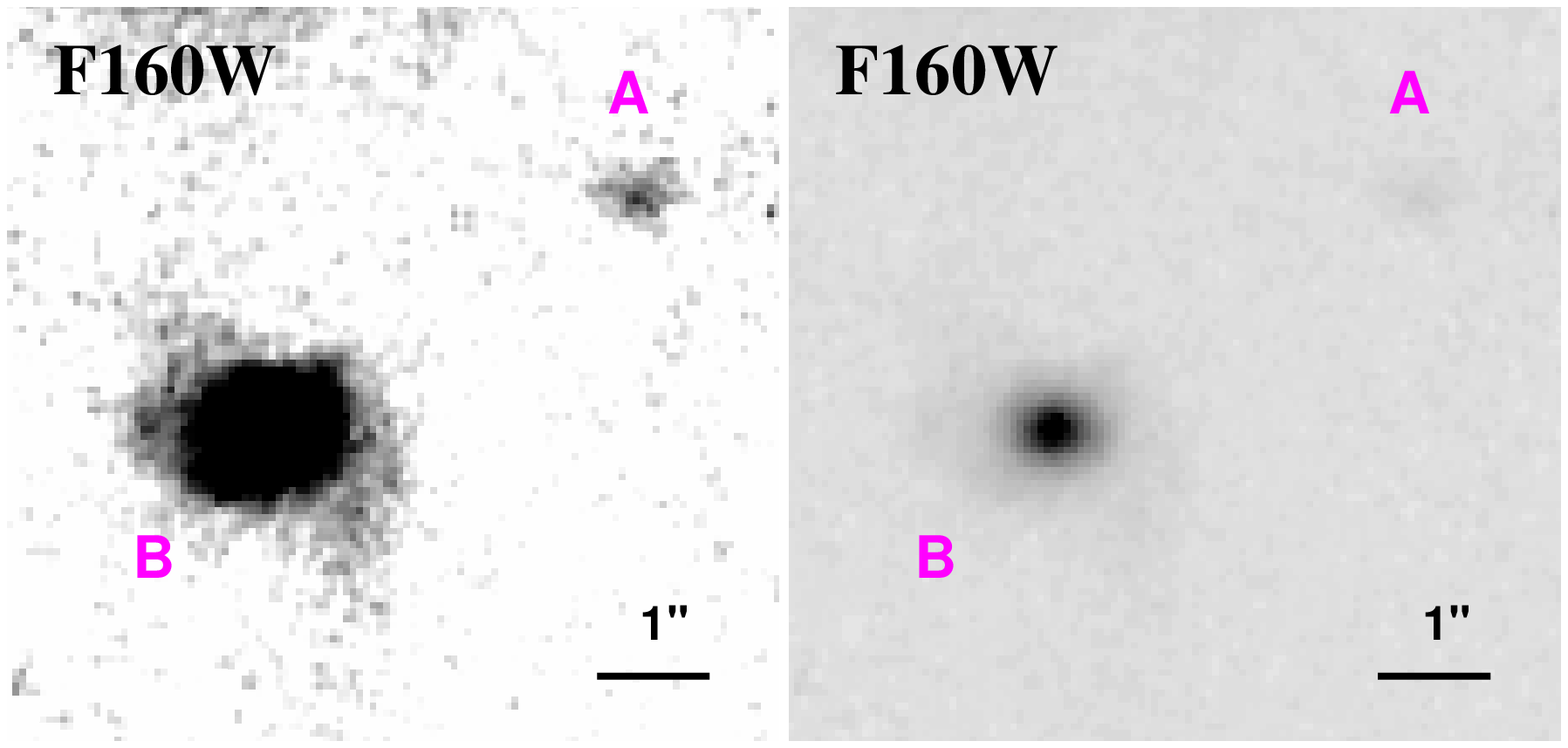}
  \includegraphics[width=0.5\textwidth] {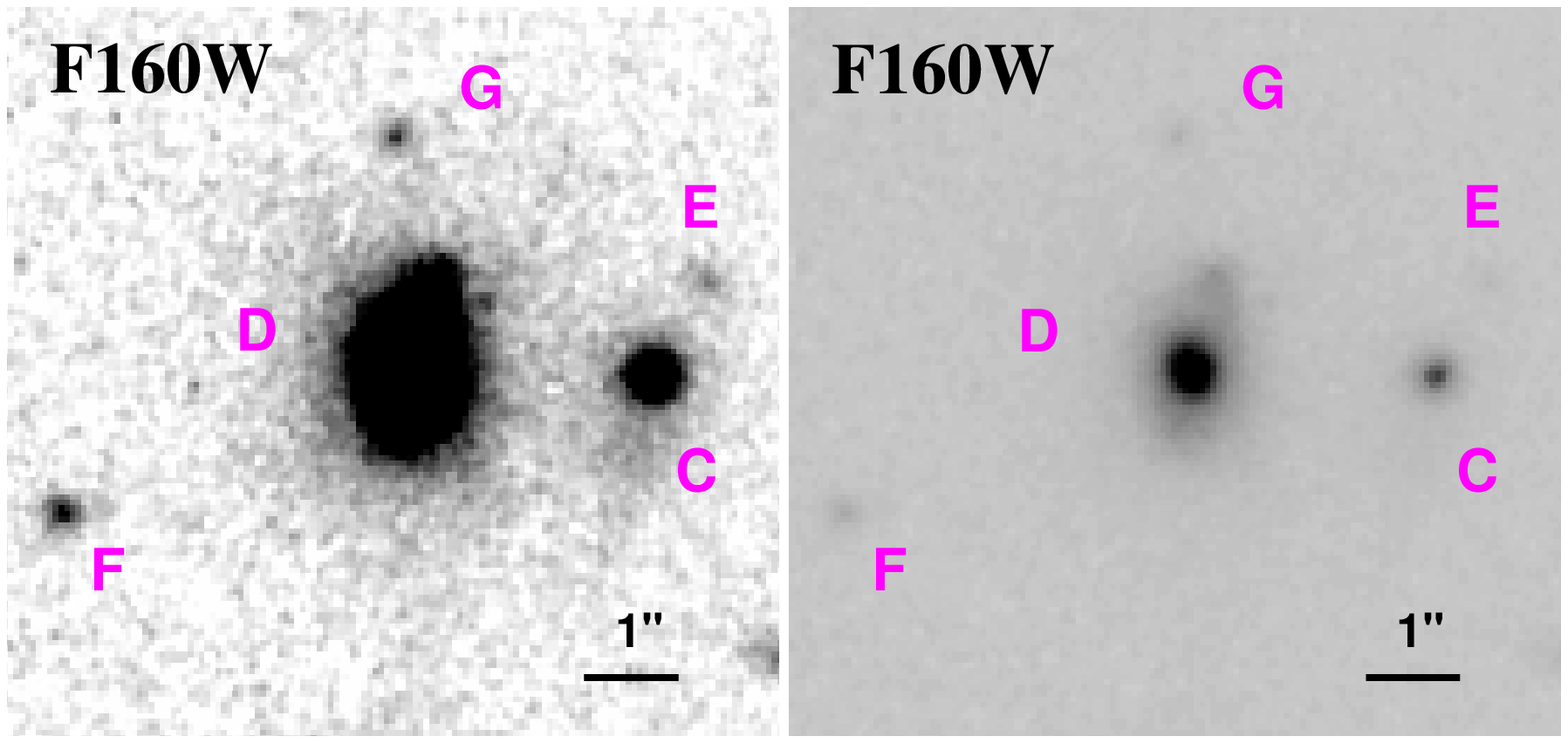}
}
\caption{NIR morphologies of the possible contributors to UDS-01, shown in
$H_{160}$ with two contrast levels to reveal different features. The left panel
is for the ``southern'' group that includes {\tt A} and {\tt B}, while the right
panel is for the ``northern'' group that includes objects from {\tt C} to
{\tt F}.
}
\end{figure*}

\subsubsection{Optical-to-near-IR SED Analysis}

   While the high-resolution WFC3 data have discerned the multiple objects in
this region as discussed above, we cannot keep the same set of objects for
the optical-to-NIR SED analysis due to the lack of the high-resolution ACS data
in the optical. As the substitution, the CFHTLS-Wide data were used for the
SED analysis. In principle, we could run TFIT on those data using the WFC3
data as the templates, however this was not practical because we discovered
that the astrometric solution of the CFHTLS-Wide data is not entirely
consistent with that of the CANDELS WFC3 data. As fixing this problem would be
beyond the scope of this paper, our approach was to carry out independent
photometry on the CFHTLS-Wide data alone and then to match with the WFC3
detections for the SED construction. For this purpose, we used SExtractor in 
dual-image mode and the $i$-band image as the detection image. {\tt C} and
{\tt E} cannot be separated in these images, and thus we take them as one 
single source and will refer to it as {\tt C/E} hereafter. {\tt E}
is $> 3$~mag fainter than {\tt C} in $J_{125}$ and $H_{160}$, and hence is
negligible for the analysis of the stellar
population.  {\tt D} and {\tt G} are taken as one single source as well for the 
same reason, and {\tt G} is $>5$~mag fainter than {\tt D} in the WFC3 and hence
likely negligible. We refer to them as {\tt D/G} hereafter. {\tt I} is not
visible in the CFHTLS data, and thus has to be excluded from this analysis. The
colors of these objects were measured in the {\tt MAG\_ISO} apertures, and the
$i$-band {\tt MAG\_AUTO} magnitudes were used as the reference to convert the
colors to the magnitudes for the SED construction. In the near-IR, we used the
CANDELS WFC3 $J_{125}$ and $H_{160}$, and these magnitudes are obtained in the
same way as in \S 4. In the longer wavelengths, we used the IRAC 3.6 and 
4.5~$\mu$m data from the SEDS program. The objects are severely blended in the
IRAC images, and hence we had to deblend by using TFIT to obtain reliable
photometry of the individual objects. As in \S 4, the $H_{160}$ image was used
as the template for TFIT.

   The photometric redshift results are summarized in Fig. 14. {\tt A} is 
excluded because this faint object is not well detected in the CFHTLS-Wide data.
Formally it has $z_{ph}=3.5$, however this is not trustworthy because of the
large photometric errors. As shown in the decomposition below, this object is
irrelevant.  {\tt B} has
$z_{ph}=1.05$, which agrees very well with its $z_{spec}=1.042$. {\tt C/E} and
{\tt D/G} have $z_{ph}=0.98$ and 1.09, respectively, and in terms of
$\Delta z/(1+z)$ they agree to $z_{spec}$ of {\tt B} to 0.02 and 0.03, 
respectively, well within the accuracy of $z_{ph}$ that can be achieved. The
peaks of their $P(z)$ distributions overlap significantly. All this means that 
these three objects are very likely at the same redshifts and 
associated. {\tt C/E} and {\tt D/G} are separated by $2\farcs52$, and {\tt B}
is $5\farcs25$ away from them. At $z=1.042$, these corresponds to 20.5 and
42.7~kpc, respectively, and are within the scale of galaxy groups. 

   The object of the highest stellar mass is {\tt D/G}, which
has $8.3\times 10^{10} M_\odot$. It has a prolonged SFH with $\tau=0.5$~Gyr,
and its best-fit age is $T=1.0$~Gyr. It has $SFR_{fit}=39.0 M_\odot/yr$. All
this suggests that the stellar mass assembly in {\tt D/G} is a slow, gradual 
process. In contrast, {\tt B} is best explained by a young SSP with an age of
130~Myr, and the inferred stellar mass is $4.6\times 10^{10} M_\odot$. This
indicates that it built up its stellar mass through a sudden onset of intense
star formation. Similarly, {\tt C/E} is best fit by a young stellar population 
with an age of 180~Myr and a short episode of SFH that has $\tau=50$~Myr.
However, its stellar mass is almost one magnitude lower at 
$5.4\times 10^{9} M_\odot$.

\begin{figure*}[tbp]
\centering
\includegraphics[width=\textwidth] {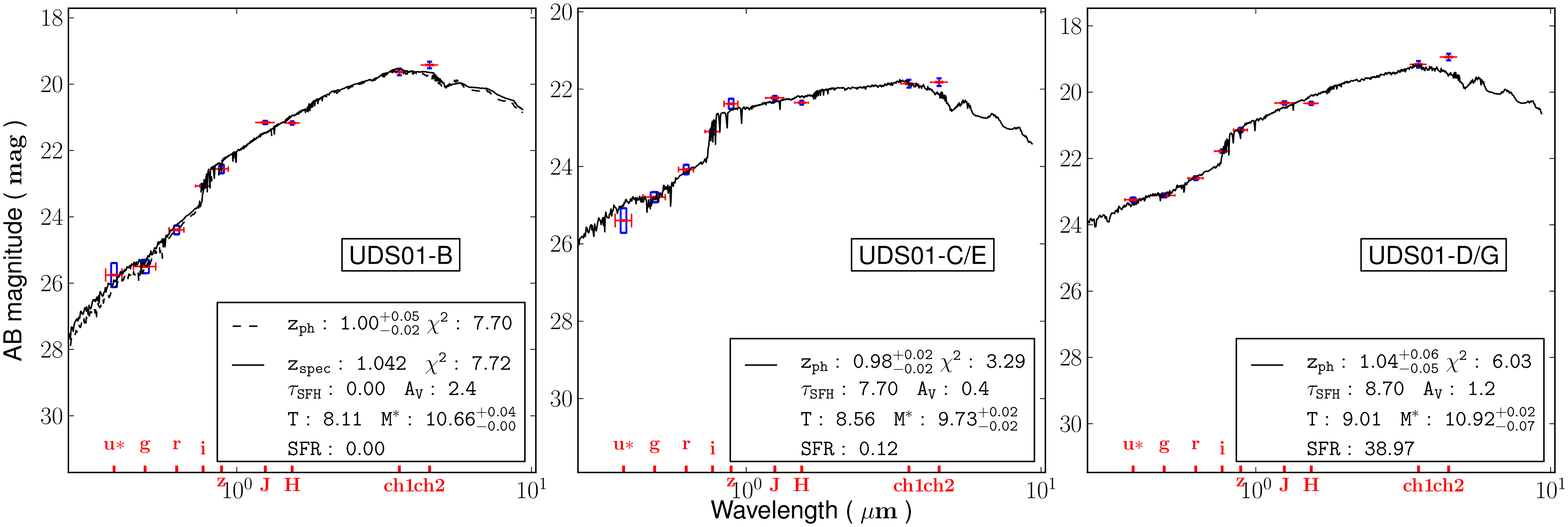}
\includegraphics[width=0.5\textwidth] {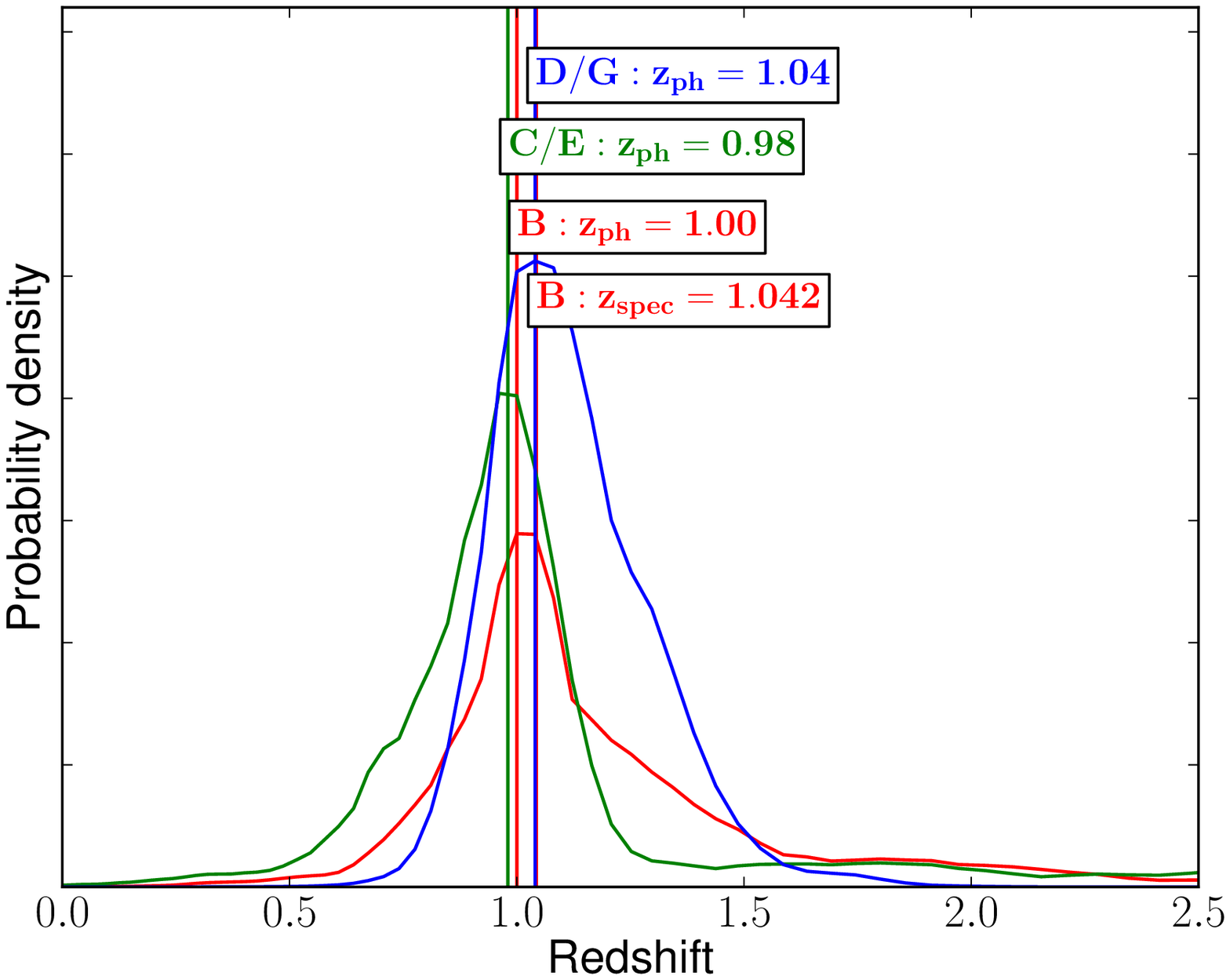}
\caption{Optical-to-NIR SED analysis of the possible contributors to UDS01. 
Legends are the same as in Fig. 4. {\tt A} is excluded here because it is not
well detected in CFHTLS-Wide data. It is highly plausible {\tt C/E}
and {\tt D/G} are at the same redshift as {\tt B}, which has $z_{spec}=1.042$.
}
\end{figure*}

\subsubsection{Decomposition in Mid-to-Far-IR}

   We used the position priors in $H_{160}$ for the decomposition. Not
surprisingly, the automatically iterative fit in 24~$\mu$m converged on only
{\tt B} and {\tt D}, the latter of which is by far the most dominant object in
24~$\mu$m. In 250~$\mu$m, however, it settled on {\tt B} and {\tt C} with
the resulted flux ratio of 55.5\%:44.5\%, 
and rejected {\tt D} because of its negligible flux. 

\begin{figure*}
\includegraphics[width=\textwidth]{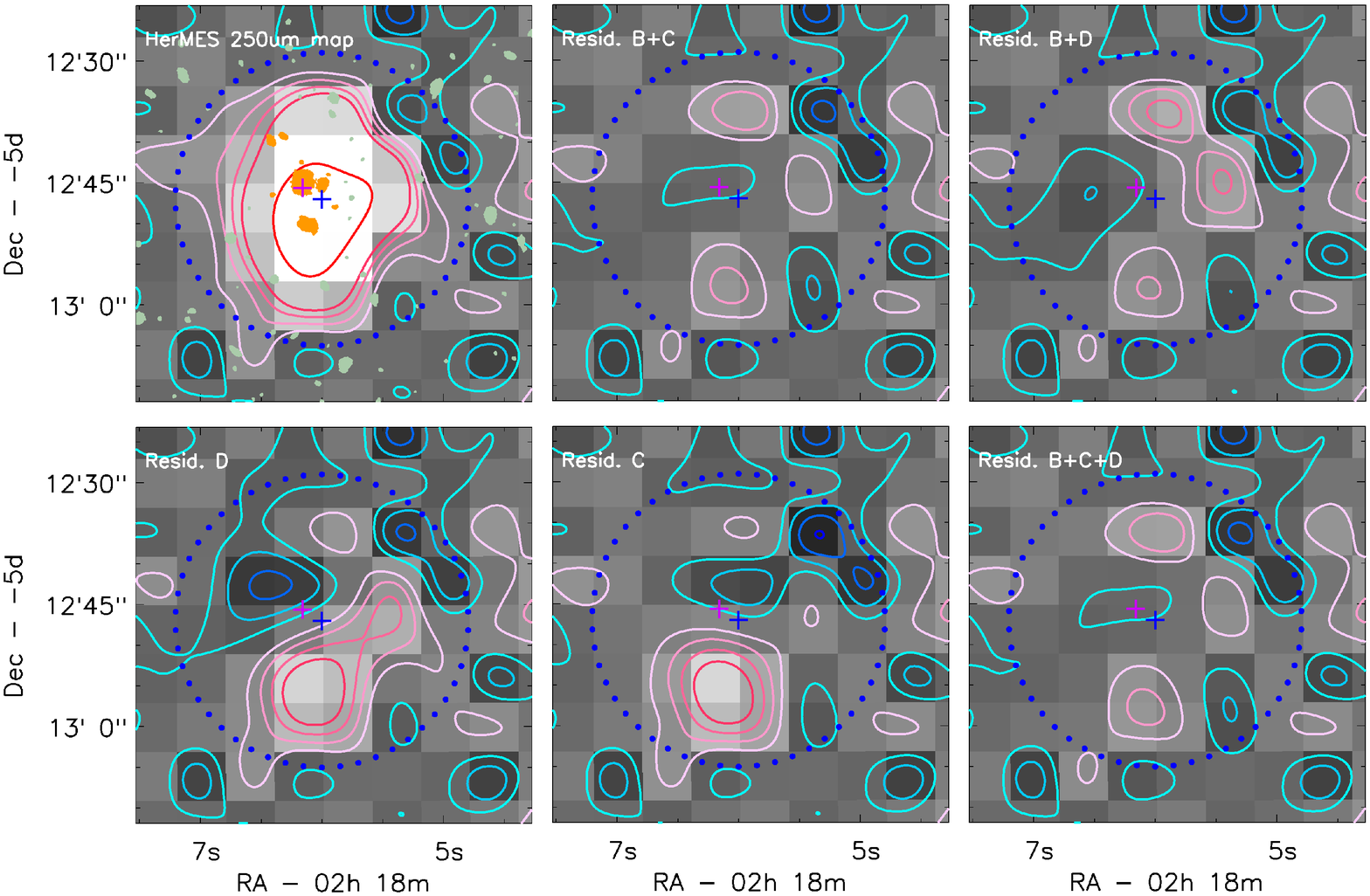}
\caption{Demonstration of the decomposition in 250~$\mu$m for UDS01. The first
panel shows the original 250~$\mu$m image, while the others show the residual
maps of the different decomposition schemes where different input sources are
considered (labeled on top). Legends are the same as in Fig. 5. This is a
degenerate case. While automatically iterative fit settles on {\tt B+C}, 
the trial-and-error run using {\tt B+D} still produces satisfactory results. 
However, the fit without the inclusion of {\tt B} are not acceptable.
}
\end{figure*}

   Although in the case of GOODSN63 we have learned that the strongest 
24~$\mu$m source is not necessarily the major contributor to the FIR emission,
it seems unusual that a FIR source would have no 24~$\mu$m counterpart. We thus
checked for degeneracy using the trail-and-error method. This confirmed that
indeed only {\tt B}, {\tt C} and {\tt D} could be relevant. The results are
summarized in Fig. 15. The forced fit with {\tt B} and {\tt D} left slightly
larger residual than the automatically iterative result (``{\tt B+C}''),
however it is not unacceptable. Fitting with only {\tt C} or {\tt D}, on the
other hand, left obvious residual at the location of {\tt B}, clearly 
indicating that a reasonable solution must include {\tt B}. Forcing the fit
to include {\tt B}, {\tt C} and {\tt D} produced the same result as the
{\tt B+C} case. 
Therefore, our decomposition slightly favors {\tt B} and {\tt C} as the
contributors to the FIR emission, however we cannot definitely rule out the
possibility that it actually is {\tt B} and {\tt D} that are responsible.

   One would hope that the radio data from the 100~mJy VLA 1.5~GHz survey
(Simpson et al. 2006) could break this degeneracy. Their catalog contains only
one strong radio source in the area occupied by UDS01, with 
$S_{1.5GHz}=0.185\pm 0.030$~mJy. The radio position is $RA=2^h18^m06^s.16$, 
$DEC=-5$\arcdeg12\arcmin45$\farcs61$ (J2000), which is right on {\tt D} (only
offset by $0\farcs 68$). While this VLA
survey used the B and C configurations and thus have worse angular resolution
than that of Morrison et al.'s data in the GOODS-N, its positional accuracy is
still good to sub-arcsec level for the high S/N radio sources such as this one.
For this reason, {\tt D} seems to be a more plausible FIR contributor
than {\tt C}. However, it is puzzling that {\tt B}, which is the strongest in
250~$\mu$m among all, does not have a radio counterpart in the catalog.

   The flux densities of {\tt B} and {\tt D}, based on the decomposition
results using the {\tt B+D} case, are summarized in Table 2. The flux
densities of {\tt B} in the {\tt B+C} case only differ by $<6$\%.

\subsubsection{Total IR Emission and Stellar Populations}

   None of the above components are in the SXDS X-ray catalog of Ueda et al.
(2008), which has the sensitivity limit of $6\times 10^{-16}$~$erg/s/cm^2$ in
its most sensitive 0.5--2~keV band. At $z\sim 1$, this corresponds to an
upper limit of $\sim 3.2\times 10^{42}$~$erg/s$ in restframe 1--4~keV.
While this cannot rule out the possibility of an AGN contamination, there is
no strong evidence suggesting that there could be an AGN.
Therefore, we take it that the FIR emission of UDS01 is all due to star
formation. 

   Regardless of the exact counterparts, it is clear that the FIR emission is
due to the interacting system that includes {\tt B}, {\tt C} and {\tt D}. To
estimate $L_{IR}$, we adopted $z_{spec}=1.042$ of {\tt B} as their common
redshift, and fit the decomposed 250~$\mu$m and 24~$\mu$m fluxes to
the SK07 models. As we only had two data points, we did not apply scaling to
the templates but simply read off the luminosities of the best-fit templates
(Fig. 16). In the case of {\tt B+C}, we obtained
$L_{IR}=1\times 10^{12} L_\odot$ for {\tt B}, and did not estimate for
{\tt C} because it only has one data point (250~$\mu$m). In the case of
{\tt B+D}, the best-fit template is the same for both objects, and is the same
as in the previous case. This is shown in the upper panel in Fig. 16. It thus
seems robust that {\tt B} is an ULIRG with $L_{IR}=1\times 10^{12}L_\odot$.
We calculated $L_{IR}^{ext}=5.2\times 10^{11}L_\odot$ and
$L_{IR}^{blk}=4.8\times 10^{11}L_\odot$, which implies 
$SFR_{IR}^{blk}=48$~$M_\odot/yr$. The SED fitting in \S 5.1.2 shows that its
exposed stellar population is an SSP and thus $SFR_{fit}=0$. Therefore,
we get $SFR_{tot}=48$~$M_\odot/yr$. In addition, {\tt D} could also be an ULIRG
of the same $L_{IR}=1\times 10^{12}L_\odot$. It has
$L_{IR}^{ext}=3.8\times 10^{11}L_\odot$ and
$L_{IR}^{blk}=6.2\times 10^{11}L_\odot$, which implies
$SFR_{IR}^{blk}=62$~$M_\odot/yr$. Its exposed region has 
$SFR_{fit}=39$~$M_\odot/yr$, and thus the whole system of {\tt D} has
$SFR_{tot}=101$~$M_\odot/yr$.

\begin{figure*}[tbp]
\centering
\subfigure{
  \includegraphics[width=0.5\textwidth]{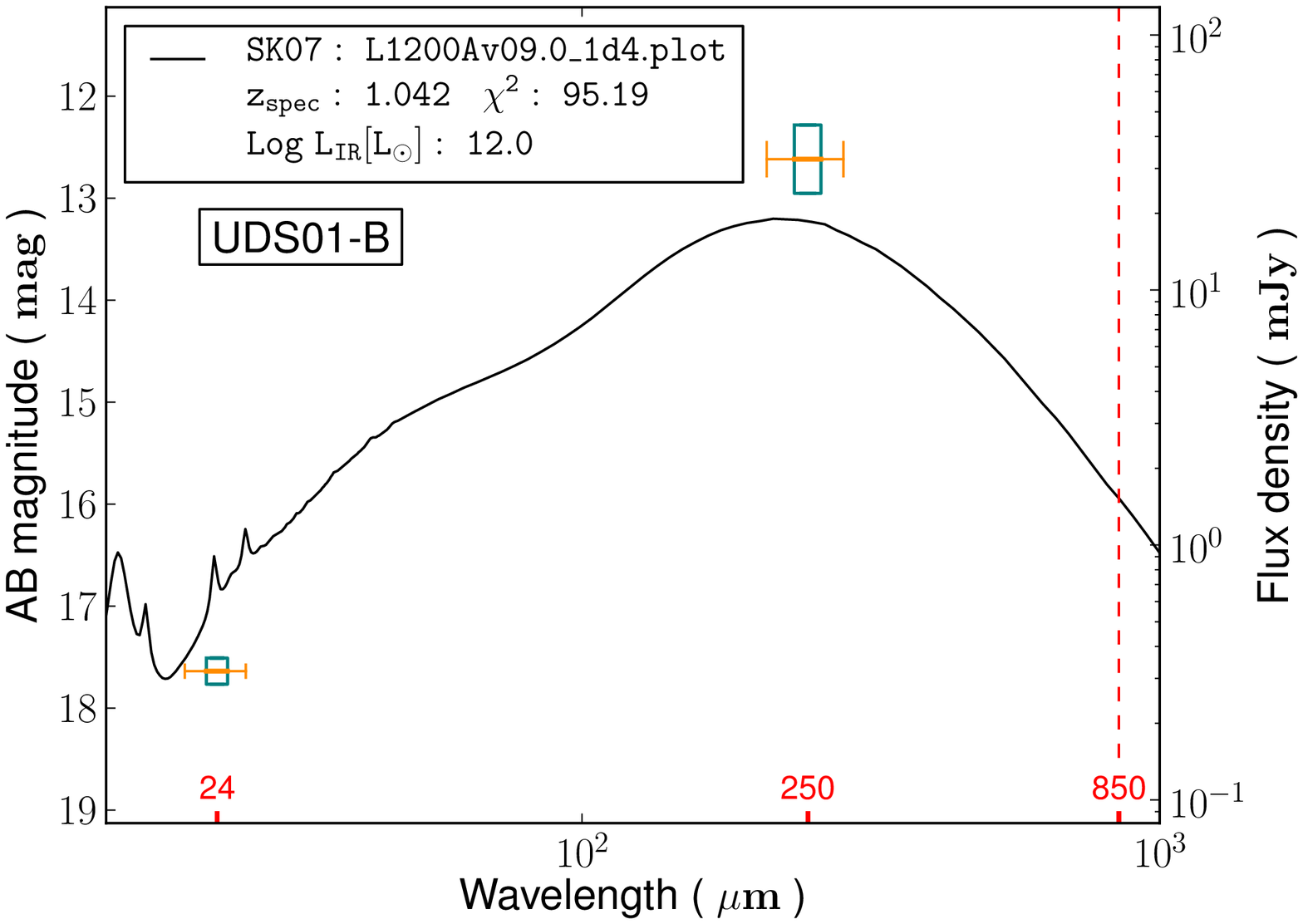}
  \includegraphics[width=0.5\textwidth]{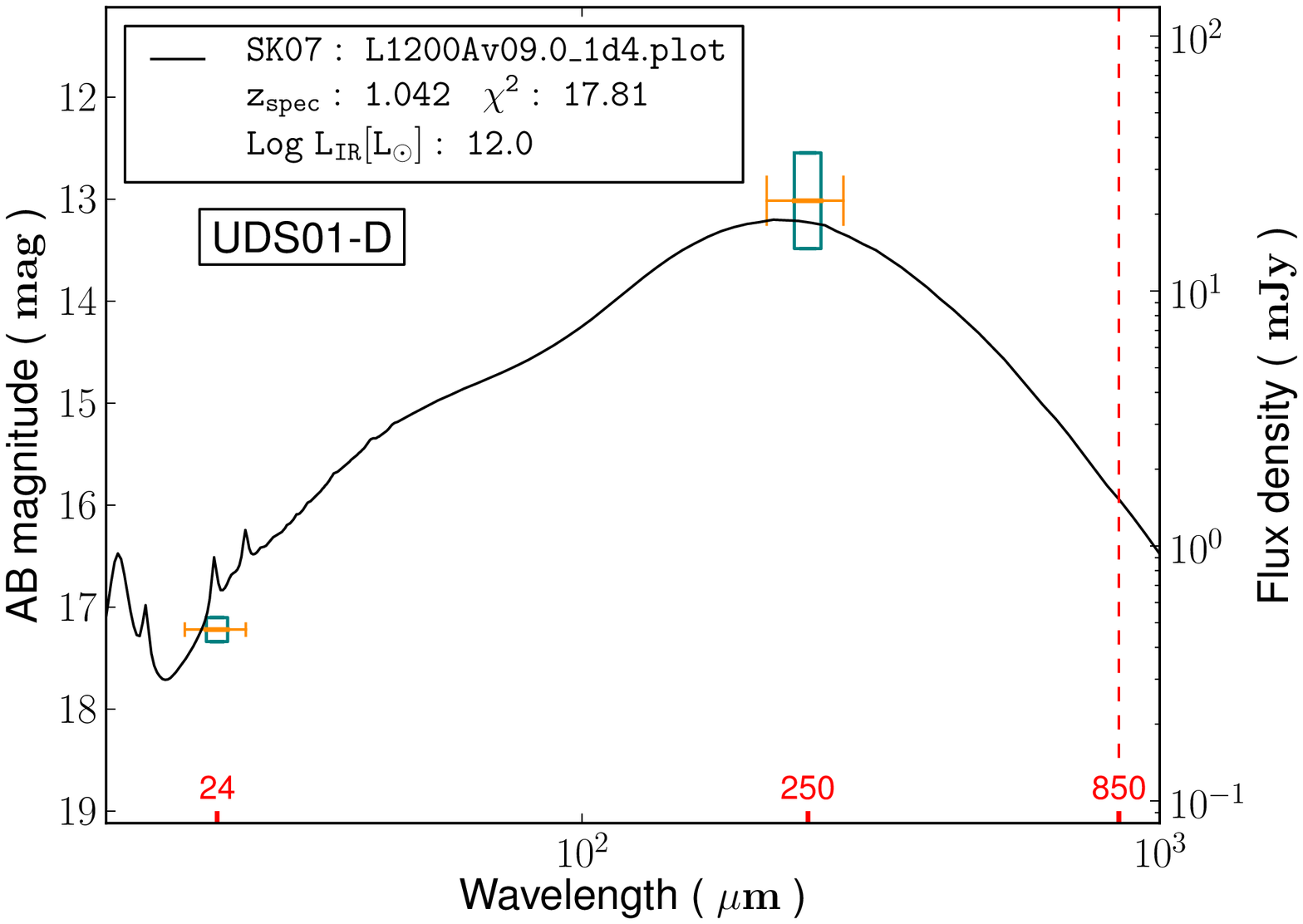}
}
\subfigure{
  \includegraphics[width=0.5\textwidth]{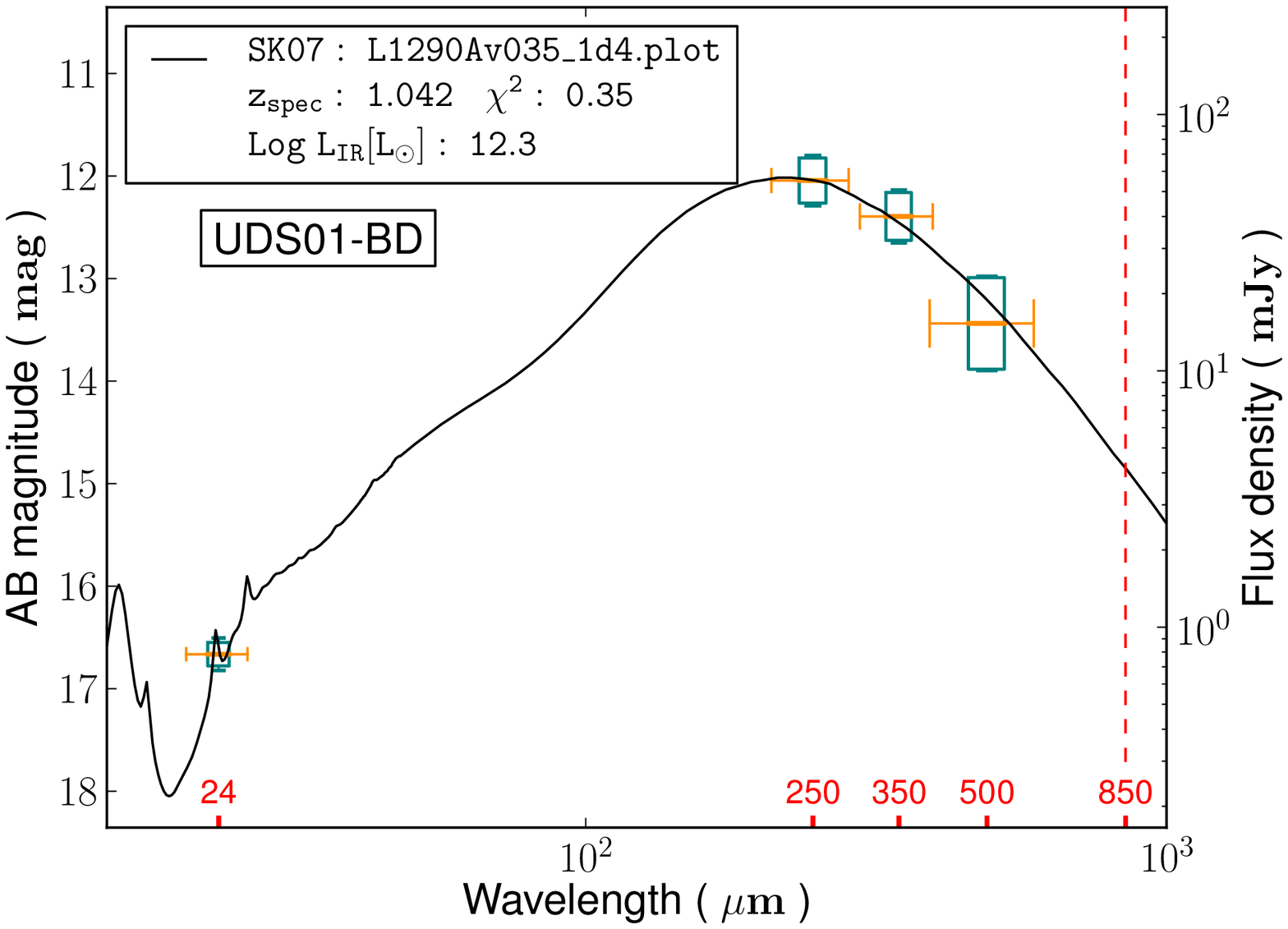}
  \includegraphics[width=0.5\textwidth]{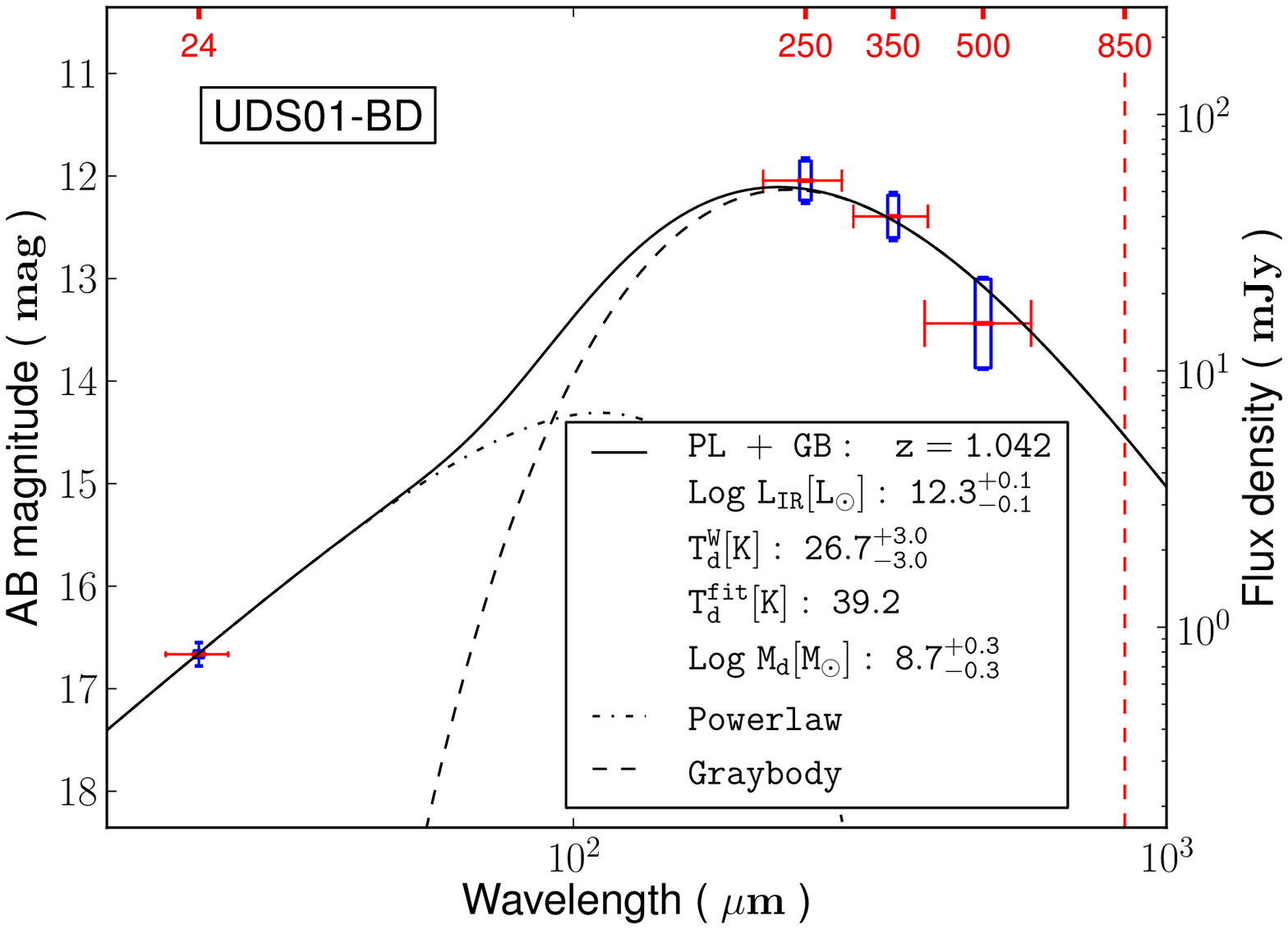}
}
\caption{Mid-to-FIR SED fitting for UDS01 at $z=1.042$. Legends are the same as
in Fig. 6. The two figures on the top show the fits to the SK07 models in the
case where {\tt B} and {\tt D} are the counterparts. No scaling is applied
because there are only two data points available (see text). In the degenerate
case where {\tt B} and {\tt C} are the counterparts, the result for {\tt B} is
exactly the same as shown here, however it is impossible to solve for {\tt C}
as it has only one data point available (lacking 24~$\mu$m). The two figures
on the bottom row show the fitting results when this system is treated as a
whole such that the 350 and 500~$\mu$m data can also be used without doing 
decomposition: the one to the left is the fit to the SK07 models, while the
one to the right is the fit to the power-law + graybody models.
}
\end{figure*}

   As all the possible counterparts could be at the same redshift, we also
considered the case where they were combined. In this case, we used the
reported 350 and 500~$\mu$m flux densities for UDS01 as a whole without doing
decomposition. The fit, as shown in the lower panel of Fig. 16, gives 
$L_{IR}=2.0\times 10^{12} L_\odot$, exactly the value if we were to combine the
luminosities obtained in the case of {\tt B+D} for the two objects individually.

   The stellar mass of {\tt B} obtained in \S 5.1.2 is 
$4.6\times 10^{10} M_\odot$, and therefore it has $SSFR=1.0$~Gyr$^{-1}$,
$T_{db}^{tot}=T_{db}^{blk}=1.0$~Gyr. This is a factor of 7.7 longer than the
age (129~Myr) of the stellar population in the exposed region and is also much
longer than the typical lifetime of an ULIRG. The stellar mass
of {\tt D} is $8.3\times 10^{10}M_\odot$, and therefore in the {\tt B+D} case
it has $SSFR_{IR}=1.2$~Gyr$^{-1}$, $T_{db}^{tot}=822$~Myr and
$T_{db}^{blk}=1.3$~Gyr. The latter is comparable to $T=1.0$~Gyr of the stellar
population in the exposed region, however is much longer than the typical 
lifetime of an ULIRG. We also 
derived $T_d^{fit}=39.2$~K and $M_d=5.0\times 10^8 M_\odot$ for the whole,
combined system. The latter implies the total gas mass of 
$M_{gas}=7.0\times 10^{10} M_\odot$, which would be sufficient to fuel {\tt B}
to double its mass but would fall short for {\tt D}. Overall speaking,
the ULIRG phase of this whole system will only play a minor role in
assembling its stellar mass.

   We can check on the FIR-radio relation for {\tt D}, assuming that the real
solution is the {\tt B+D} case. We obtained $q_{IR}=2.00$, significantly lower
than the mean of $2.40\pm 0.24$. There is no radio source in the catalog of
Simpson et al. (2006) that corresponds to {\tt B}. We can place an upper limit
of 0.1~mJy at 1.5~GHz, which is the flux density limit of this catalog. This
would then result in $q_{IR}<1.91$ for {\tt B}.

\subsection{UDS04 (UDS-J021731.1-050711)}

   This source does not have WFC3 IR data because it is outside of the
CANDELS WFC3 footprint. Therefore, we used UKIRT $JHKs$ data for the SED
analysis in the near-IR. Fortunately, it has CANDELS ACS data for morphological
information. 

\subsubsection{Morphologies and Potential Components}

   Fig. 17 shows the image of UDS04.
the UKIDSS $H$, and the ACS $I_{814}$. 
We searched for its potential 
components in the $I_{814}$ image, and found 50 objects within 18\arcsec\, to 
the 250~$\mu$m position that have $S/N>5$ in $I_{814}$ and $S/N>3$ in $V_{606}$.
The extra detection criterion in $V_{606}$ is to reject any cosmic-ray and
image defect residuals.

    From Fig. 17, it is obvious that UDS04 corresponds to only two possible
sources in 24~$\mu$m. The more dominant one, which coincides better with the
UDS04 centroid, is actually made of eight tightly packed objects detected in
$I_{814}$. The detail of this region is shown in the bottom-left panel of
Fig. 17, and it is clear that these
eight objects are actually the result of a very disturbed system being
resolved. From its morphology alone, it is consistent with being a disc system
undergoing a violent disk instability (VDI; e.g., Dekel et al. 2009).
We designate this system as object {\tt A}. As it turns out, it
has spectroscopic redshift of $z=1.267$ (Simpson et al. 2006; 
Akiyama et al., in prep.).
The other 24~$\mu$m source is much further away from
the 250~$\mu$m centroid, and is made of three $I_{814}$ objects. The
$I_{814}$-band image shows that they are actually a normal spiral 
galaxy and two bright knots on its arms. We designate this spiral as {\tt B}.
This source does not have spectroscopic redshift.

\begin{figure*}[tbp]
\centering
\includegraphics[width=\textwidth] {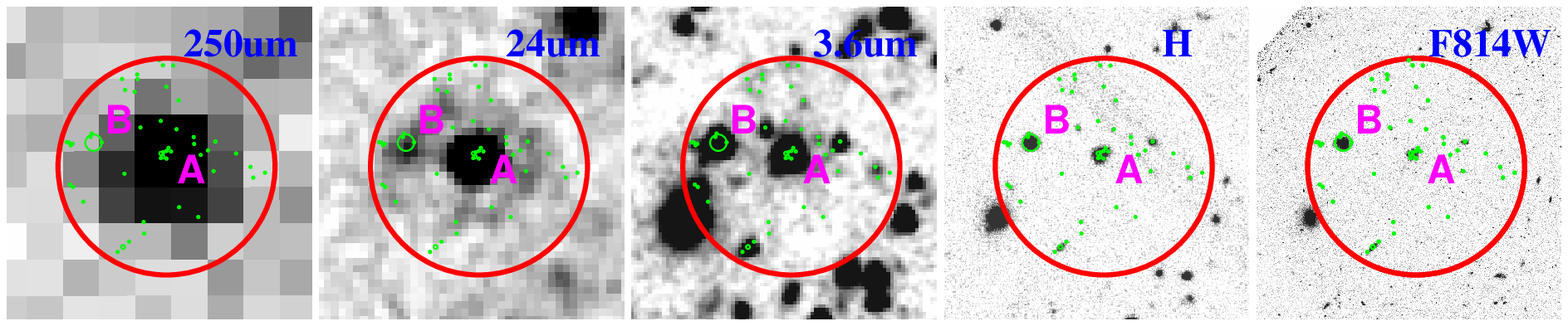}
\includegraphics[width=\textwidth] {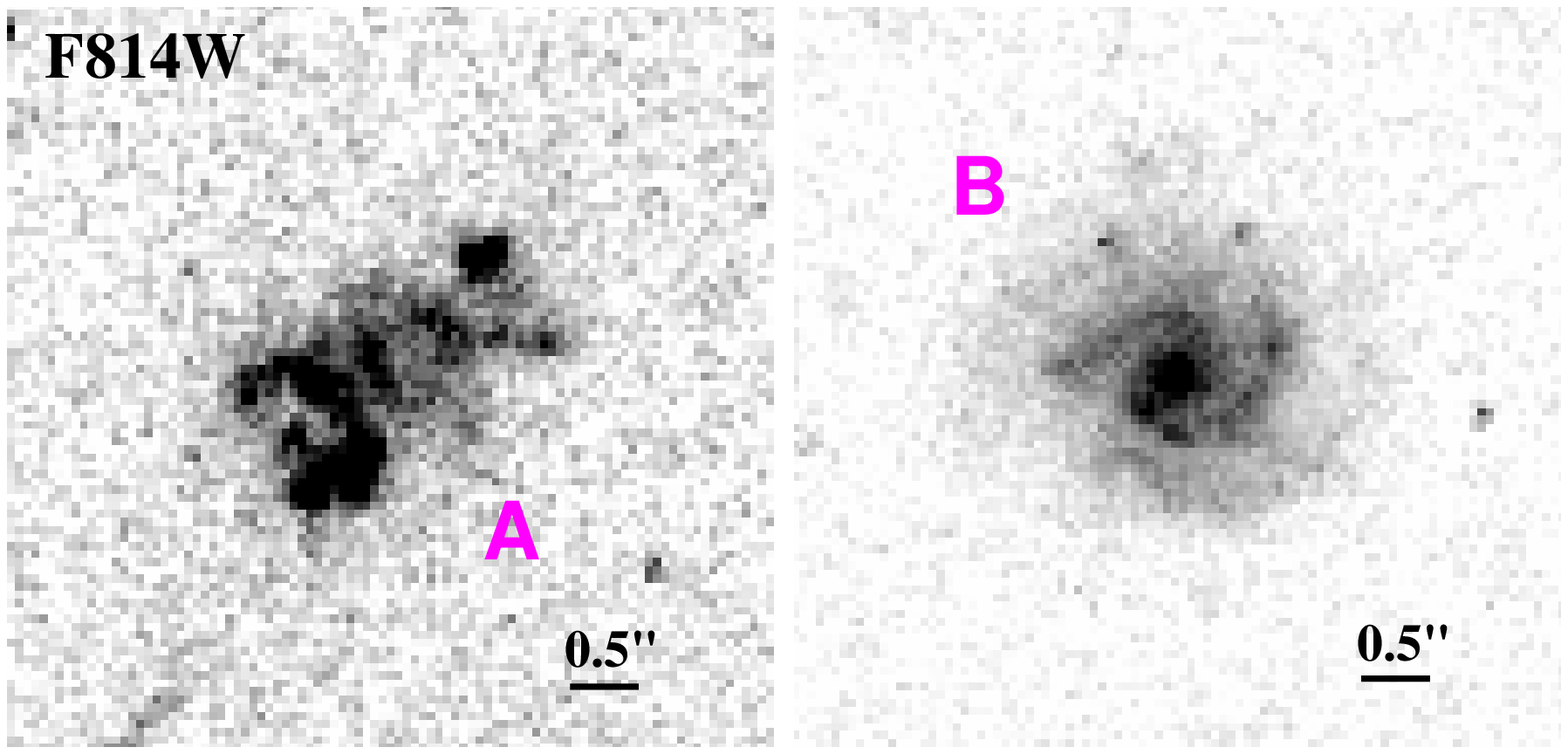}
\caption{FIR to optical images of UDS04 are shown in the top row. The legends
are the same as in Fig. 2. As there are only two potential contributors 
({\tt A} and {\tt B}) to the FIR emission, no zoomed-in view of their 
vicinity is shown. This source lacks 160, 100 and 70~$\mu$m data. The H-band
image is from UKIDSS. The morphological details of {\tt A} and {\tt B} as
revealed by the ACS $I_{814}$ image are shown in the bottom panel.
}
\end{figure*}

\subsubsection{Optical-to-near-IR SED Analysis}

   For the optical part of the SED, we only used the CFHTLS-Wide $u^*griz$ data
as in \S 5.1 but did not include the ACS $V_{606}$ and $I_{814}$ data 
because the ACS images break {\tt A} and {\tt B} into subcomponents
and are not straightforward to obtain the total light. On the other hand, 
{\tt A} and {\tt B} are detected as single objects in the CFHTLS-Wide images, 
which is more appropriate in this context. 
As both objects lack WFC3 IR data, the UKIDSS data were used for the
near-IR part of the SED. We carried out the 
photometry on the UKIDSS DR8 images, using $K_s$-band as the detection band.
Following the case in optical, we used {\tt MAG\_ISO} for colors and then used
the $K_s$-band {\tt MAG\_AUTO} magnitude for normalization. 
In the longer wavelengths, we used the same IRAC data as in \S 5.1.

   The SED fitting results are summarized in Fig. 18. For {\tt A}, we derived 
$z_{ph}=1.32$, which is in good agreement with its $z_{spec}=1.267$ 
($\Delta z/(1+z)=0.02$). Fixing the redshift at its $z_{spec}$,
the stellar mass thus derived is $1.0 \times 10^{11} M_\odot$. The best-fit
stellar population has $\tau=79.4$~Myr and the age of $T=182$~Myr. It also
gives $SFR_{fit}=208 M_\odot/yr$. Therefore, the whole life of this young 
galaxy up to this point has been in an intense star-bursting phase.

   For {\tt B} we get $z_{ph}=0.68$. Thus {\tt B} cannot be associated with
{\tt A}. 

\begin{figure*}[tbp]
\centering
\includegraphics[width=\textwidth] {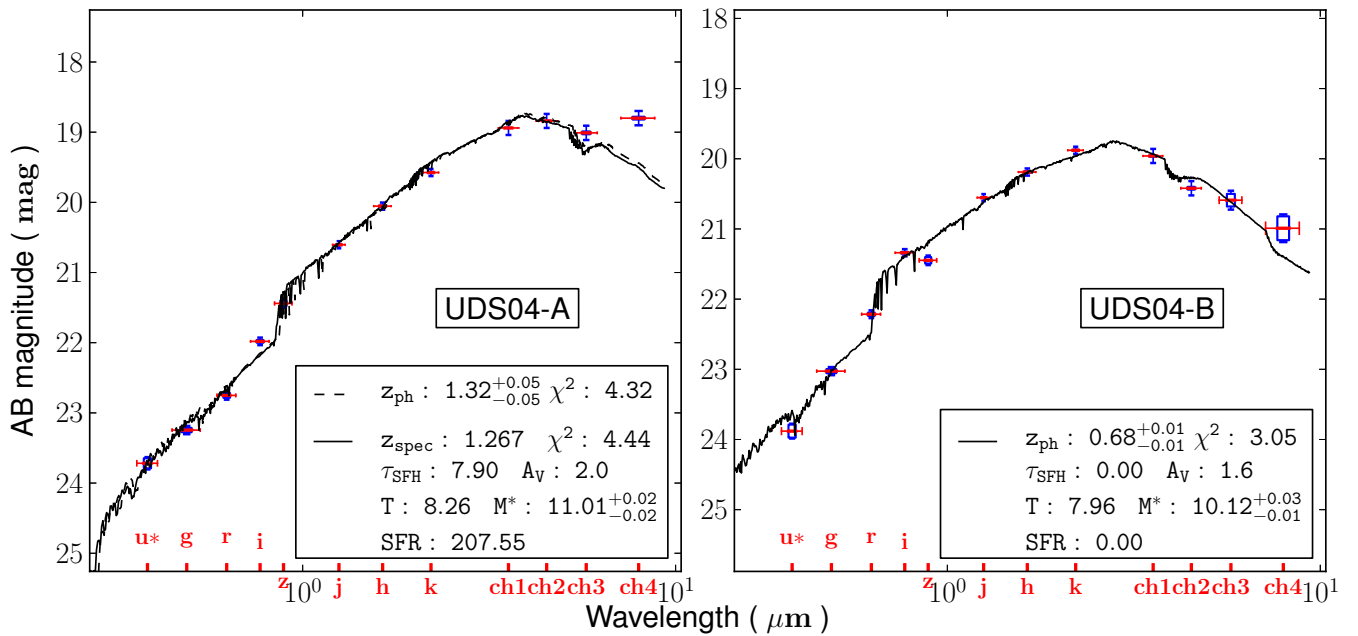}
\caption{Optical-to-NIR SED analysis of the possible contributors to UDS04.
Legends are the same as in Fig. 4. {\tt B} is unlikely to be at the same
redshift as {\tt A}.
}
\end{figure*}

\subsubsection{Decomposition in Mid-to-Far-IR}

   While aperture photometry would be sufficient in 24~$\mu$m, we still 
performed the same PSF fitting technique to derive the fluxes for {\tt A} and
{\tt B}. As both objects are resolved into multiple components in the ACS data,
we used their centroids as measured in the CFHTLS-Wide data to avoid any
ambiguity. The automatically iterative decomposition in 250~$\mu$m shows that
the contribution from {\tt B} is negligible and that {\tt A} should be
the only source responsible for the FIR emission. Thus we adopted the results
from the HerMES DR1 catalog for the three SPIRE bands. 
Table 2 lists these results.

\subsubsection{Total IR Emission and Stellar Populations}

   UDS04 is not detected in the SXDS X-ray data (Ueda et al. 2008). At
$z=1.267$, the sensitivity limit of this catalog corresponds to an upper limit
of $5.8\times 10^{42}$~$erg/s$ in restframe 1.1--4.5~keV. Based on the same
argument as in \S 5.1.4, we take it that the FIR emission of UDS04, which is
all from {\tt A}, is all due to star formation. The mid-to-far-IR SED fitting
results are summarized in Fig. 19. Similar to the case in GOODSN63 where there
is only one contributor to the FIR emission, we incorporated the SPIRE 350 and
500~$\mu$m photometry taken directly from the HerMES DR1 catalog. Using the
SK07 templates, we obtained $L_{IR}=5.0 \times 10^{12} L_\odot$ and thus it is
an ULIRG. From the result in \S 5.2.2, we also obtained
$L_{IR}^{ext}=2.55\times 10^{12} L_\odot$. Therefore, 
$L_{IR}^{blk}=2.55\times 10^{12} L_\odot$ and $SFR_{IR}^{blk}=255$~$M_\odot/yr$.
It has $SFR_{fit}=208$~$M_\odot/yr$, and thus $SFR_{tot}=463$~$M_\odot/yr$. As
derived in \S 5.2.2, the exposed region has the stellar mass of
$1.0\times 10^{11} M_\odot$, which leads to $SSFR=4.6Gyr^{-1}$,
$T_{db}^{tot}=216$~Myr and $T_{db}^{blk}=392$~Myr. The exposed region has 
the best-fit age $T=182$~Myr and the declining time scale $\tau=79$~Myr. In this
sense, UDS04-{\tt A} is similar to GOODSN06-{\tt A}: this young system
is in a star-bursting phase ever since its birth, and while the star formation
in the exposed region is quickly winding down, it is still sustained in the
dust-blocked region. As shown in the right panel of Fig. 19, we derived
$T_d^{fit}=38.2$~K and $M_d=1.0\times 10^9 M_\odot$. The latter implies 
$M_{gas}=1.4\times 10^{11} M_\odot$, which would be sufficient to fuel the
dust-blocked region for the next 550~Myr to more than double the stellar mass.

\begin{figure*}[tbp]
\centering
\subfigure{
  \includegraphics[width=0.5\textwidth]{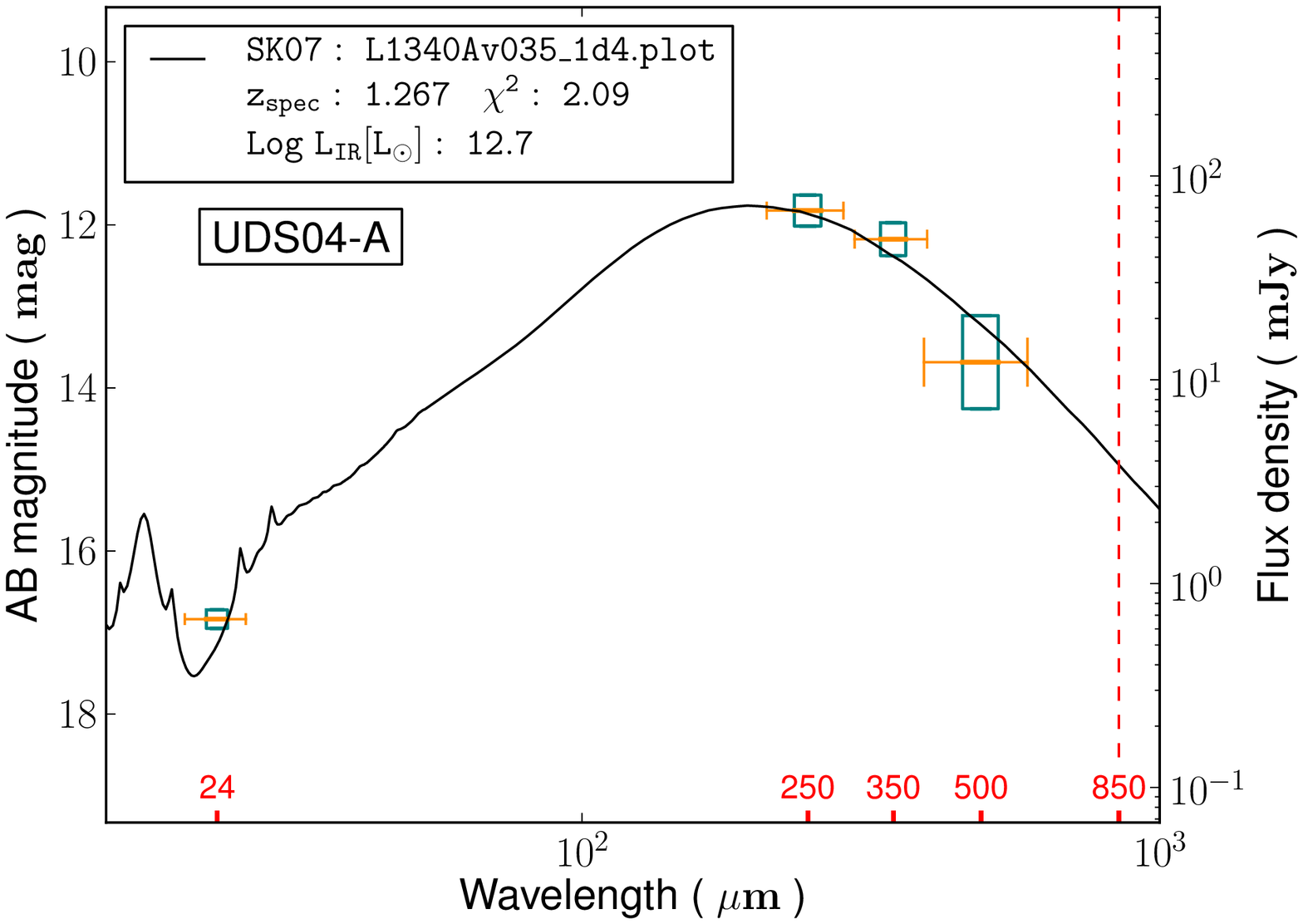}
  \includegraphics[width=0.5\textwidth]{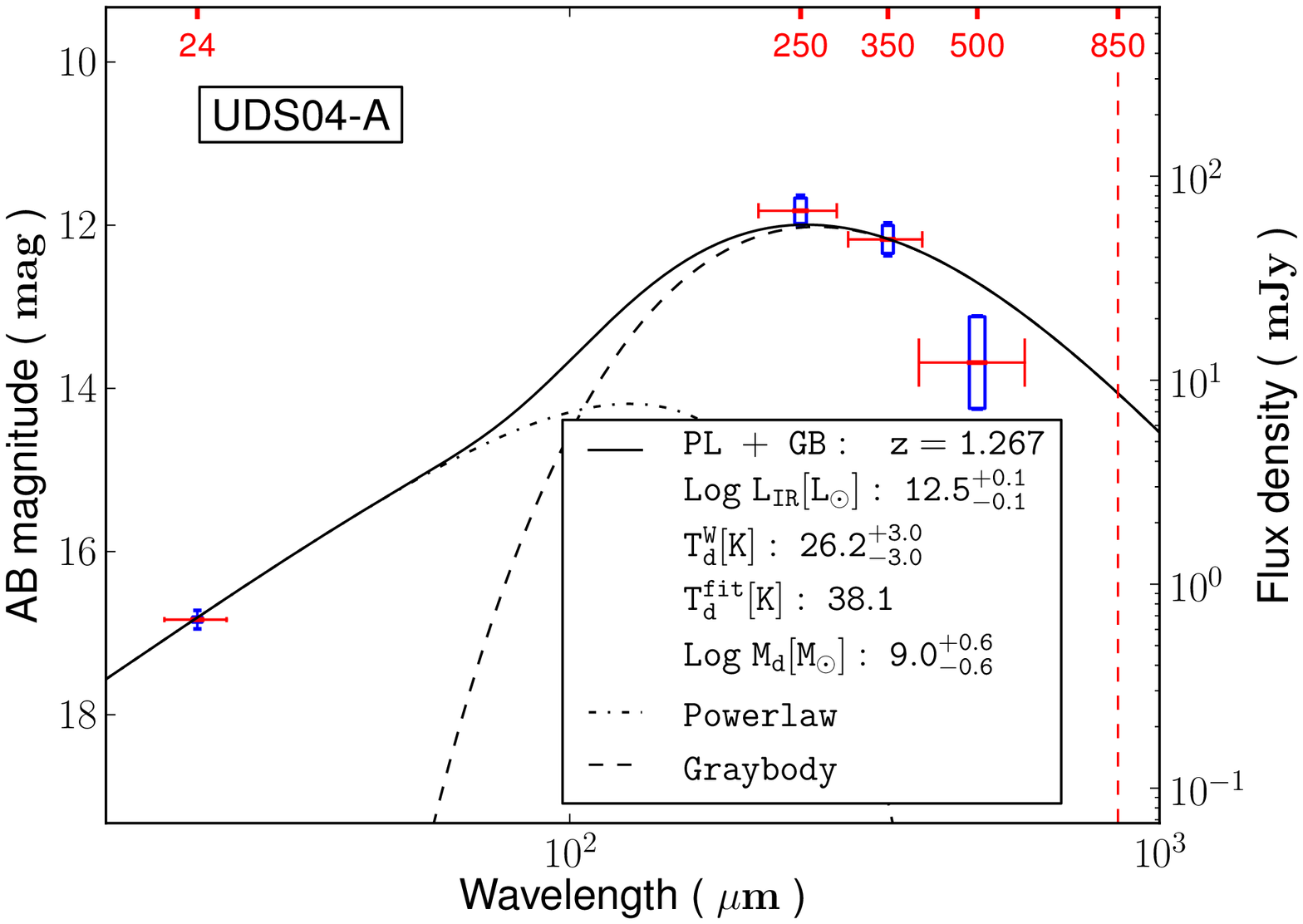}
}
\caption{Mid-to-FIR SED fitting for UDS04-{\tt A} at $z_{spec}=1.267$. The SED
incorporates the SPIRE 350 and 500~$\mu$m photometry directly from the HerMES 
catalog because {\tt A} is the sole contributor at 250~$\mu$m and longer 
wavelengths and hence no decomposition in these two bands are necessary. The
left panel shows the fit to the SK07 models, and the right panel shows the fit
using the power-law + graybody models. Legends are the same as in Fig. 6.
}
\end{figure*} 

   {\tt A} is detected in the 100~$\mu$Jy 1.5~GHz data (Simpson et al. 2006),
whose radio location is at $RA=2^h17^m31^s.17$, 
$DEC=-5$\arcdeg 07\arcmin$09\farcs 35$ (J2000) and is only $0\farcs 26$ away
from the centroid of {\tt A}. It has $S_{1.5GHz}=1.055\pm 0.013$~mJy and is
the strongest radio source in our sample. We derived $q_{IR}=1.74$, which is
much lower than $2.40\pm 0.24$.

\section{Sources in the EGS}

   As mentioned in \S 2, the EGS field has both the HerMES SPIRE and PEP 
PACS data. For the latter, we use the PEP DR1 ``blind'' (i.e., constructed
without using the positions priors from the MIPS 24~$\mu$m data) 100~$\mu$m and
160~$\mu$m catalogs. When this work started, the CANDELS WFC3 observations in
this field was not yet finished, and none of the sources in this field had 
WFC3 data. Therefore, we relied on the ACS $V_{606}$ and $I_{814}$ images from
the AEGIS program to study their morphologies. To define the possible
components, on the other hand, we used the CFHTLS-Deep $u^*griz$ images. We 
made this choice because the CFHTLS-Deep data are significantly deeper than the
AEGIS ACS images. The CANDELS WFC3 data have just become available, and they
are incorporated in the SED analysis. The possible contributors to the FIR
sources remain unchanged. 

\subsection{EGS07 (EGS-J141900.3+524948)}

\subsubsection{Morphologies and Potential Components}

   Fig. 20 shows the images of EGS07. Within 18\arcsec\, of the 250~$\mu$m 
source centroid, there are 55 objects detected in the CFHTLS Deep data that 
have $S/N\geq 5$ in the $i$-band (measured in the MAG\_ISO aperture).
Our inspection identified 8 objects as the
potential contributors to the 250~$\mu$m flux, which we label as object 
``{\tt 1}'' to ``{\tt 8}'' in increasing order of the distance to the
250~$\mu$m source centroid. To emphasize the choice of the ground-based images
for the counterpart identification, the labeling scheme here is numerical 
instead of alphabetical.  In 24~$\mu$m image, these objects are all blended
together as one single source and cannot be separated. This is also true in the
100 and 160~$\mu$m images. 

    The central source, {\tt 1}, has spectroscopic redshift of $z=1.497$ from 
the DEEP2 galaxy redshift survey (Newman et al. 2013). In the ACS images, this
source shows a very disturbed morphology that is indicative of strong 
interaction. This suggests that it could be the dominant contributor.
Running SExtractor on the $I_{814}$ image, this source is broken
into five components extending over $\sim 4.0$\arcsec, for which we label as
{\tt 1-a} to {\tt 1-e} (see Fig. 21). The component {\tt 1-e} is compact
and consistent with being point-like, and the other are all diffuse and
irregular in shape. While they cannot be separated by
SExtractor on the CFHTLS Deep images, most of these components are still
visually discernible in these images with the exception that {\tt 1-e} is 
swamped by {\tt 1-a} due to the coarser spatial resolution and thus cannot be
separated. As the CFHTLS data are deeper than the ACS images, 
{\tt 1-b} is better detected (so is {\tt 2}). The centroids of the
250 and 160~$\mu$m images are between {\tt 1-a}, {\tt 1-d} and {\tt 1-e}, while
the 100~$\mu$m centroid is $\sim 2\farcs 5$ to the north.

     Other objects do not have obvious features. {\tt 4}, {\tt 5} and {\tt 8}
are compact and point-like, while {\tt 2}, {\tt 3}, {\tt 6}, and {\tt 7} are 
extended. 

\begin{figure*}[tbp]
\centering
\includegraphics[width=\textwidth] {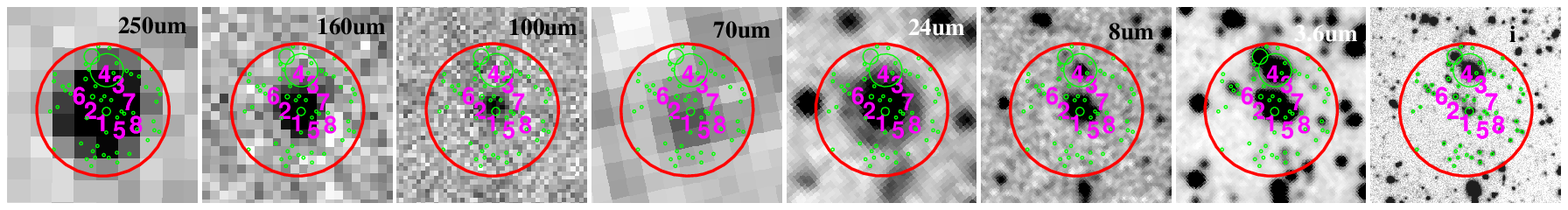}
\includegraphics[width=\textwidth] {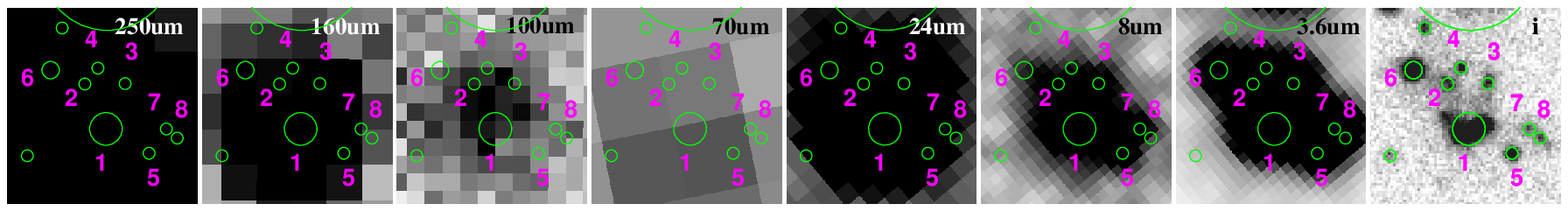}
\caption{FIR to optical images of EGS07. The legends and the organization of
the panels are the same as in Fig. 2. The $i$-band image is from 
CFHTLS-Deep.
}
\end{figure*}

\begin{figure*}[tbp]
\centering
\includegraphics[width=\textwidth] {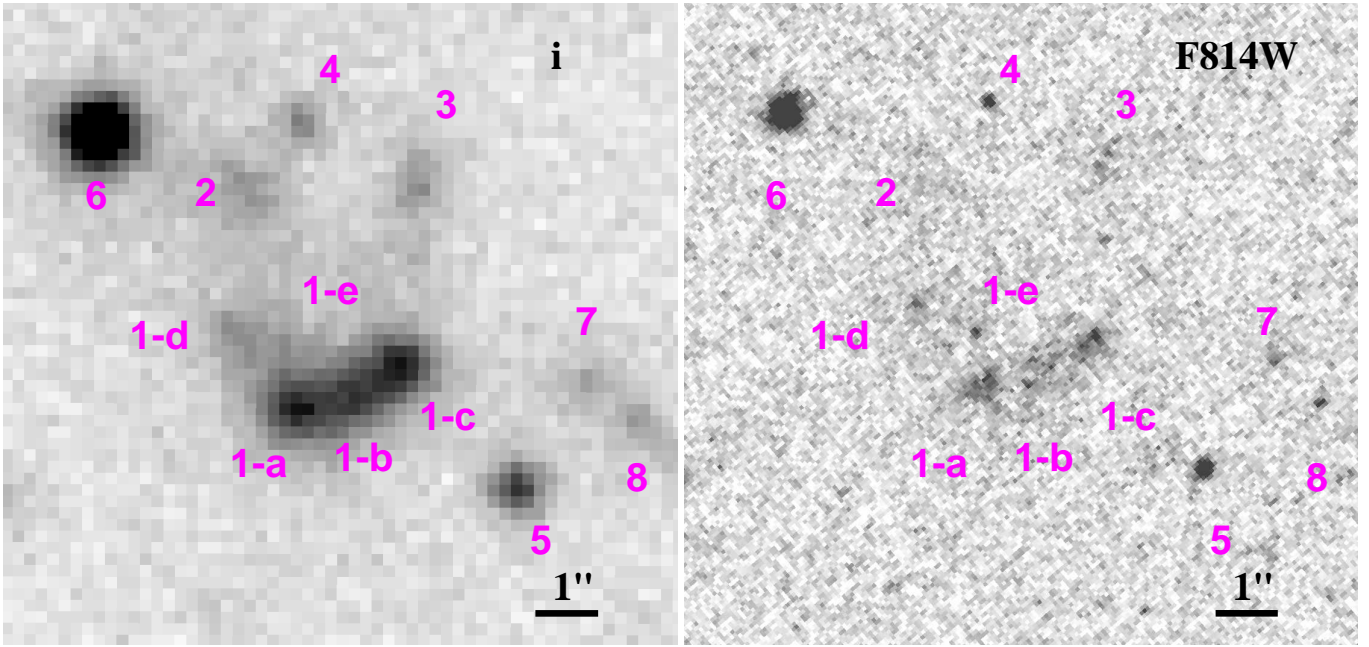}
\caption{Morphological details of the potential contributors to EGS07. The left
panel is the $i$-band image from the CFHTLS-Deep program, while the right panel
is the $I_{814}$-band image from the AEGIS ACS program. The CFHTLS images are
deeper that the ACS images, but the latter show more morphological details.
}
\end{figure*}

\subsubsection{Optical-to-near-IR SED Analysis}

   The optical-to-near-IR SED were constructed using the CFHTLS-Deep $u^*griz$
and the CANDELS WFC3 $J_{125}$ and $H_{160}$ data, following the same
procedures as in \S 5.1.2. The AEGIS ACS data were not used for this purpose 
because they are not as deep as the CFHTLS data. Unfortunately, the TFIT
procedure using the WFC3 $H_{160}$ does not produce satisfactory results in
IRAC (with bad residual maps) due to some unknown reasons, and therefore we had
to exclude the IRAC data in this analysis.

  The results are summarized in Fig. 22. Object {\tt 1} has $z_{ph}=1.46$, 
which agrees with its $z_{spec}=1.497$ very well ($\Delta z/(1+z)=0.01$).
Object {\tt 6} has $z_{ph}=1.48$, and from the $P(z)$ distribution it seems
very likely that it is at the same redshift as {\tt 1}. They are $3\farcs 4$
apart, corresponding to 28.9 kpc, and hence could be in the same group. Their
physical properties are rather different, however. {\tt 1} has a high stellar
mass of $6.9\times 10^{10} M_\odot$, a moderate age of 724~Myr, and an extended
SFH with $\tau=7$~Gyr (i.e., almost constantly star forming as compared to its
age). The inferred current SFR is 137~$M_\odot/yr$. {\tt 6}, on the other hand,
is significantly less massive ($M^*=8.3\times 10^9 M_\odot$) and much younger
($T=182$~Myr), and has assembled most of its current stellar mass through a 
short, modest episode of star formation ($\tau=60$~Myr) that
leaves the current SFR of $\sim 10 M_\odot/yr$.

   Another possible group consists of objects {\tt 2}, {\tt 4}, {\tt 5},
and {\tt 8}, which have $z_{ph}=1.01\pm 0.07$. In terms of stellar mass,
this group is not significant, as the most massive one, {\tt 5}, has only
$M^*=3.2\times 10^9 M_\odot$.

   Object {\tt 3} and {\tt 7} have $z_{ph}=1.23$ and 1.89, respectively, and
thus are unlikely associated with either of the two possible groups.

\begin{figure*}[tbp]
\centering
  \includegraphics[width=\textwidth]{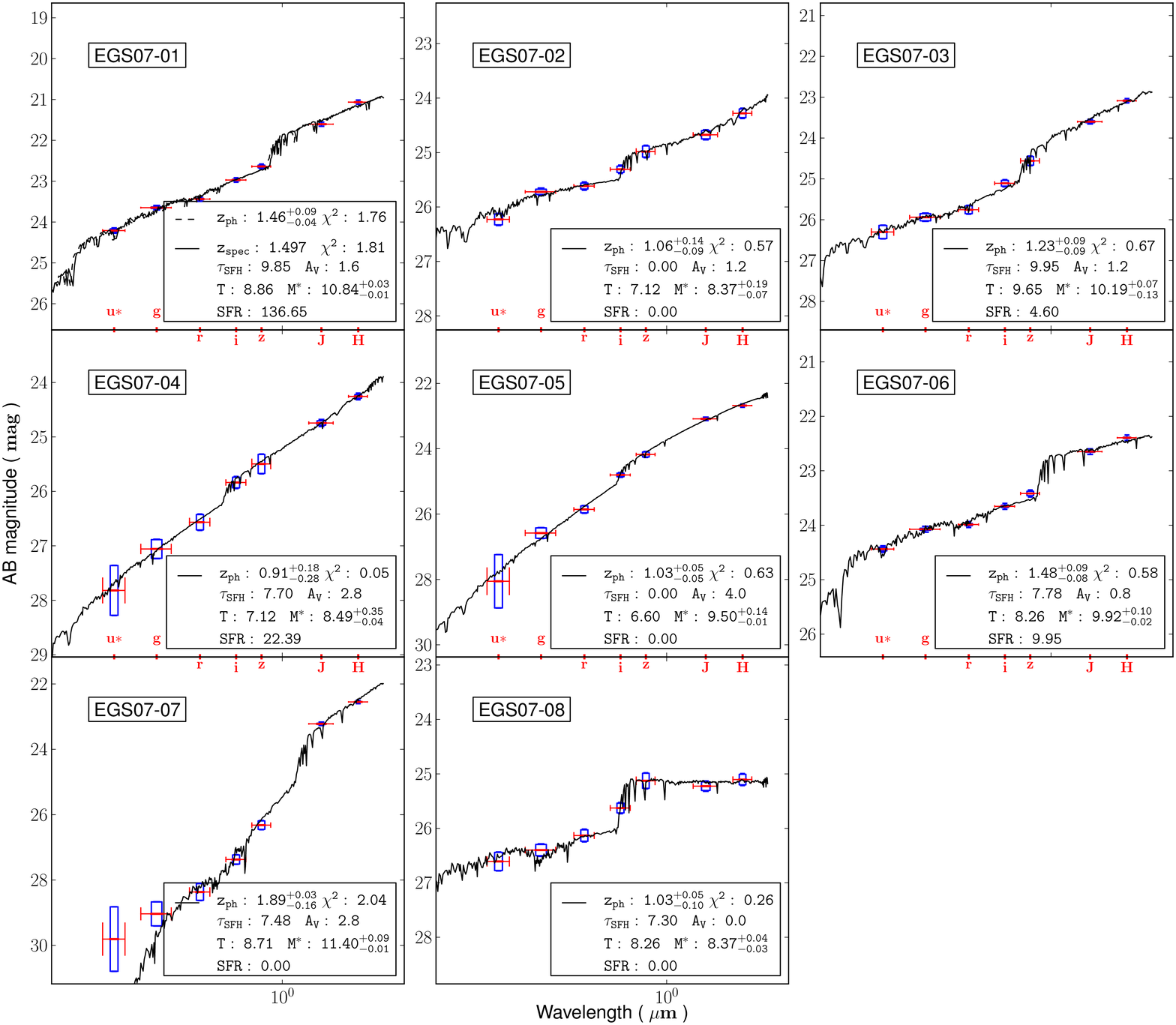}
  \includegraphics[width=0.5\textwidth]{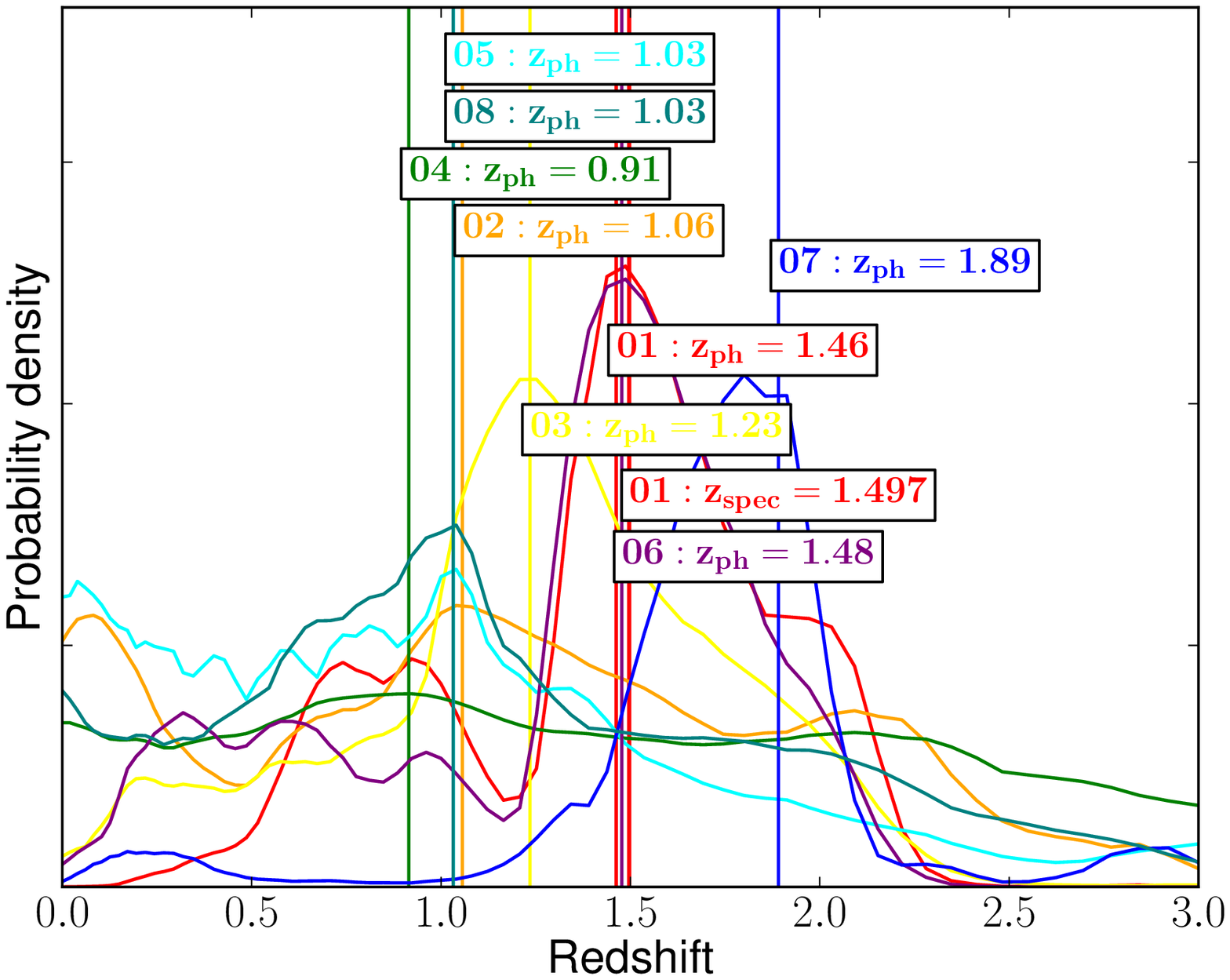}
\caption{Optical-to-NIR SED analysis of the possible contributors to EGS07.
Legends are the same as in Fig. 4. Objects {\tt 1} and {\tt 6} are likely at
the same redshift. Another possible group consists of {\tt 2}, {\tt 4}, {\tt 5},
and {\tt 8} at $z_{ph}=1.01$.
}
\end{figure*}

\subsubsection{Decomposition in Mid-to-Far-IR}

  In 24~$\mu$m, the automatically iterative fit settled on objects {\tt 1}, 
{\tt 2} and {\tt 3}. These three objects account for $\sim$ 61\%, 32\% and 6\%
of the total flux, respectively. For 70~$\mu$m, it converged on only {\tt 1},
although this leaves visible residual. In 100~$\mu$m, {\tt 1} was still the 
dominant component and account for $\sim$ 67\% of the total flux, however the 
secondary components changed to {\tt 4} and {\tt 6}, which are responsible for
15\% and 17\% of the total flux, respectively. In 160~$\mu$m, the automatically
iterative fit failed, and we had to use the trial-and-error method. The only
sensible results were obtained when the fit used {\tt 1} and {\tt 4} or just
{\tt 1}, and we adopted the former case because {\tt 4} seems to be 
non-negligible in 100~$\mu$m, although this reason is not overwhelming. The
flux density of {\tt 1} would increase by $\sim 10$\% if we were to choose
otherwise. Similarly, we had to use the trial-and-error approach to fit
250~$\mu$m, and the only sensible results were obtained when using {\tt 1} and
{\tt 4} or just {\tt 1}, and we again adopted the former case. If we were to
chose otherwise, the flux density of {\tt 1} would increase by 24\%. The
decomposition in 250~$\mu$m in demonstrated in Fig. 23.
Regardless of the exact counterparts in each band, it is obvious
that {\tt 1} is the dominant contributor to the mid-to-far-IR emission. The
flux densities of this object are summarized in Table 2.

\begin{figure*}[tbp]
\centering
  \includegraphics[width=\textwidth]{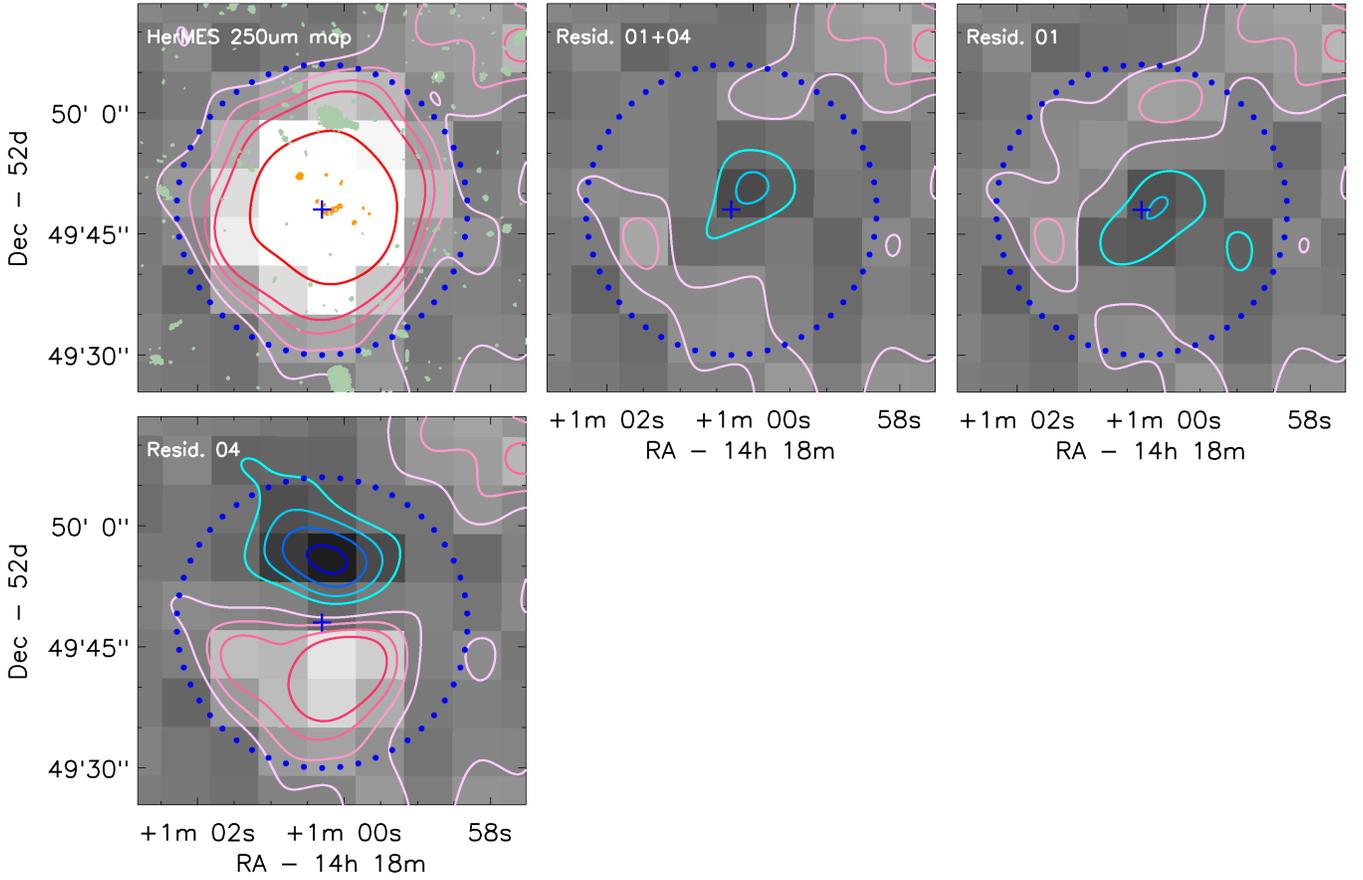}
\caption{Demonstration of the decomposition in 250~$\mu$m for EGS07. The first
panel shows the original 250~$\mu$m image, while the others show the residual
maps of the different decomposition schemes where different input sources are 
considered (labeled on top). Legends are the same as in Fig. 5. 
The ``{\tt 1+4}'' case is adopted as the final solution. See text for details.
}
\end{figure*}

\subsubsection{Total IR Emission and Stellar Populations}
   
   Fig. 24 summarizes the analysis of the FIR emission of EGS07-{\tt 1}. As it
turns out, the SK07 models can provide a good fit from 70~$\mu$m to 250~$\mu$m,
but the 24~$\mu$m data point cannot be satisfactorily explained (left
panel of Fig. 24). Nevertheless, the derived 
$L_{IR}=4.0\times 10^{12} L_\odot$ agrees reasonably well with 
$L_{IR}=5.0\times 10^{12} L_\odot$ inferred from the 
analytic fit (right panel), and hence it seems robust that this object is 
an ULIRG.

   Object {\tt 1} has an X-ray counterpart in
the 800~Ks AEGIS Chandra data (Nandra et al. in prep.; Laird et al. 2009)
\footnote{Object {\tt 5} also has an X-ray counterpart that has the full-band
0.5--10~keV fluxes of $12.78^{+1.95}_{-1.79}\times 10^{-16}$~$erg/s/cm^2$.
However we do not discuss it further as it does not seem to be a significant
contributor to the FIR flux.},
with the full-band fluxes in 0.5--10~keV of
$7.05^{+1.55}_{-1.37}\times 10^{-16}$~$erg/s/cm^2$.
At $z=1.497$, this implies a total X-ray luminosity in restframe 1.2--25~keV
of $1.0\times 10^{43}$~$erg/s$, which is in the AGN regime.
   Thus the FIR emission of {\tt 1} might have an AGN contribution,
which might be the reason that the 24~$\mu$m data point is not
well fitted. Its X-ray position has a very good accuracy of $0\farcs 56$, 
and is consistent with the sub-component {\tt 1-e}. Among all the
sub-components of {\tt 1}, {\tt 1-e} is unique in its point-like morphology and
its absence from the $V_{606}$-band. This raises a possibility that {\tt 1-e}
might not be part of {\tt 1} but actually be a background quasar, which could
be at $z\sim 5$ given its being a dropout from $V_{606}$. If true, its X-ray
luminosity would be $4.9\times 10^{44}$~$erg/s$. However, the FIR emission of
EGS07 would not be from {\tt 1-e} alone, otherwise the SPIRE flux densities
would peak at 500~$\mu$m instead of between 250 and 350~$\mu m$. Estimating the
IR luminosity for {\tt 1-e} only is hardly possible in this case, as {\tt 1-e}
is so close to other subcomponents that it cannot be decomposed in the FIR. 
Therefore, while it is clear that {\tt 1} has an AGN and that the AGN 
could contribute significantly to the FIR emission, the nature of object
{\tt 1} remains inconclusive at this point. Taking $L_{IR}$ at its face value,
one would infer an upper limit of $SFR_{IR}=400 M_\odot/yr$. 
We refrain from discussing the SFR of this system in the usual way.

\begin{figure*}[tbp]
\subfigure{
  \includegraphics[width=0.5\textwidth]{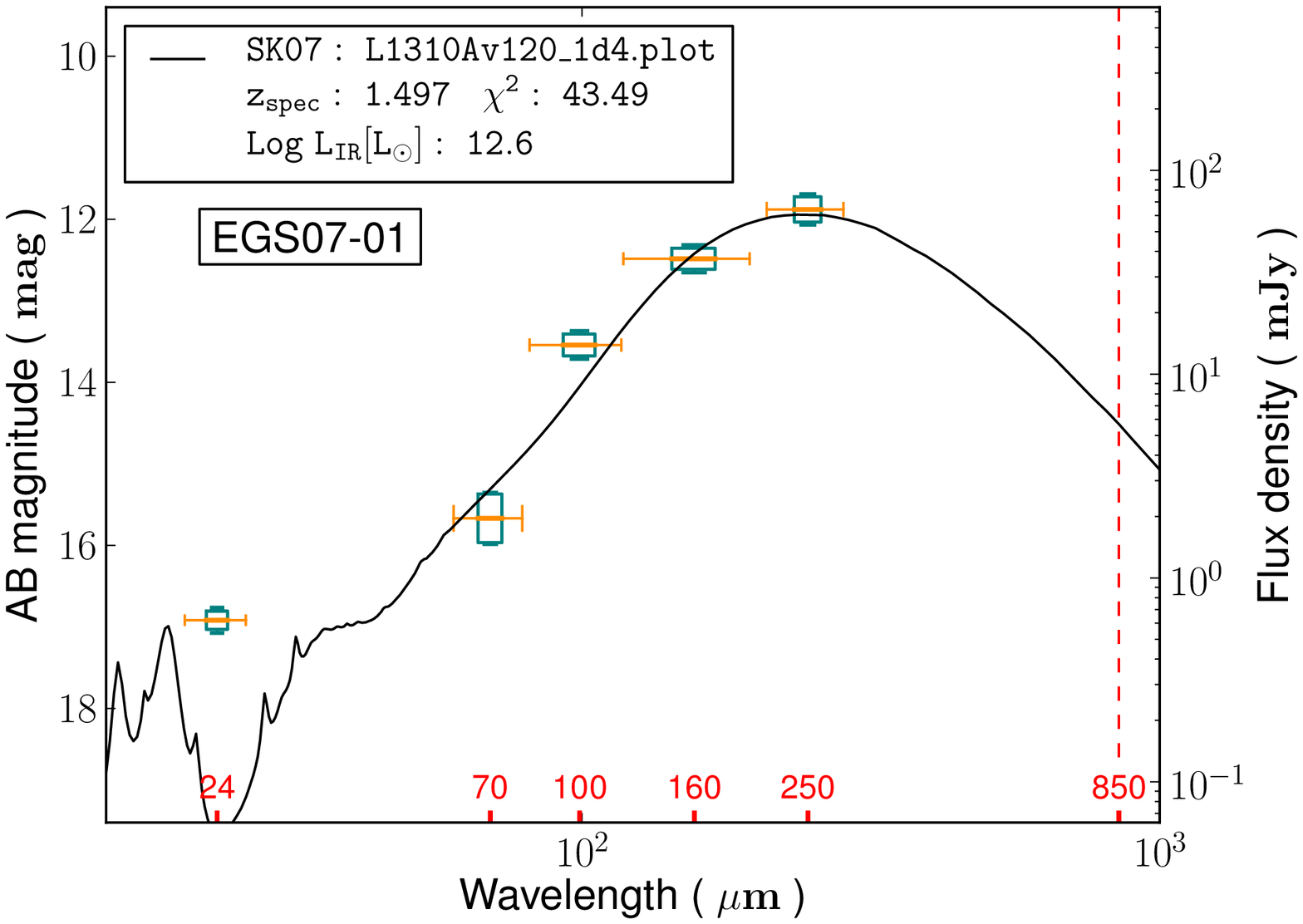}
  \includegraphics[width=0.5\textwidth]{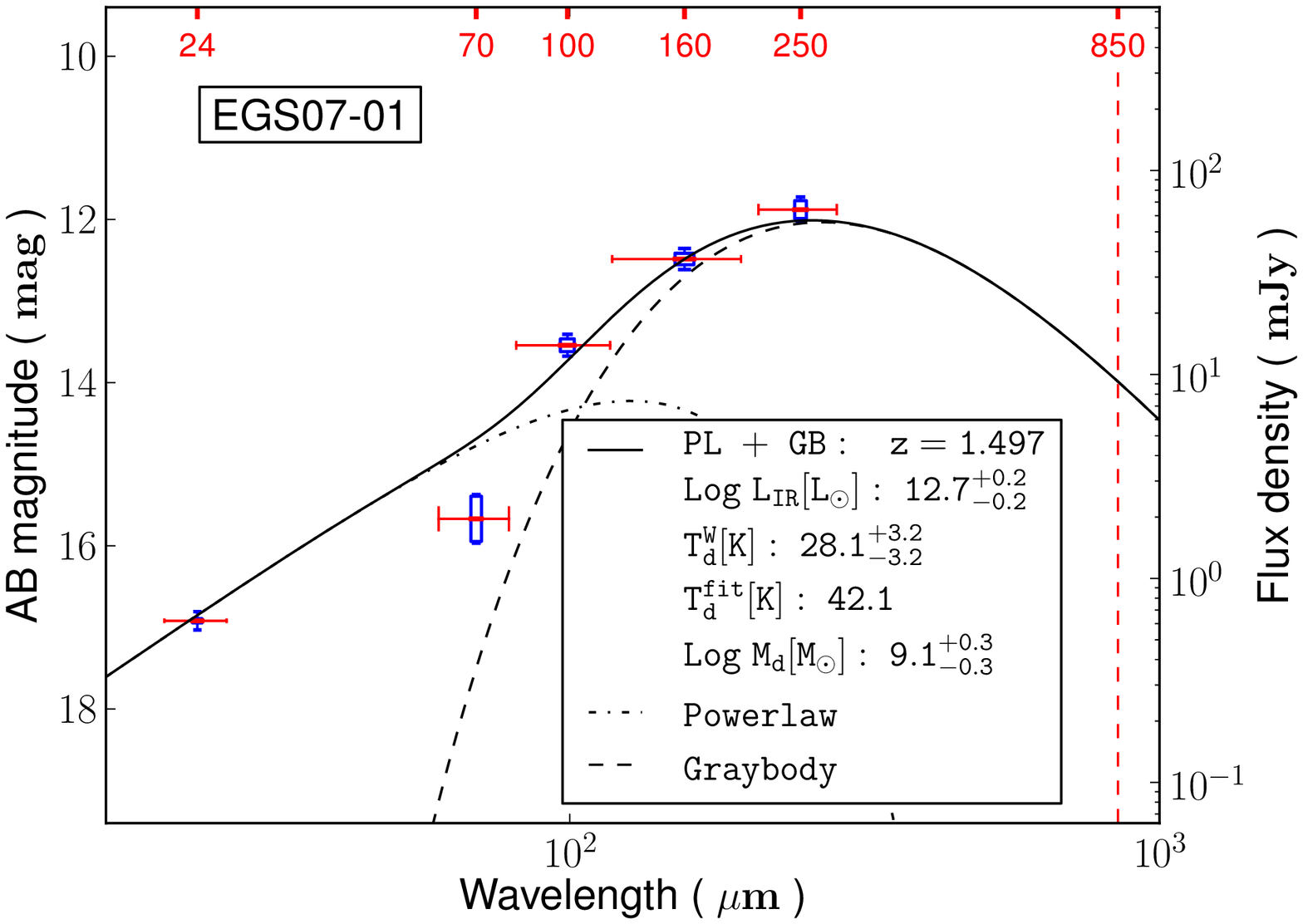}
} 
\caption{Mid-to-FIR SED analysis for EGS07. Legends are the same as Fig. 6.
The 24~$\mu$m data point cannot be well explained by the best-fit SK07 models
(left panel). Nevertheless, the derived $L_{IR}$ is in good agreement with the
result from the fit to the powerlaw + graybody models (right panel).
}
\end{figure*}

    Finally, we investigate the FIR-radio relation for this source. The
catalog of Ivison et al. (2010) include a strong radio source at the
position of {\tt 1-e}, which has $S_{1.4GHz}=0.316$~mJy. We obtained
$q_{IR}=2.02$, which is significantly lower than $2.40\pm 0.24$. 

\subsection {EGS14 (EGS-J142025.9+525935)}

\subsubsection{Morphologies and Potential Components}

    Fig. 25 shows the images of EGS14.
Within 18\arcsec\, of the 250~$\mu$m source centroid, there are 58 objects
detected in $i$ that have $S/N\geq 5$. Based on inspection, 
we identified 10 objects as potential contributors to the 
250~$\mu$m flux, which we label as ``{\tt 1}'' to ``{\tt 10}''. In this area,
only two sources are discernible in the 24~$\mu$m image, which seem to be
dominated by {\tt 5} and {\tt 9}, respectively. While it is difficult see
from the 100 and 160~$\mu$m images, the PEP DR1 catalog identifies two sources
in this area as well, and their positions are also consistent with {\tt 5} and
{\tt 9}, respectively. All this suggests that {\tt 5} and {\tt 9} are the
major contributors to the 250~$\mu$m flux. 

    The ACS images reveal that both {\tt 5} and {\tt 9} have interesting
morphologies (Fig. 26). They are $4\farcs 39$ and $5\farcs 67$ from the
250~$\mu$m centroid, respectively, and are on the opposite sides ($10\farcs0$
apart). {\tt 5} is a very disturbed, curvy and knotty system whose shape
resembles a scorpion. {\tt 9} is also highly irregular, having a bright central
core and at least two satellite features around it.
Among other objects, {\tt 1}, {\tt 4} and {\tt 8} are compact, and the
rest are extended. {\tt 2} and {\tt 10} seem to be disc systems with dust
lanes, {\tt 3} and {\tt 6} are amorphous, and {\tt 7} actually consists of
one central core and two smaller objects on each side. 

\begin{figure*}[tbp]
\centering
  \includegraphics[width=\textwidth] {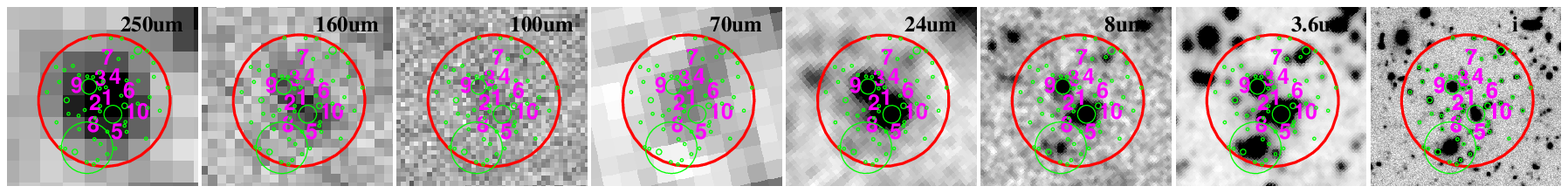}
  \includegraphics[width=\textwidth] {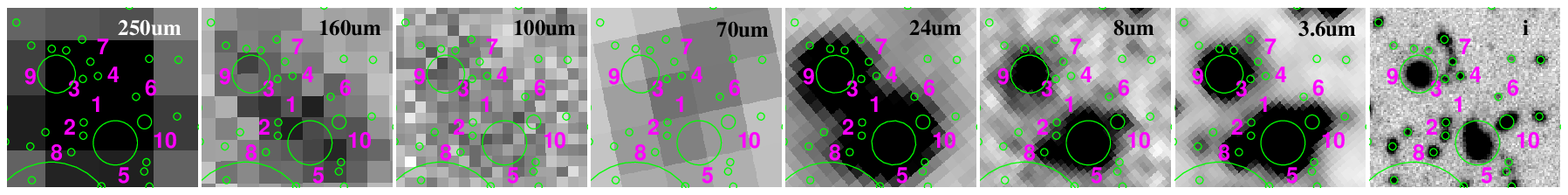}
\caption{FIR to optical images of EGS14. The legends and the organization of
the panels are the same as in Fig. 2. The $i$-band image is from CFHTLS-Deep.
}
\end{figure*}

\begin{figure*}[tbp]
\centering
\includegraphics[width=\textwidth] {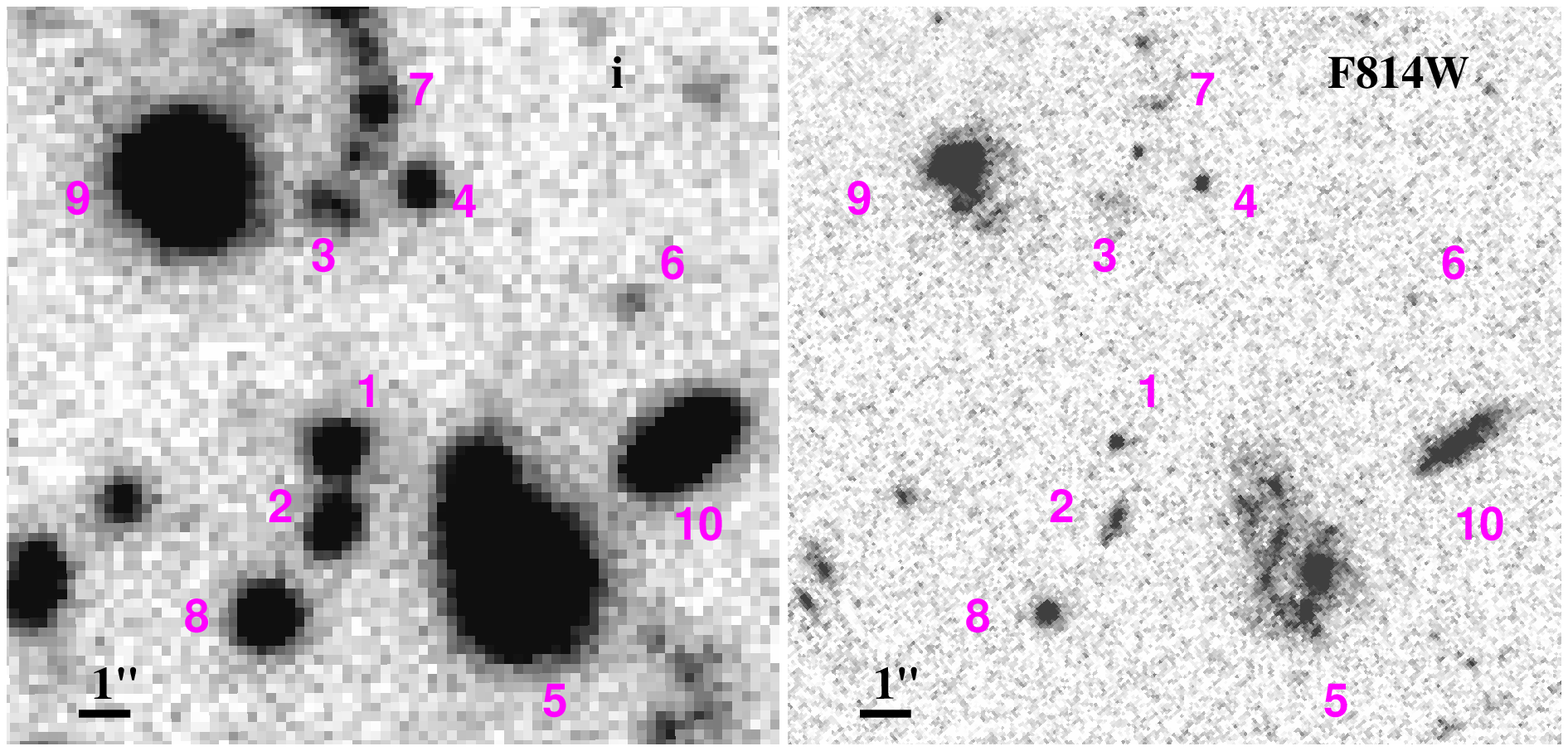}
\caption{Morphological details of the potential contributors to EGS14 in
$i$-band from CFHTLS-Deep (left) and in $I_{814}$-band from the AEGIS ACS
(right).
}
\end{figure*}

\subsubsection{Optical-to-near-IR SED Analysis}

   None of these ten objects have spectroscopic redshifts, and we derived their
$z_{ph}$ in the usual way. The results are summarized in Fig. 27.
The optical-to-near-IR SED are different from those
in \S 6.1.2 in that we are able to incorporate the SEDS IRAC 3.6 and 4.5~$\mu$m
data, as the TFIT procedure produces good results comparable to those in 
\S 4 and 5. Object {\tt 6} has to be excluded, as it is only significantly 
detected in $r$ and $i$. This leaves nine objects in total. 
The fits to objects {\tt 1}, {\tt 3} and {\tt 7} are
poor, but these three sources are likely irrelevant to the FIR emission
(see below in \S 6.2.3.).
Most of these objects are likely at $z\approx 1$. In particular,
objects {\tt 5} and {\tt 9}
have their $z_{ph}$ agrees extremely well, at 1.11 and 1.12, respectively.
{\tt 2}, {\tt 4} and {\tt 8} all have $\Delta z/(1+z)\leq 0.05$ with respect
to {\tt 5} and {\tt 9}, and the peaks of their $P(z)$ distributions coincide
well with each other. This suggests that EGS14 could be in a rich group
environment.
Objects {\tt 5} and {\tt 9} are by far the most dominant members of this
possible group. However, their stellar populations are distinctly different.
{\tt 5} has a very high stellar mass of $3.0\times 10^{11} M_\odot$ and an old
age of 5.5~Gyr, comparable to the age of the universe at this redshift. It has
a prolonged SFH with $\tau=4.0$~Gyr, which is comparable to its age, and a 
modest on-going $SFR=45.5 M_\odot/yr$. If {\tt 5} indeed is a merging system
as its morphology suggests, its subcomponents must have been persistently and 
gradually assembling their stellar masses right after the Big Bang, at a
nearly constant rate of $\sim 50 M_\odot/yr$ (combining over all its
subcomponents). Using the conversion from SFR to UV luminosity as in
Madau et al. (1998), one can see that the progenitor of this entire system
would be quite readily visible by $z\approx 10$, with a total magnitude of
$H_{AB}\approx 25.2$ if there were not much dust at such an early stage. 
{\tt 9}, on the other hand, has a much lower stellar mass of 
$5.0\times 10^{10} M_\odot$ and a young age of 129~Myr. Its SFH is an intense,
short burst with $\tau=40$~Myr, which still leaves an on-going
$SFR=72 M_\odot/yr$.

\begin{figure*}[tbp]
\centering
  \includegraphics[width=\textwidth]{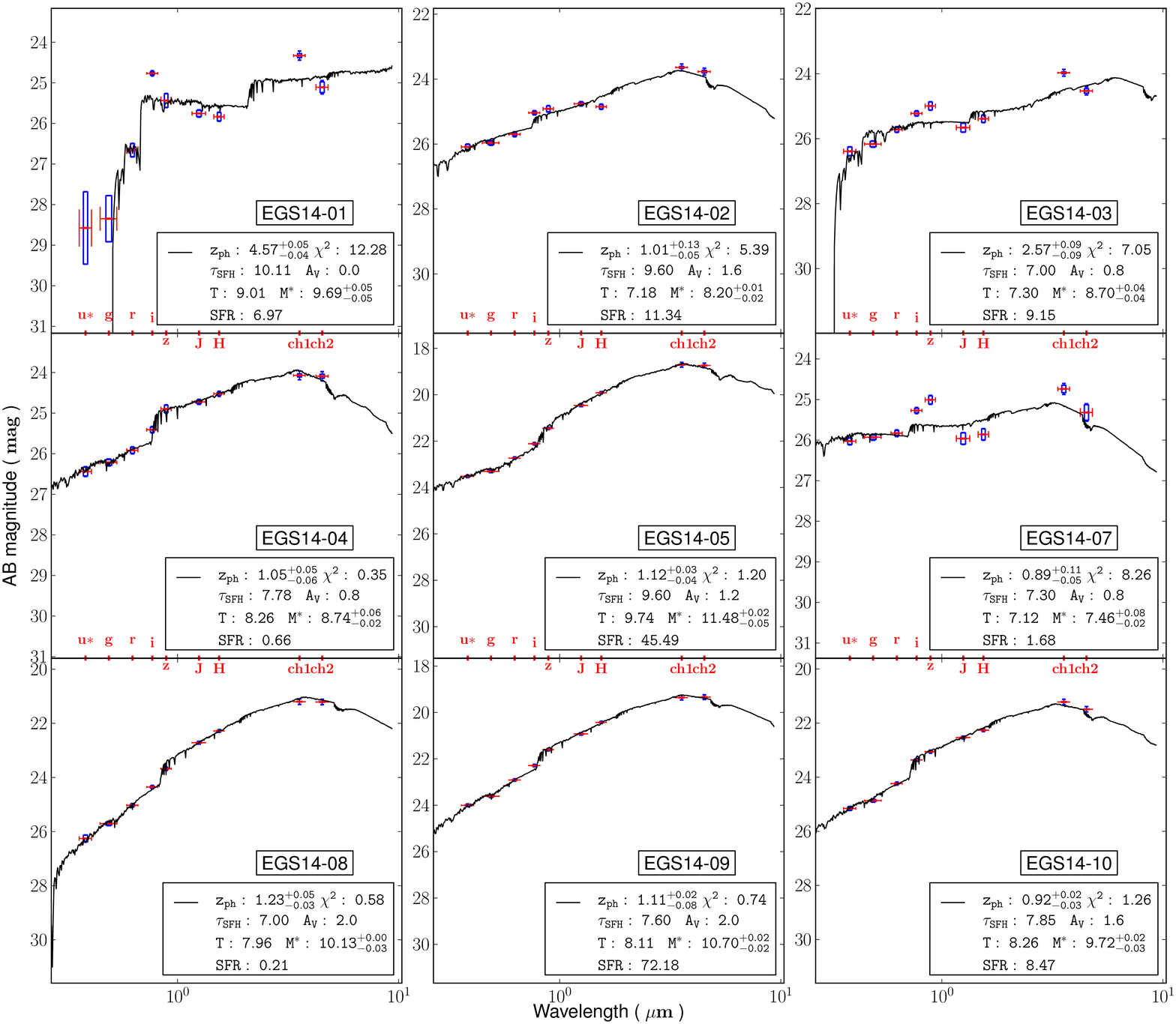}
  \includegraphics[width=0.5\textwidth]{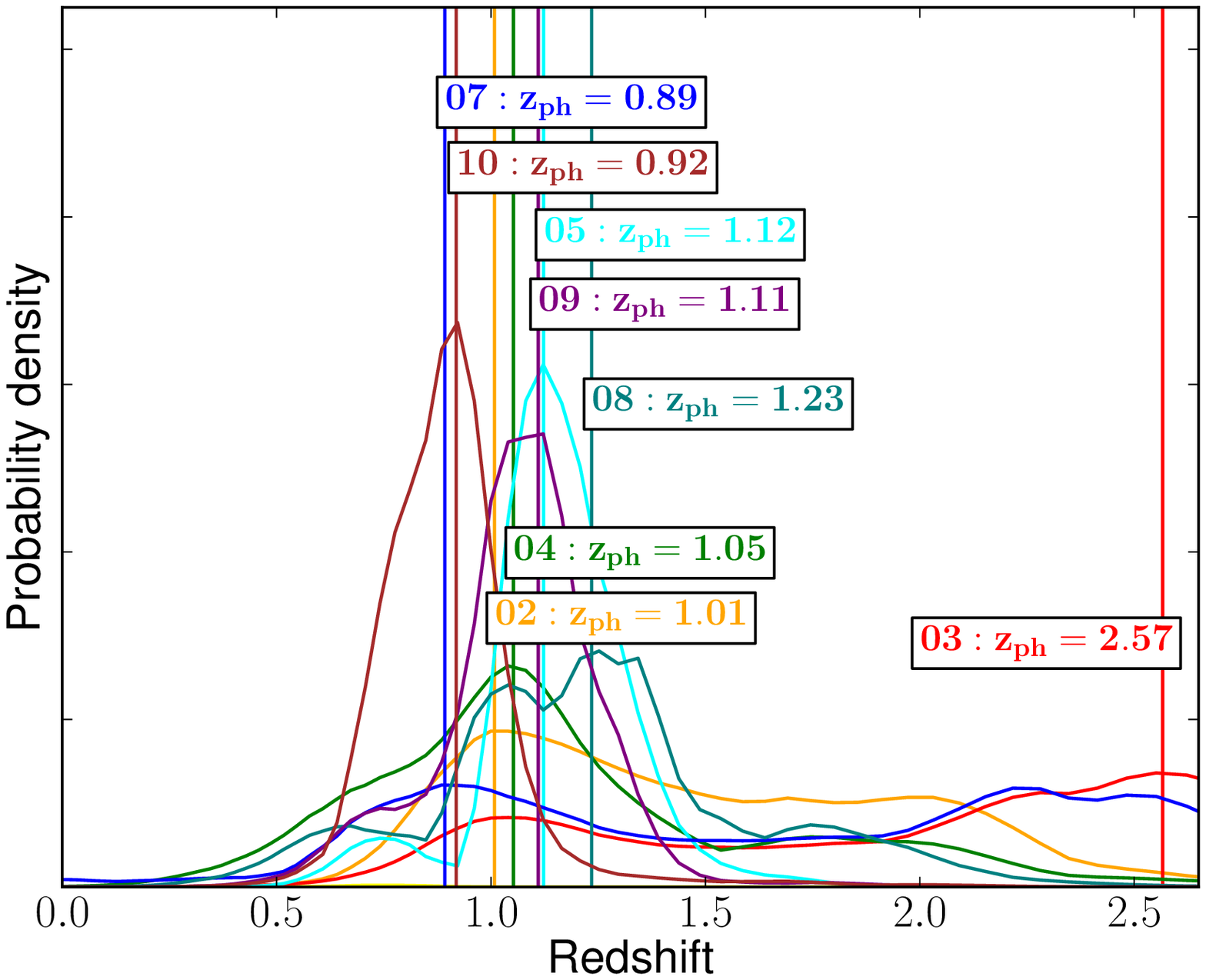}
\caption{Optical-to-NIR SED fitting results for EGS14. Legends are the same as
in Fig. 4. The two most massive objects, {\tt 5} and {\tt 9}, are likely at
the same redshift ($z_{ph}=1.11$--1.12) and associated, at objects {\tt 2},
{\tt 4} and {\tt 8} could also belong to this group.
}
\end{figure*}

\subsubsection{Decomposition in Mid-to-Far-IR}

   The decomposition is carried out for the nine objects using their centroids
determined from the CFHTLS-Deep $i$-band. 

   The automatically iterative fit succeeded in the 24~$\mu$m and 250~$\mu$m,
the latter of which is shown in Fig. 28 for demonstration.
For the 24~$\mu$m, the fit converged on {\tt 2}, {\tt 5}, {\tt 9}, and
{\tt 10}. Objects {\tt 5} and {\tt 9} almost equally split $\sim 82$\% of the
total light in this area. For the 250~$\mu$m, the fit converged on {\tt 5},
{\tt 7}, {\tt 8}, {\tt 9}, and {\tt 10}. However {\tt 7}, {\tt 8} and {\tt 10}
were formally rejected because they all had negligible fluxes, and hence the
output only included {\tt 5} and {\tt 9}. Fitting only {\tt 5} or only {\tt 9}
results in obvious residuals at the position of the other object, and hence is
not acceptable. 

   Unfortunately, the decomposition in 70~$\mu$m using either the automatically 
iterative method or the trail-and-error method could not converge on any 
object, apparently because of the low S/N of the data. Therefore we had to skip
this band. The decomposition in both 100 and 160~$\mu$m also failed due to the
low S/N of the data. Nevertheless, the PEP DR1 catalog includes two sources
whose position coincide with {\tt 5} and {\tt 9}. As the decomposition of both
the 24~$\mu$m and the 250~$\mu$m images shows that {\tt 5} and {\tt 9} are
by far the most dominant objects in both bands, we adopted the PEP catalog
values for these two objects in 100 and 160~$\mu$m and assumed that all other
objects are negligible in these two bands.
   Table 2 summarizes these results for the major components {\tt 5} and
{\tt 9}.

\begin{figure*}[tbp]
\centering
  \includegraphics[width=\textwidth]{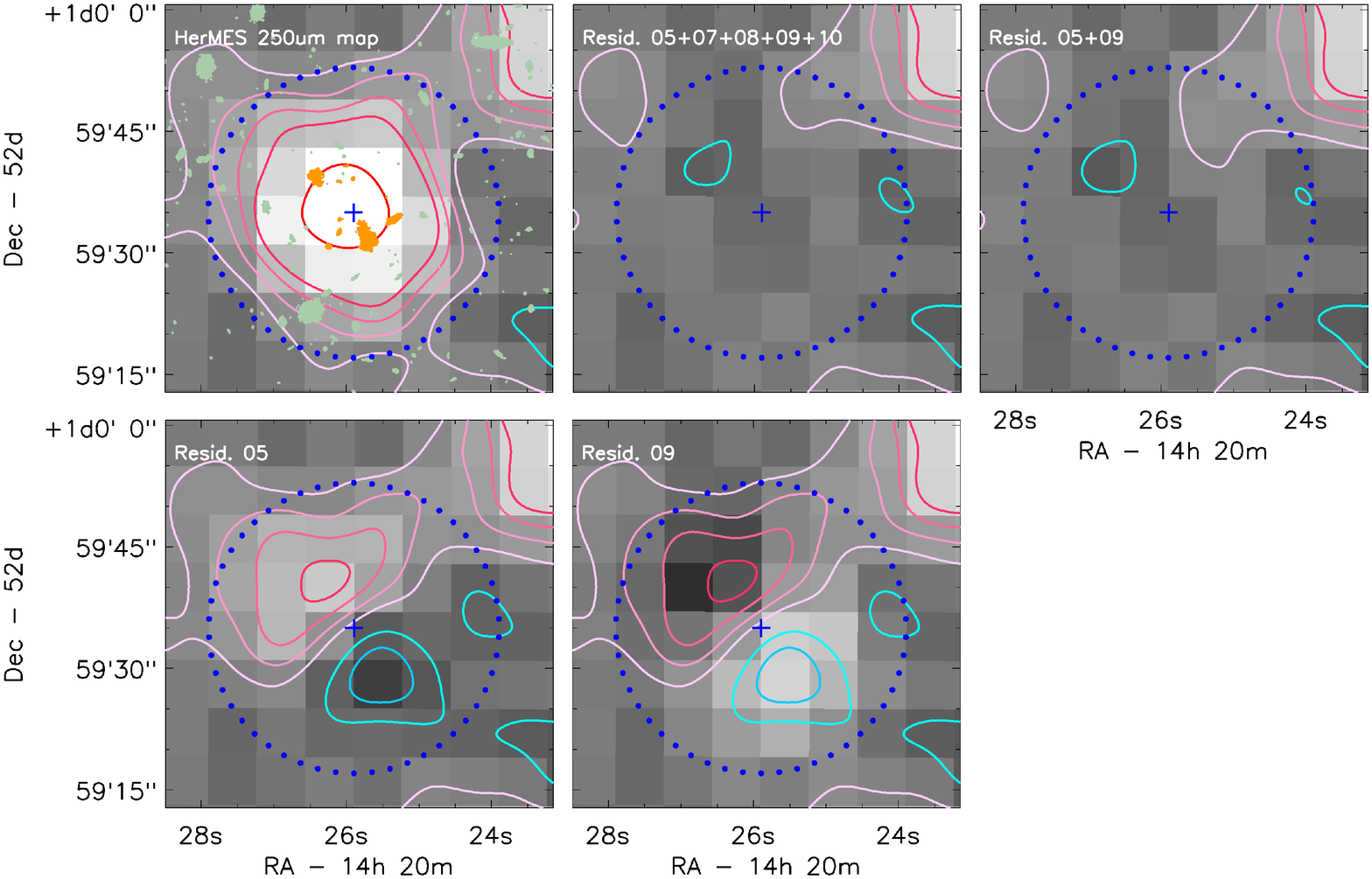}
\caption{Demonstration of the decomposition of EGS14 in 250~$\mu$m. The first
panel shows the original 250~$\mu$m image, while the others show the residual
maps of the different decomposition schemes where different input sources are 
considered (labeled on top). Legends are the same as in Fig. 5. 
The ``{\tt 5+9}'' case is the adopted final solution.
}
\end{figure*}

\subsubsection{Total IR Emission and Stellar Populations}

   The FIR emission of EGS14 should be mostly from {\tt 5} and {\tt 9}. Based
on their $z_{ph}$ estimates, we adopted $z_{ph}=1.12$ as their common redshift. 
Based on the AEGIS X-ray catalog, there is no X-ray source detected in the
EGS14 region. The limit of this catalog in the most sensitive 0.5--2~keV band
is $\sim 3\times 10^{-17}$~$erg/s/cm^2$, which corresponds to an upper limit of
$2.1\times 10^{41}$~$erg/s$ in the restframe 1--4~keV at $z=1.12$. Therefore,
we believe that its FIR emission is most likely due to star formation.
Fitting their mid-to-far-IR SED separately to the SK07
models, we got $L_{IR}=1.3\times 10^{12}$ and $1.0\times 10^{12} L_\odot$ for
{\tt 5} and {\tt 9}, respectively, and therefore they are both ULIRG. This is
shown in the left panels of Fig. 29. From the results in \S 6.2.2., we also
got $L_IR^{ext}=4.7\times 10^{11} L_\odot$ for {\tt 5}, which then implies 
$L_IR^{blk}=8.3\times 10^{11} L_\odot$ and $SFR_IR^{blk}=83$~$M_\odot/yr$ for
this object. It has $SFR_{fit}=46$~$M_\odot/yr$, and hence we get
$SFR_{tot}=129$~$M_\odot/yr$. For {\tt 9}, we got 
$L_{IR}^{blk}=1.2\times 10^{12} L_\odot$, which is even slightly larger than
$L_{IR}$. In this case, we take it that the observed IR emission can be fully
explained by the extinction in the exposed region.

   From the analytic fits,
we got $T_d^{fit}=36.9$ and 37.2~$K$, and $M_d=4.0\times 10^8$ and 
$3.2\times 10^8 M_\odot$ for {\tt 5} and {\tt 9}, respectively. All this
suggests that the dust properties of these two ULIRGs are very similar. 
However, as discussed in \S 6.2.2, the stellar populations in the exposed 
regions of these two objects are vastly different in their stellar masses,
ages and SFHs: {\tt 5} has $M^*=3.0\times 10^{11} M_\odot$, $T=5.5$~Gyr and
$\tau=4.0$~Gyr, while {\tt 9} has $M^*=5.0\times 10^{10} M_\odot$, $T=129$~Myr
and $\tau=129$~Myr.
Therefore, these imply $SSFR=0.4$ and 1.4~Gyr$^{-1}$ for {\tt 5} and
{\tt 9}, respectively. For {\tt 5}, one can also get $T_{db}^{tot}=2.3$~Gyr and
$T_{db}^{blk}=3.6$~Gyr. For {\tt 9}, $T_{db}^{tot}=694$~Myr, and $T_{db}^{blk}$
is not applicable.
The inferred total gas masses for these two objects from the above dust masses
are $5.6\times 10^{10} M_\odot$ and $4.4\times 10^{10} M_\odot$, respectively.
Therefore, the gas reservoir for {\tt 5} would only allow it to add $<20$\% to
its existing stellar mass even if it could turn all the available gas
into stars. On the other hand, the situation for {\tt 9} is somewhat different
in that its current ULIRG phase would be
able to add $\sim 88$\% to its existing stellar mass if it could turn all
the gas into stars, however this would require a much longer time than the
typical lifetime of an ULIRG. Considering that its on-going star formation is
all in the exposed region and has a sharp declining SFH, this is not likely to
happen.

\begin{figure*}[tbp]
\centering
\subfigure{
  \includegraphics[width=0.5\textwidth]{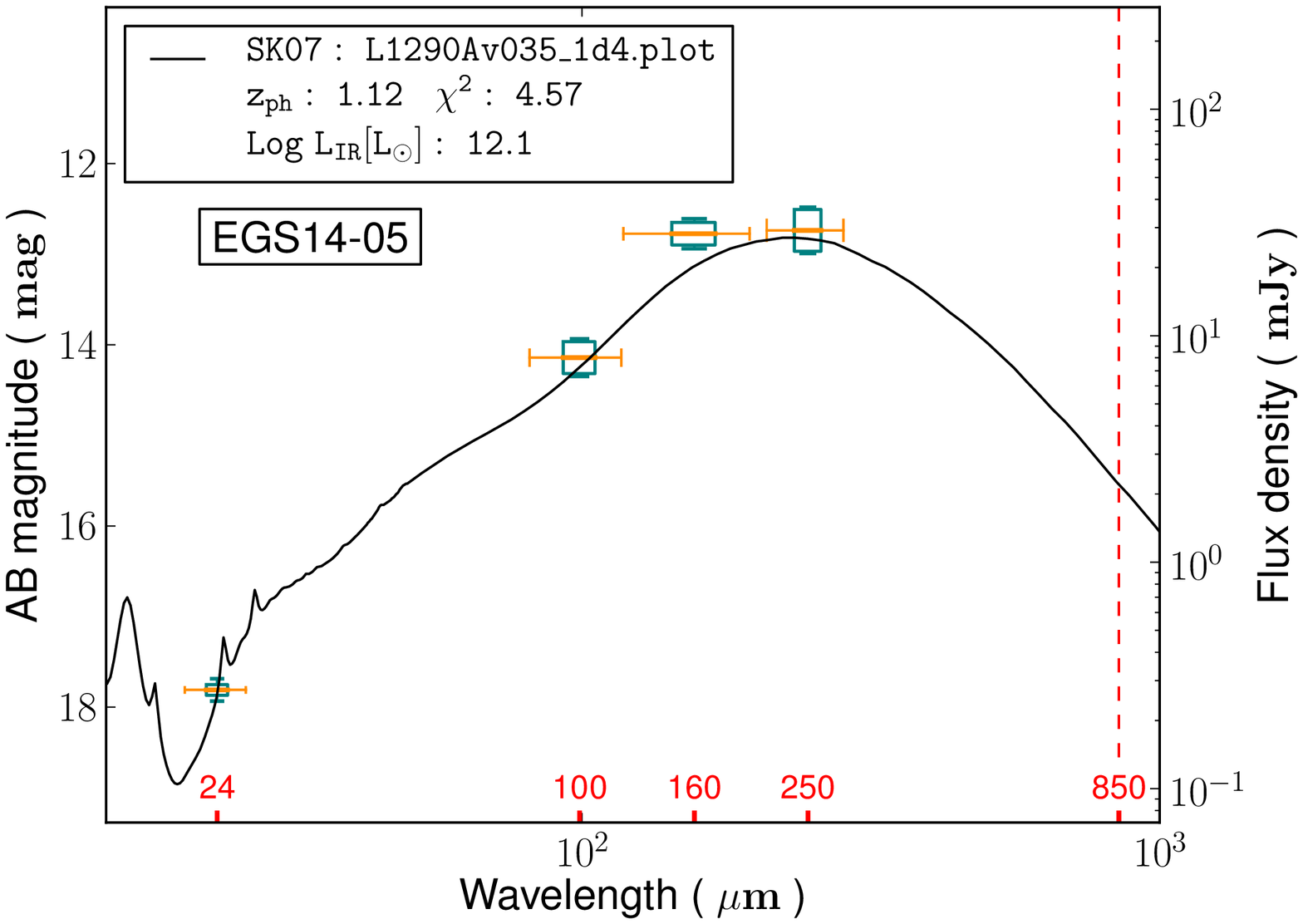}
  \includegraphics[width=0.5\textwidth]{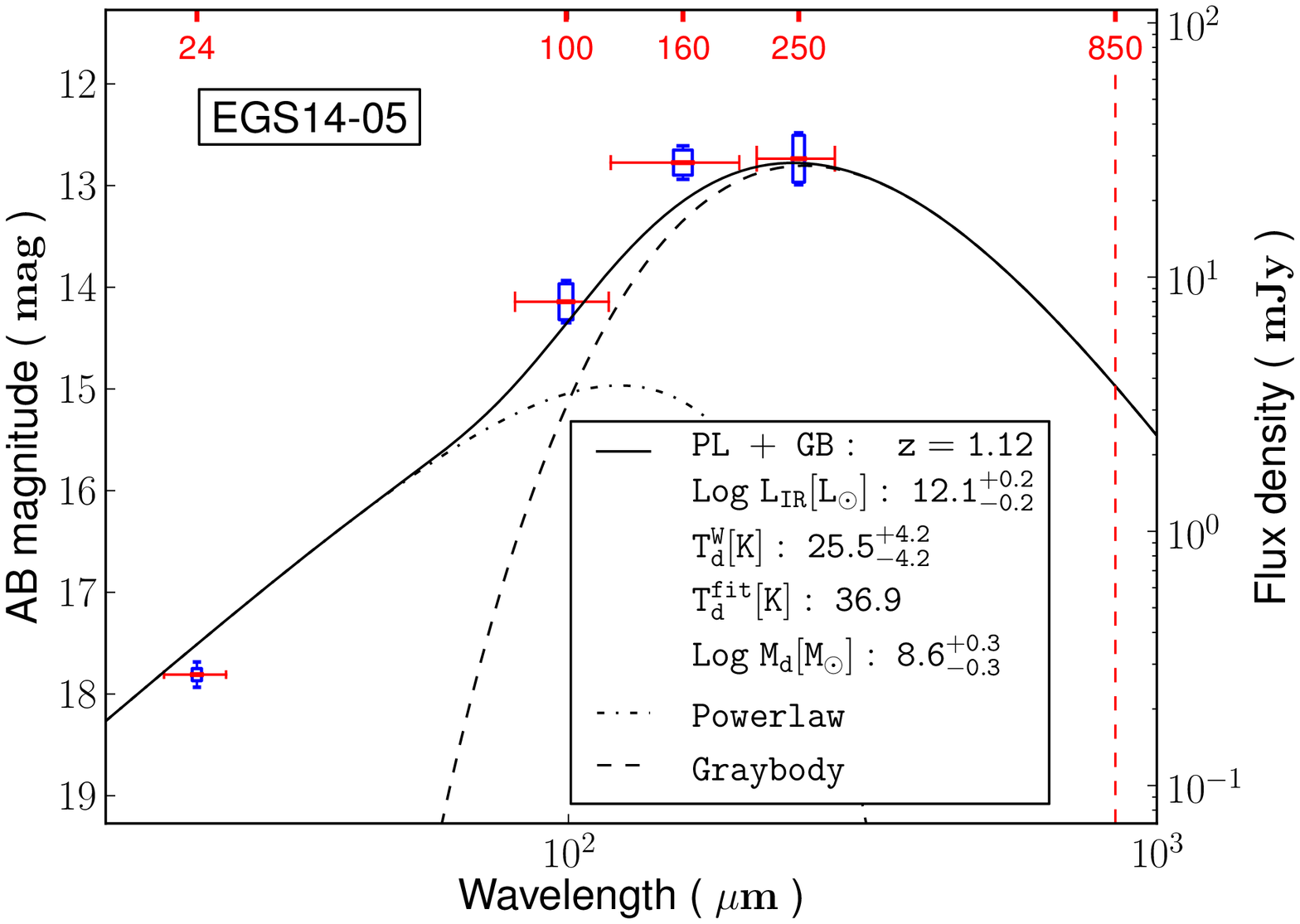}
}
\subfigure{
  \includegraphics[width=0.5\textwidth]{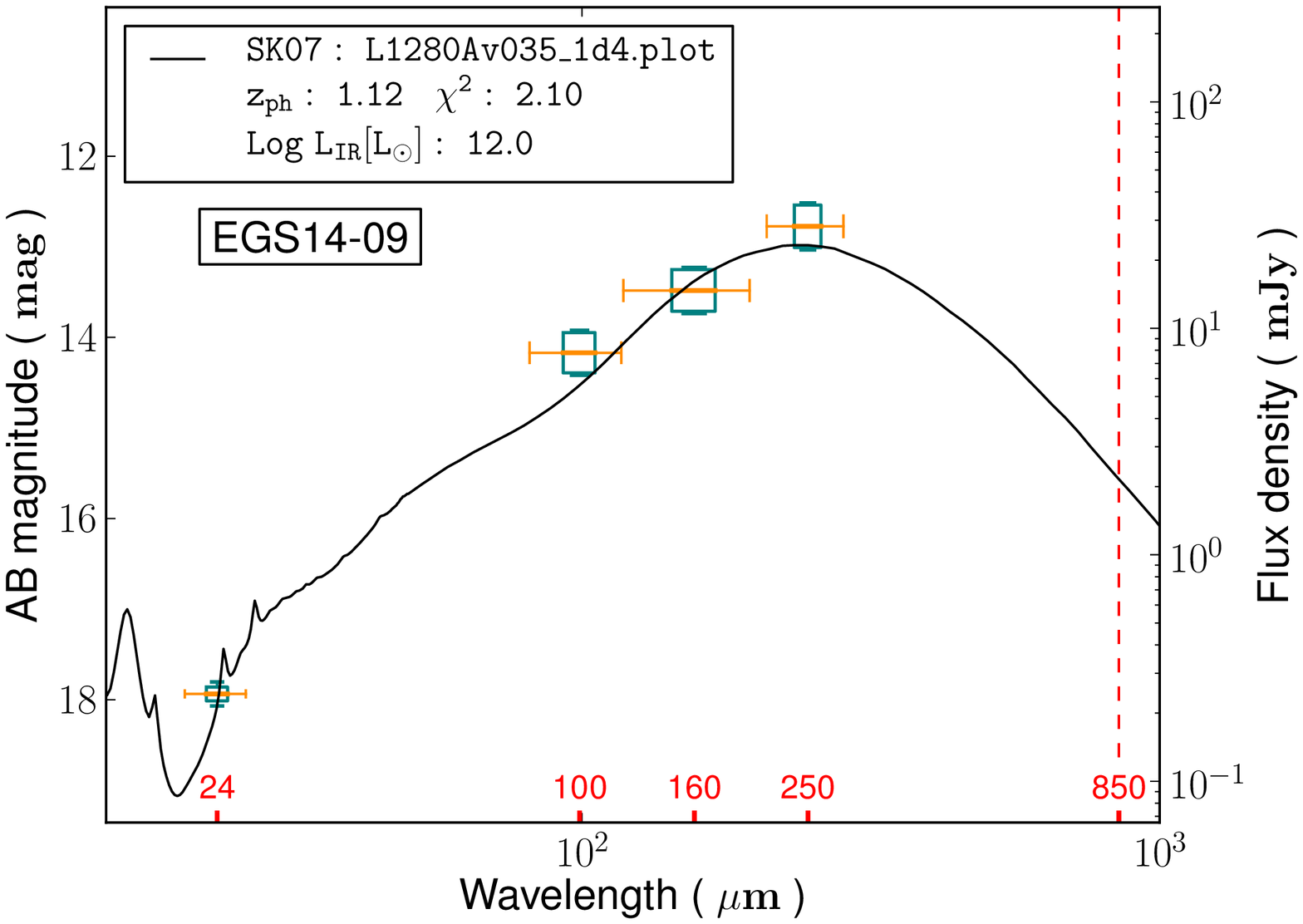}
  \includegraphics[width=0.5\textwidth]{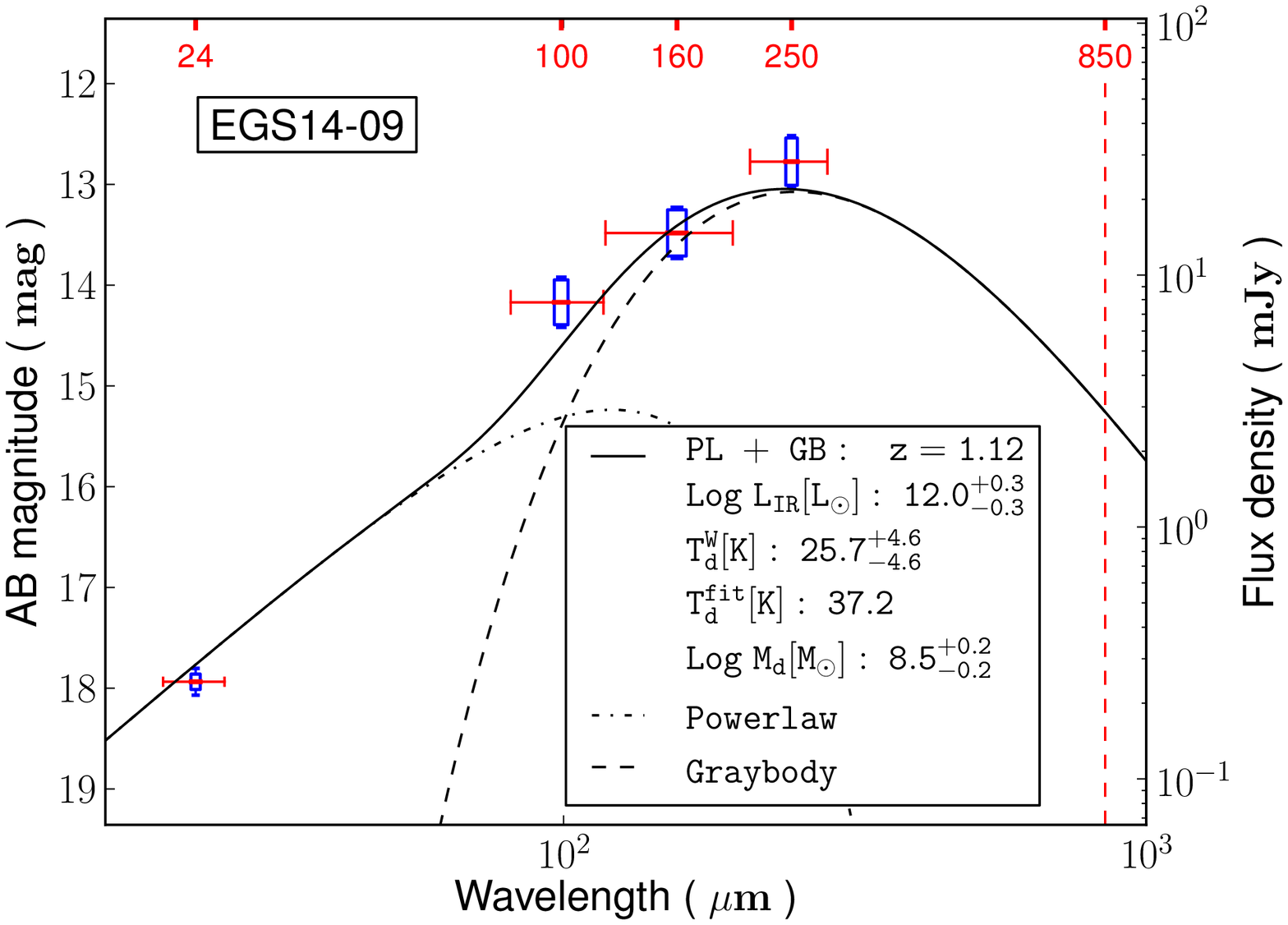}
}
\caption{Mid-to-FIR SED analysis for EGS14-{\tt 5} (top) and {\tt 9}
(bottom). Legends are the same as in Fig. 6. The panels to the left are the
results from the fit to the SK07 models, while the panels to the right are
the results from the fit to the powerlaw + graybody models.}
\end{figure*}

   Finally we investigate the FIR-radio correlation for EGS14.
The catalog of Ivison et al. (2010) does not include any strong radio 
source in the area of EGS-14. However, from the radio map (Ivison, priv. comm.)
we detect two moderate sources at the exact locations of {\tt 5} and {\tt 9},
which have $S_{1.4GHz}=0.043\pm 0.012$ and $0.071\pm 0.013$~mJy, respectively.
From these we obtained $q_{IR}=2.63$ and 2.34, respectively, which are
consistent with the mean of $2.40\pm 0.24$ in Ivison et al. (2010).

\subsection {EGS19 (EGS-J141943.4+525857)}

  This source is different from the previous two in that it is outside of the 
CANDELS WFC3 area. At its 250~$\mu$m source location, the PEP catalog
reports three sources. From north to south, their flux densities are
$6.4\pm 1.2$, $9.9\pm 1.2$ and $6.0\pm 1.2$~mJy in 100~$\mu$m, respectively,
and $19.7\pm 3.3$, $22.0\pm 3.4$ and $17.1\pm 2.9$~mJy, respectively.

\subsubsection {Morphologies}

\begin{figure*}[tbp]
\centering
\includegraphics[width=\textwidth] {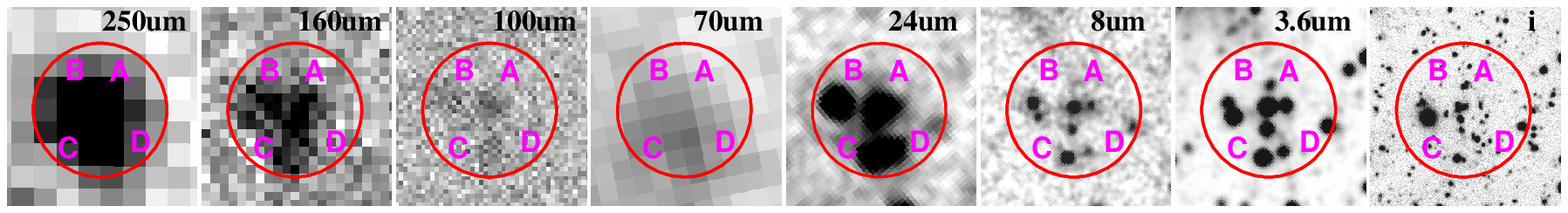}
\includegraphics[width=\textwidth] {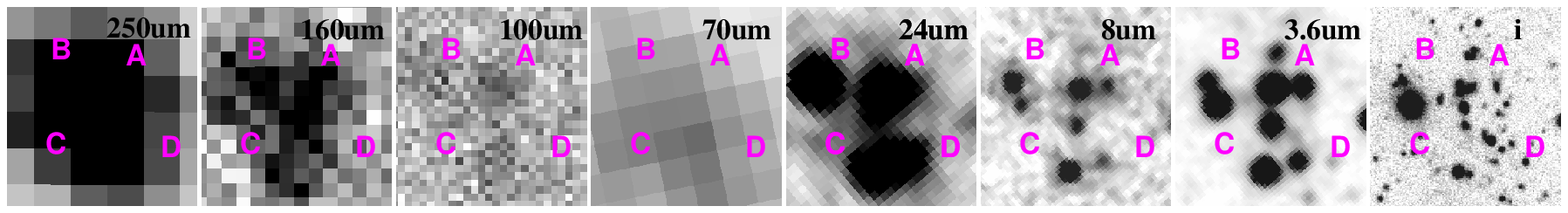}
\caption{FIR to optical images of EGS19. The legends and the organization of
the panels are the same as in Fig. 2. The $i$-band image is from CFHTLS-Deep.
}
\end{figure*}

   Within $r=18$\arcsec, there are 46 objects detected in the CFHTLS-Deep data
that have $S/N\geq 5$ in the $i$-band measured in the MAG\_ISO aperture. Among
these objects, more than 20 of them could be possible contributors, most of
which segregate into groups. The sub-components within each group are so close
to each other that using them directly for the decomposition would crash the 
process. Therefore, we took a slightly different approach in analyzing this
source. To emphasize this difference, we use yet another different labeling 
scheme in referring to the possible counterparts. From the 24~$\mu$m image, we 
identified four possible contributors, which are labeled from 
$\mathbb A$ to $\mathbb D$. Each of these 24~$\mu$m sources could be made of
several components as revealed by the CFHTLS Deep images, for which we label
numerically as shown in Fig. 31 to 33.

   Component $\mathbb A$ is made of six subcomponents. In the higher resolution
ACS images, $\mathbb A-1$ seems like a dusty disk system viewed edge-on. 
$\mathbb A-2$, which is only $1\farcs 56$ away from $\mathbb A-1$, is a small
but resolved galaxy whose major axis is almost perpendicular to $\mathbb A-1$.
$\mathbb A-3$ consists of two irregular objects separated by $0\farcs 69$, which
seem to be embedded in a somewhat extended halo. $\mathbb A-4$ is an irregular
object with a small core. $\mathbb A-5$ consists of two objects separated by
$0\farcs 67$, one resolved and the other unresolved. $\mathbb A-6$ is
irregular and does not have a well defined core.

\begin{figure*}[tbp]
\centering
\includegraphics[width=\textwidth] {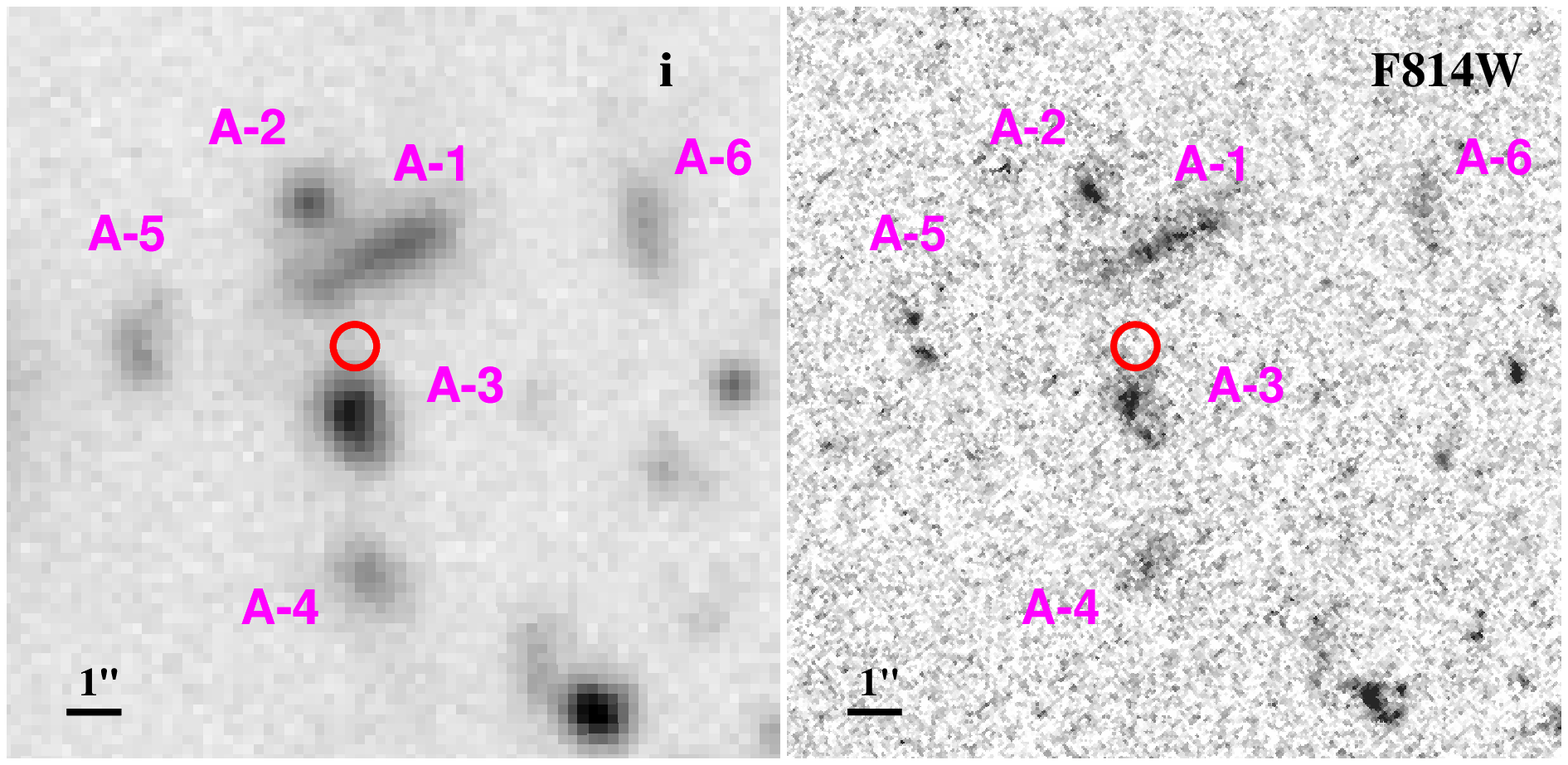}
\caption{Morphological details of EGS-19, $\mathbb A$ component in the
CFHTLS-Deep $i$-band (left) and the AEGIS $I_{814}$-band (right). The red circle
is $0.4\arcsec$ in radius and indicates the position of {\tt A$^\prime$}, which
is the geometric center of $\mathbb A-1$, 2 and 4 (see \S 6.3.2).
}
\end{figure*}

   Component $\mathbb B$ consists of four possible subcomponents. As revealed
by the ACS images, $\mathbb B-1$, which is closest to the 24~$\mu$m source
centroid, is made of three irregular objects
\footnote{The additional, compact ``object'' to the south-west is actually
due to bad pixels.}
that stretch over $\sim 1$\arcsec.
While currently we do not have any additional data to determine if these three
objects are at the same redshift, we assume that they are associated and
take them as one single object.
$\mathbb B-2$ and $\mathbb B-3$ are two small, compact objects that are
$2\farcs 3$ and $4\farcs 7$ away from $\mathbb B-1$, respectively.
$\mathbb B-4$ is a regular spheroidal. While it is the brightest in optical
among all, it is far away from the 24~$\mu$m source center ($\sim 4$\arcsec) 
and likely only contributes a minimal amount to the 24~$\mu$m flux.

\begin{figure*}[tbp]
\centering
\includegraphics[width=\textwidth] {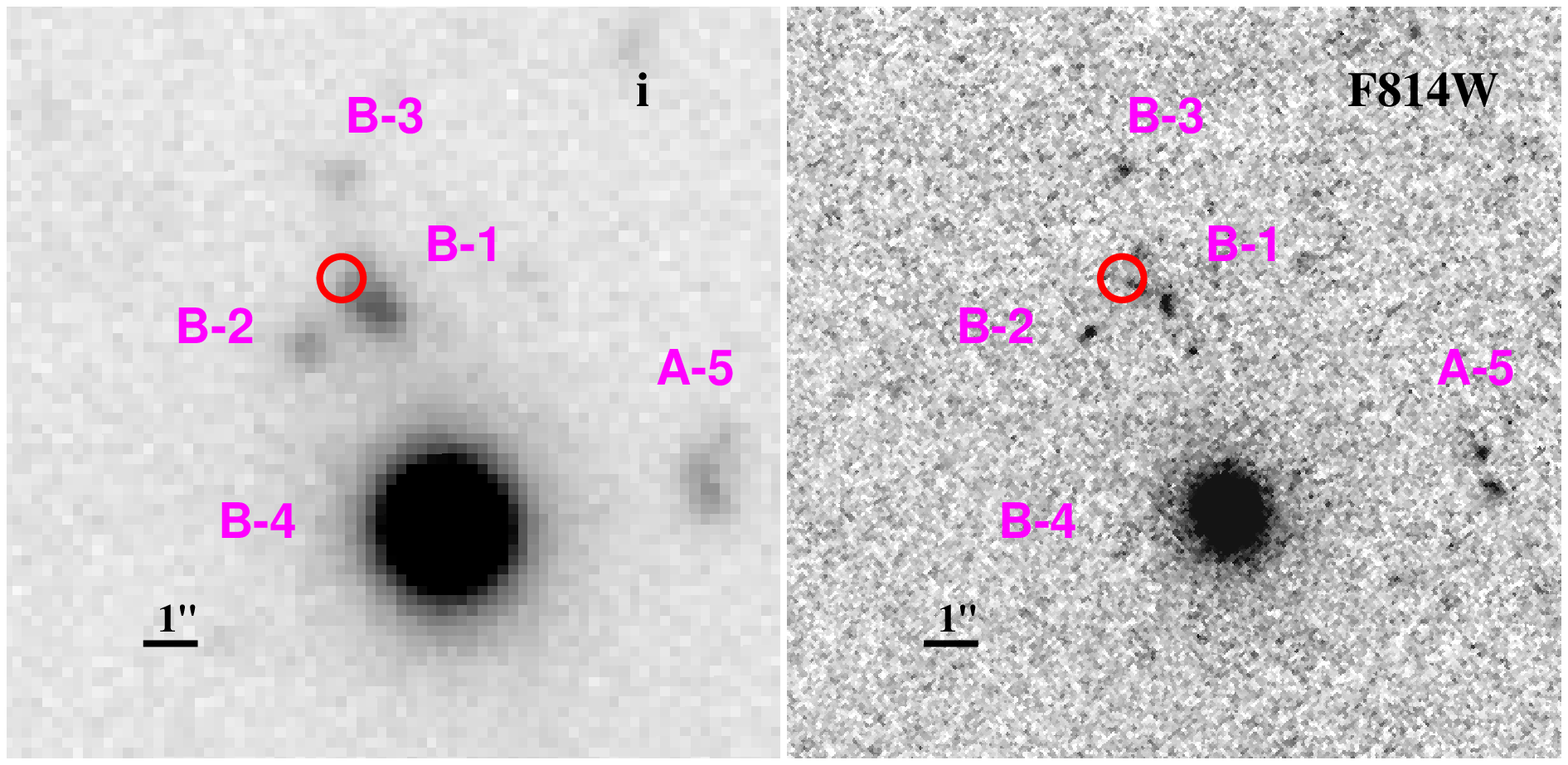}
\caption{Similar to Fig. 31, but for the $\mathbb B$ component of EGS19.
The red circle indicates the position of {\tt B$^\prime$}, which
is the geometric center of $\mathbb B-1$, 2 and 3 (see \S 6.3.2).}
\end{figure*}

    Component $\mathbb C$ consists of two objects. $\mathbb C-1$ seems to be
a regular elliptical in the CFHTLS images, however the ACS images reveal that
it is most likely a merger. It has spectroscopic redshift of 1.180 from the
Deep3 program. $\mathbb C-2$ is $1\farcs 9$ away, and is a point
source. 

    Component $\mathbb D$, which is $6\farcs 1$ away from $\mathbb C-1$, is 
blended with $\mathbb C$ in 24~$\mu$m. The CFHTLS images show that it is 
made of two subcomponents separated by $\sim 0\farcs 5$. The surface brightness
of $\mathbb D$ is quite low, and it is not detected in the AEGIS images. 

\begin{figure*}[tbp]
\centering
\includegraphics[width=\textwidth] {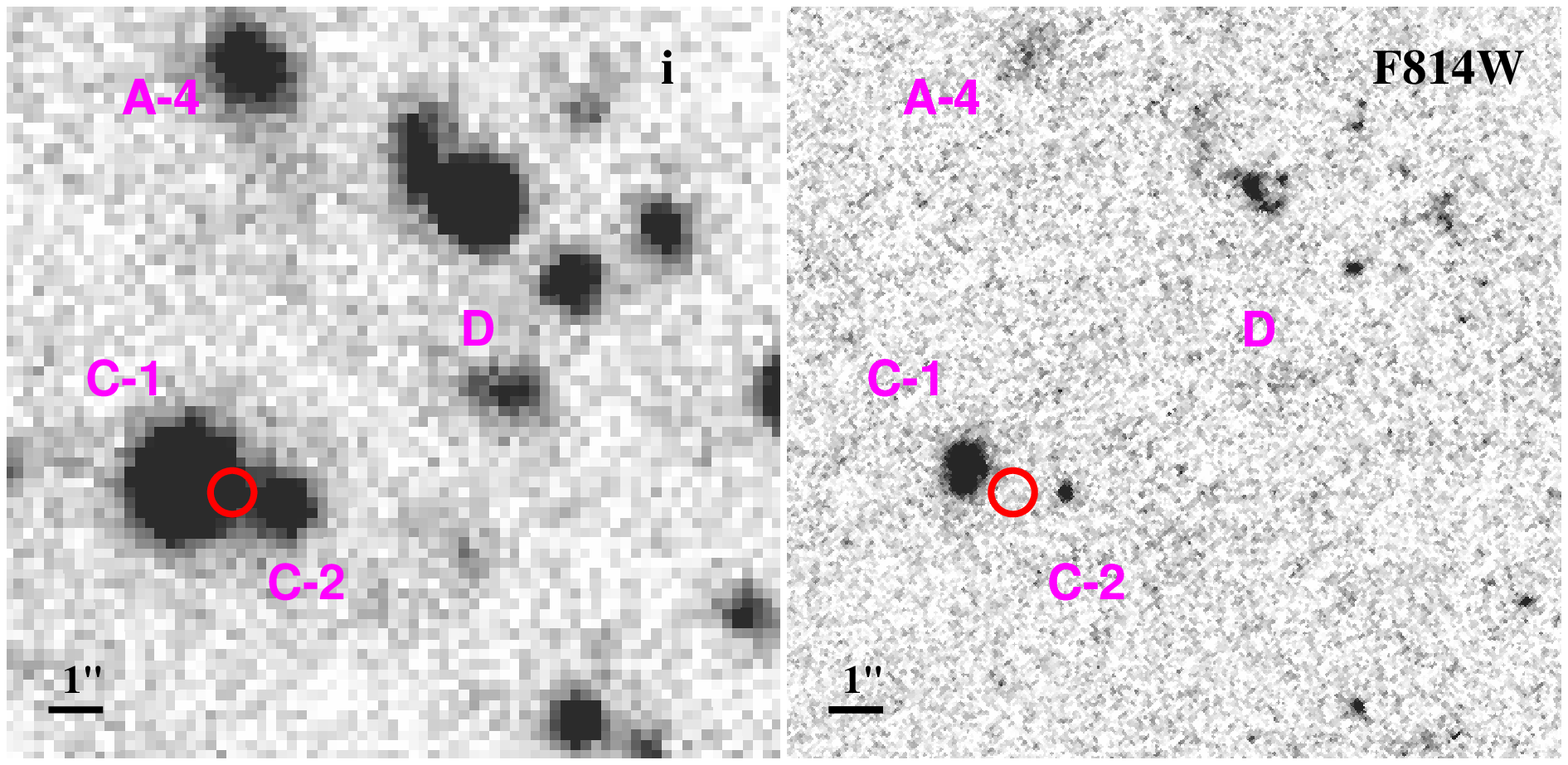}
\caption{Similar to Fig. 31, but for the $\mathbb C$ and $\mathbb D$ 
components of EGS19. The red circle indicates the position of {\tt C$^\prime$},
which is the geometric center of $\mathbb C-1$ and 2 (see \S 6.3.2).
}
\end{figure*}

\subsubsection {Optical SED Analysis and Redshifts}

   To proceed with our analysis, the first step is to determine if the
subcomponents in each of the four 24~$\mu$m clumps could be at the same
redshifts.  As EGS-19 does not have WFC3 images to be used as the morphological
templates, we refrained from doing TFIT of the IRAC image. Therefore, we only
rely on the optical images from the CFHTLS Deep program to carry out the SED
analysis. The photometry of the CFHTLS-Deep data was done in the same way as
described in the previous sections.

   The SED fitting results are summarized in Fig. 34 for $\mathbb A$ and in
Fig. 35 for $\mathbb B$, $\mathbb C$ and $\mathbb D$. From the best-fit models
and the $P(z)$ distributions, we concluded that $\mathbb A-1$, 2 and 4 could
be at the same redshift and associated. Taking the average, we adopted 
$z_{ph}=1.06\pm 0.07$ as their common redshift. For the decomposition purpose
below, we took the geometric center of these three objects as the common center
of the system, and denoted this new ``object'' as {\tt A$^\prime$}. Our
assumption is that {\tt A$^\prime$} is the part from the $\mathbb A$ clump that
have significant FIR contribution, and that all other components in this clump
can be ignored. Similarly,
$\mathbb B-1$, 2 and 3 could be at the same redshift of $z_{ph}=2.72\pm 0.07$
and associated, and the new ``object'' at their geometric center is denoted
as {\tt B$^\prime$}. The situation for $\mathbb C-1$ and 2 is somewhat
uncertain because their $z_{ph}$ differ significantly. However, since the 
$P(z)$ distribution for $\mathbb C-2$ has a wide, flat peak that contains the
sharp $P(z)$ peak of $\mathbb C-1$, it is reasonable to assume that they are
actually associated. Therefore we adopt the best-fit $z_{ph}=1.08$ for 
$\mathbb C-1$ as the common redshift for this system, which is denoted as
{\tt C$^{\prime}$} and assigned the position at the geometric center. This also
suggests that {\tt C$^{\prime}$} could be with the same group as
{\tt A$^\prime$}, however it is not appropriate to combine the two for the
decomposition as their separation is too large. Finally, we note that the
single object $\mathbb D$ might be within the same group as {\tt B$^\prime$},
however this is highly uncertain due to the lack of prominent peaks in its
$P(z)$ distribution.

\begin{figure*}[tbp]
\centering
\includegraphics[width=\textwidth]{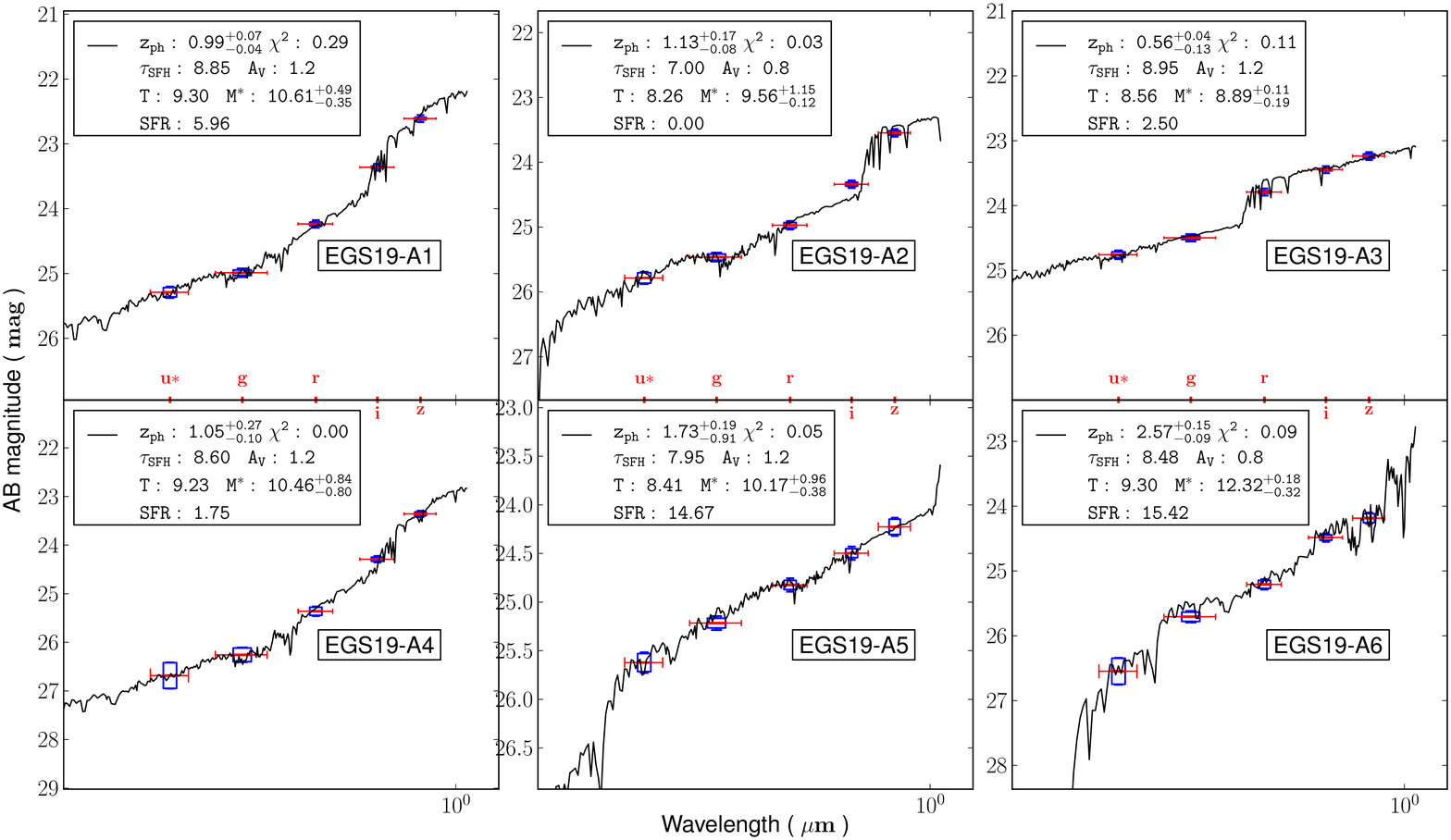}
\includegraphics[width=0.5\textwidth] {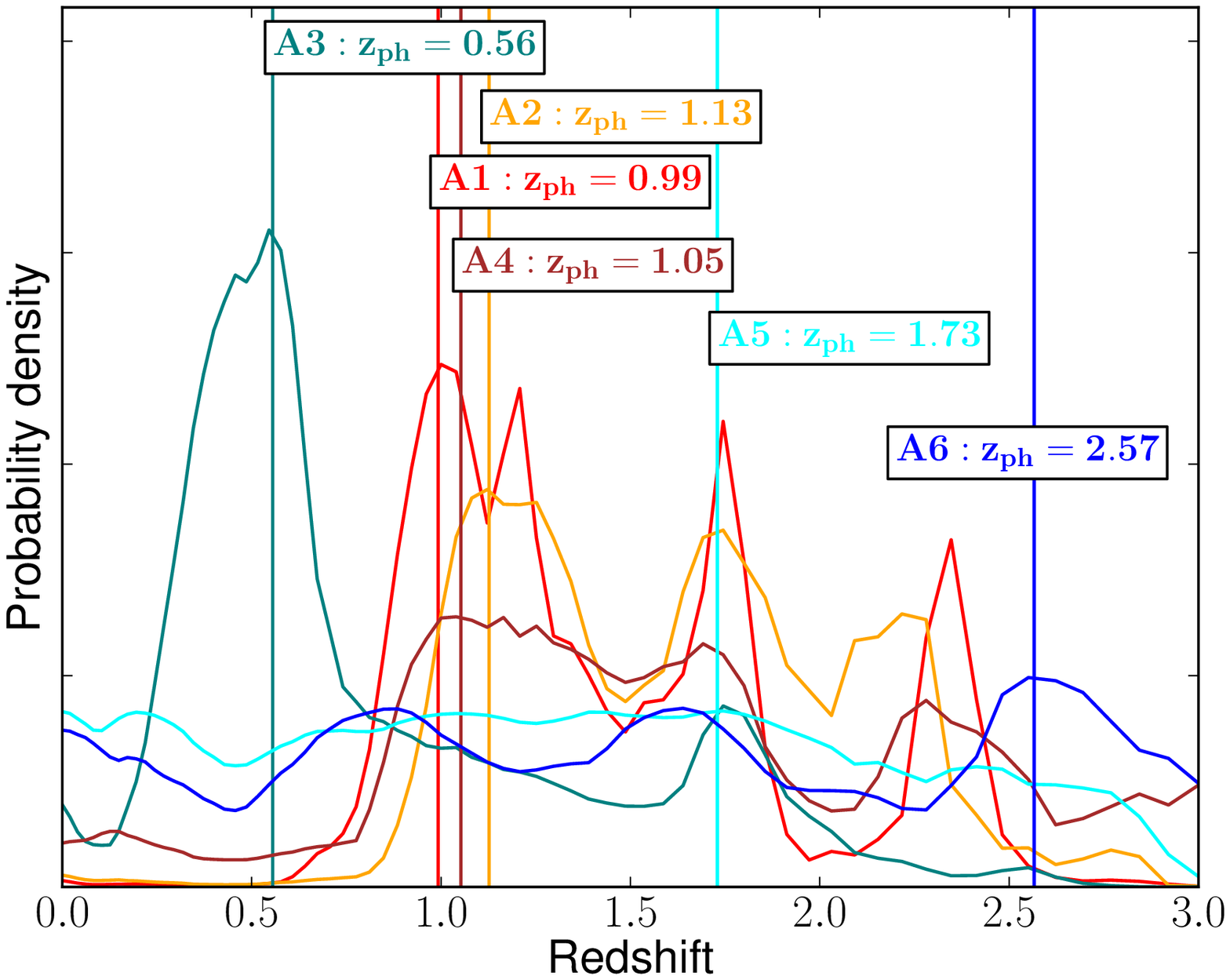}
\caption{Optical SED fitting for the subcomponents of EGS19-$\mathbb A$ clump.
Legends are the same as in Fig. 4. $\mathbb A-1$, 2 and 4 could be at the same
redshift of $z_{ph}=1.06$ and associated.
}
\end{figure*}

\begin{figure*}[tbp]
\centering
\includegraphics[width=\textwidth]{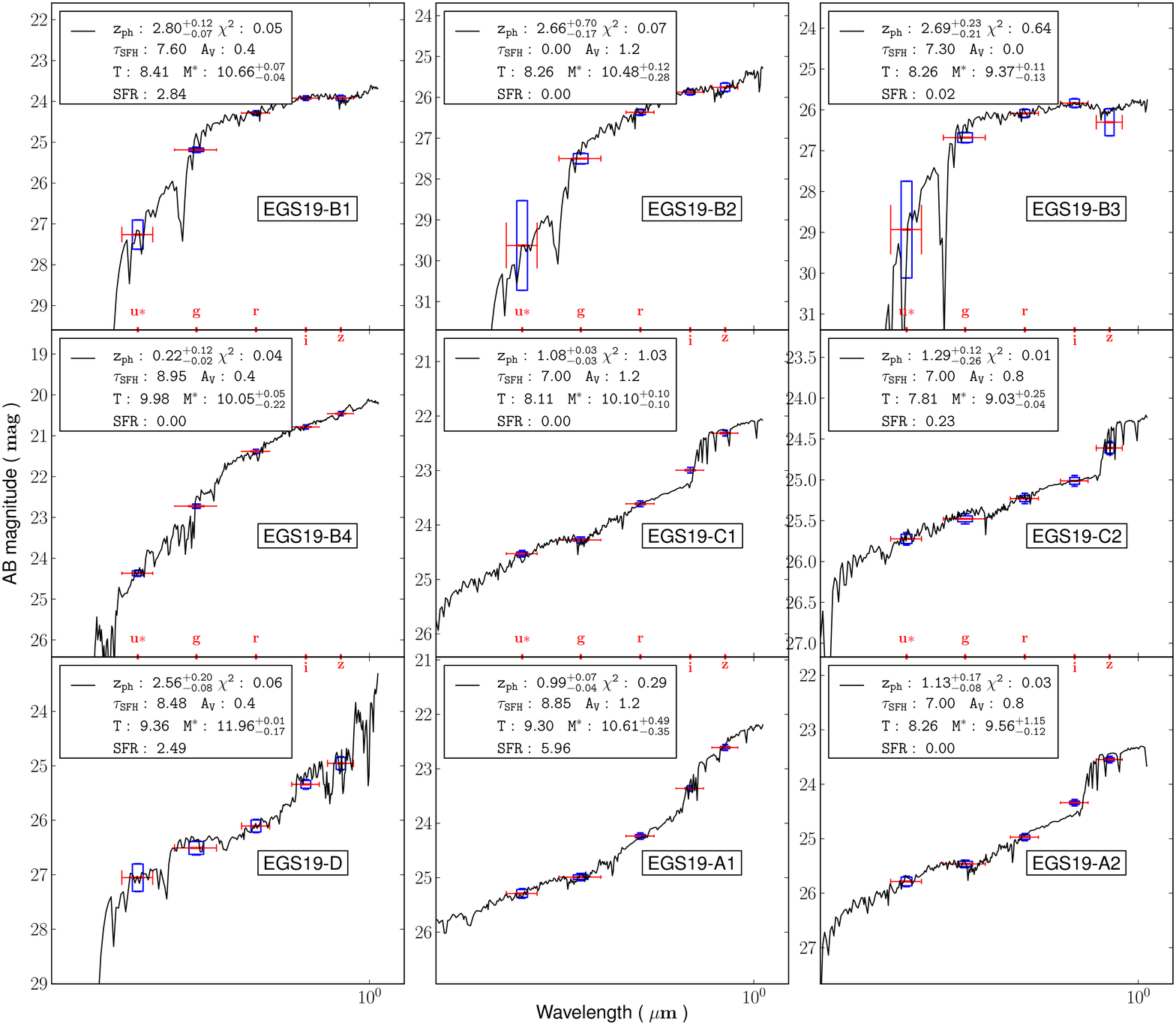}
\includegraphics[width=0.5\textwidth]{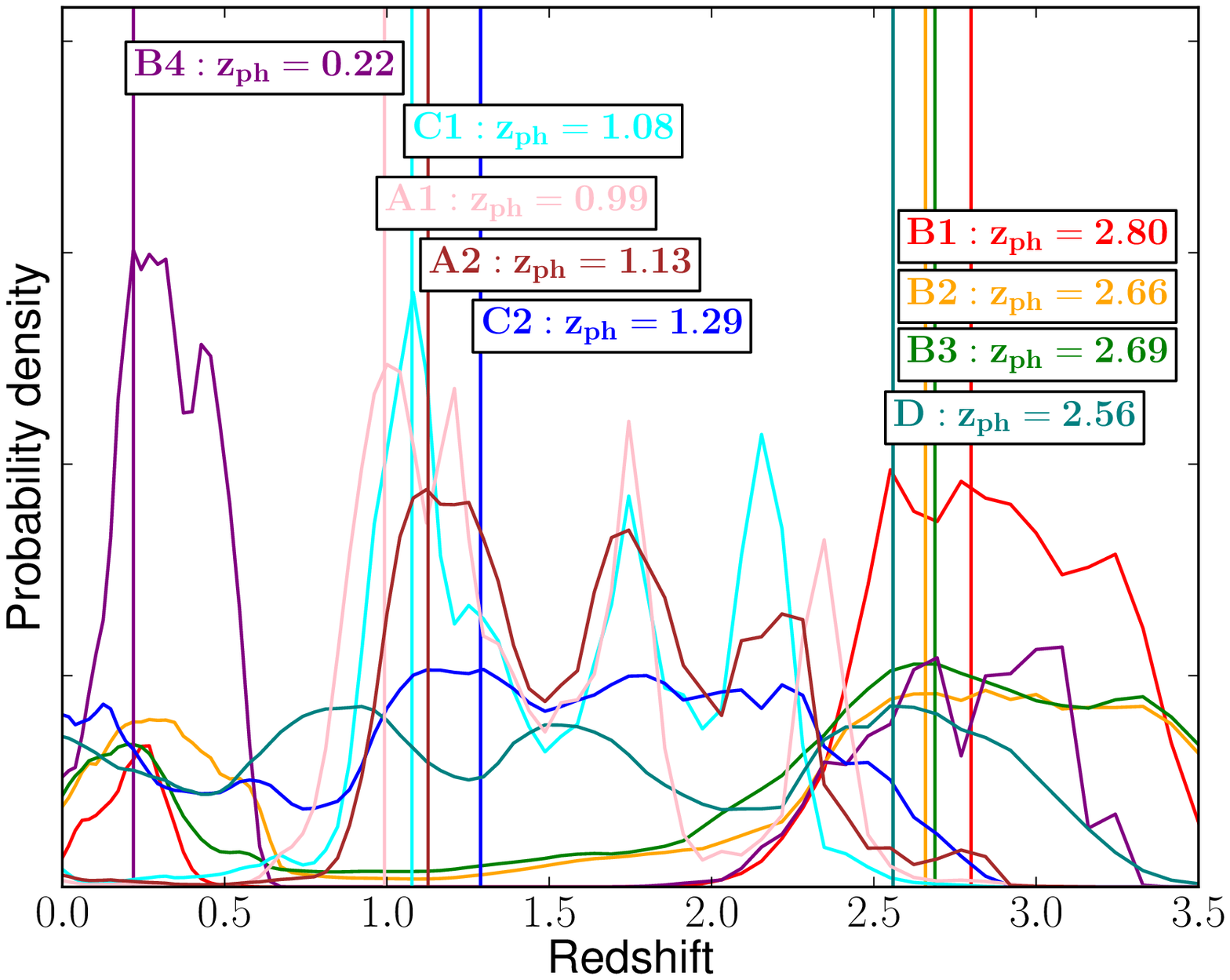}
\caption{Similar to Fig. 34, but for the EGS19-$\mathbb B$, $\mathbb C$ and 
$\mathbb D$ clumps. Legends are the same as in Fig. 4. Based on their $P(z)$
distribution, $\mathbb B-1$, 2 and 3 could be at the same redshift of 
$z_{ph}=2.72$. Similarly, $\mathbb C-1$ and 2 could also be at the same
redshift of $z_{ph}=1.08$, and might even be within the same $\mathbb A$ group.
$\mathbb D$ could be within the same $\mathbb B$ group.
}
\end{figure*}

\subsubsection{Decomposition in Mid-to-Far-IR}

    The decomposition was done at the locations of {\tt A$^\prime$}, 
{\tt B$^\prime$}, {\tt C$^\prime$}, and $\mathbb D$. At 24~$\mu$m, the
automatically iterative decomposition converged at all four positions, 
albeit with significant residuals, which indicate that the other components
that we ignored are non-negligible in this band. The decomposition failed in
70~$\mu$m due to the low S/N of the data, therefore we had to ignore this band.
The automatic decomposition failed in 100 and 160~$\mu$m, presumably
due to the insufficient S/N in these bands. The trail-and-error fit at
100~$\mu$m resulted in {\tt A$^\prime$} and {\tt C$^\prime$} as the
contributors, however in 160~$\mu$m it could only settle on {\tt A$^\prime$}
and $\mathbb D$. In both cases notable residuals and/or over subtractions could
be seen at the positions of the other objects that did not get fitted. Forcing 
the fit with other combinations of objects either resulted in worse residuals
or crashed the program. Nevertheless, the results for {\tt A$^\prime$} were
repeatable, and therefore we believed that the decomposition was successful for
{\tt A$^\prime$} in these two bands.
In 250~$\mu$m, the automatically iterative decomposition was successful for
all the four ``objects'', which is show in Fig. 36. {\tt A$^\prime$} takes
$\sim 51$\% of the total flux and hence is the major contributor. Therefore
our further discussion will only include {\tt A$^\prime$}.
Table 2 summarizes the flux densities of this major component.

\begin{figure*}[tbp]
\centering
\includegraphics[width=\textwidth]{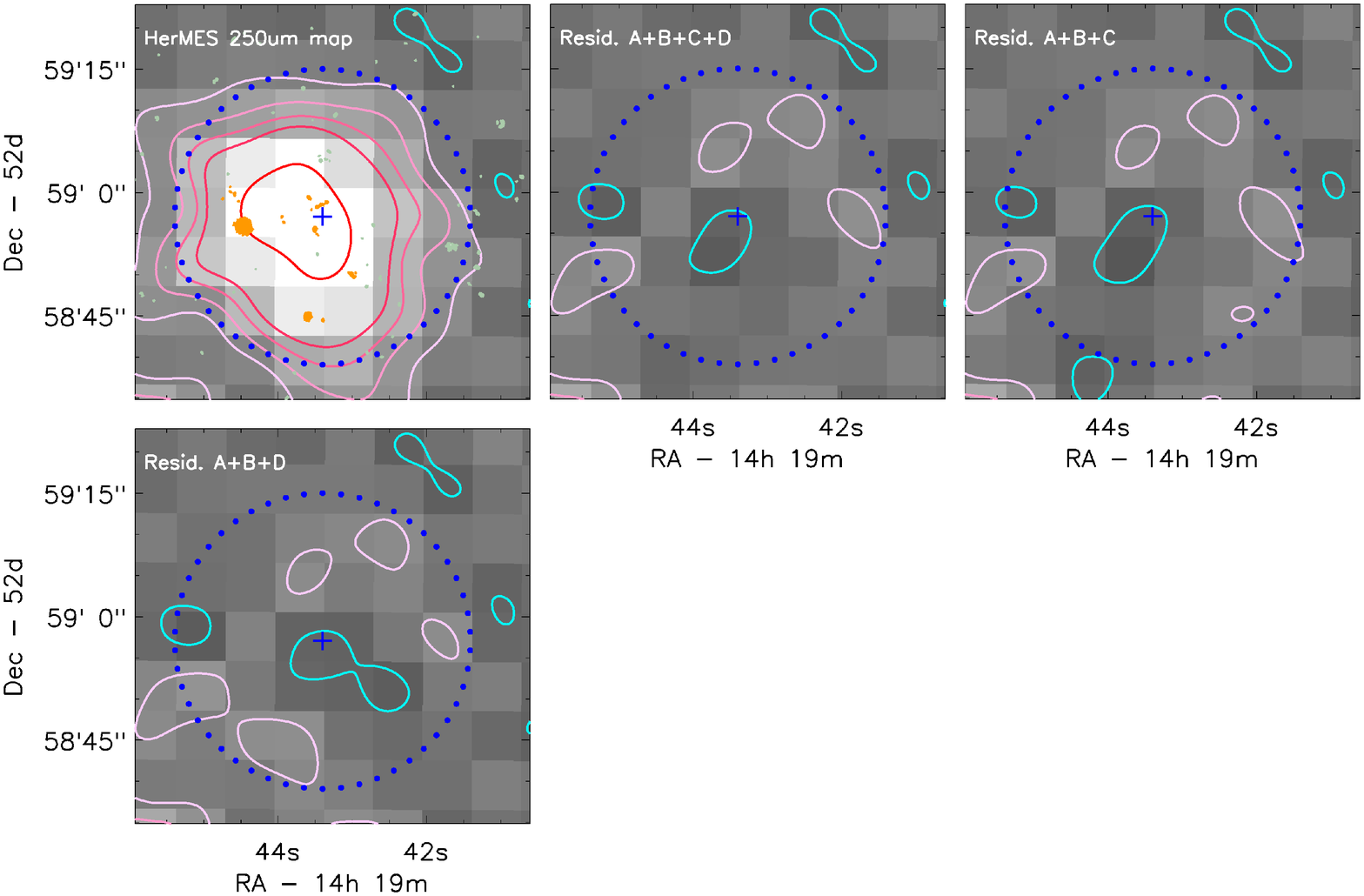}
\caption{Demonstration of the decomposition of EGS19 in 250~$\mu$m. The first 
panel shows the original 250~$\mu$m image, while the others show the residual
maps of the different decomposition schemes where different input sources are
considered (labeled on top). Legends are the same as in Fig. 5. While the 
automatically iterative fit converges on 
{\tt A$^\prime$+B$^\prime$+C$^\prime$+}$\mathbb D$, there are degenerate cases.
However, {\tt A$^\prime$} is always the dominant contributor and its extracted
flux is essentially the same in all these cases.
}
\end{figure*}

\begin{figure*}[tbp]
\centering
\subfigure{
  \includegraphics[width=0.5\textwidth]{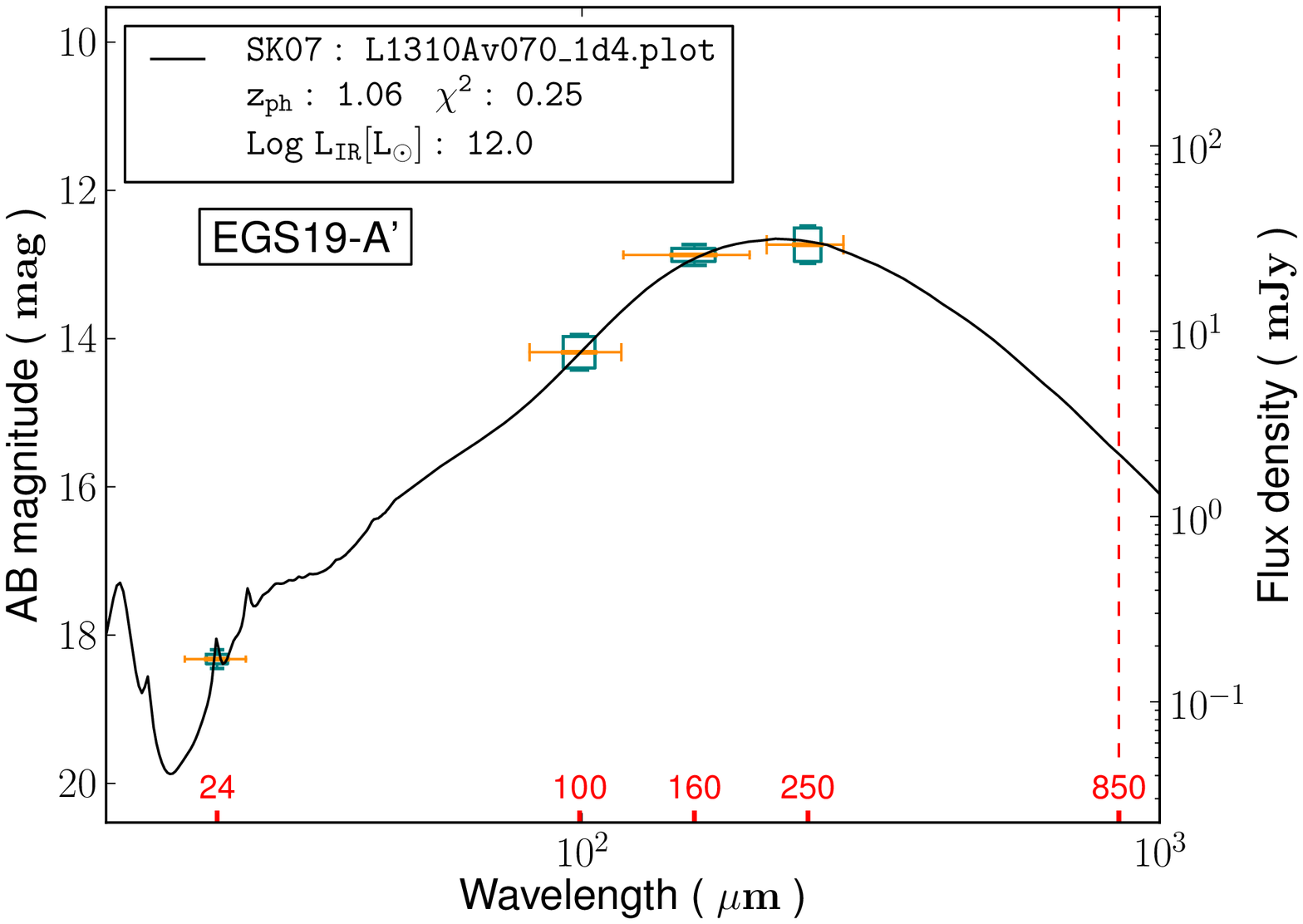}
  \includegraphics[width=0.5\textwidth]{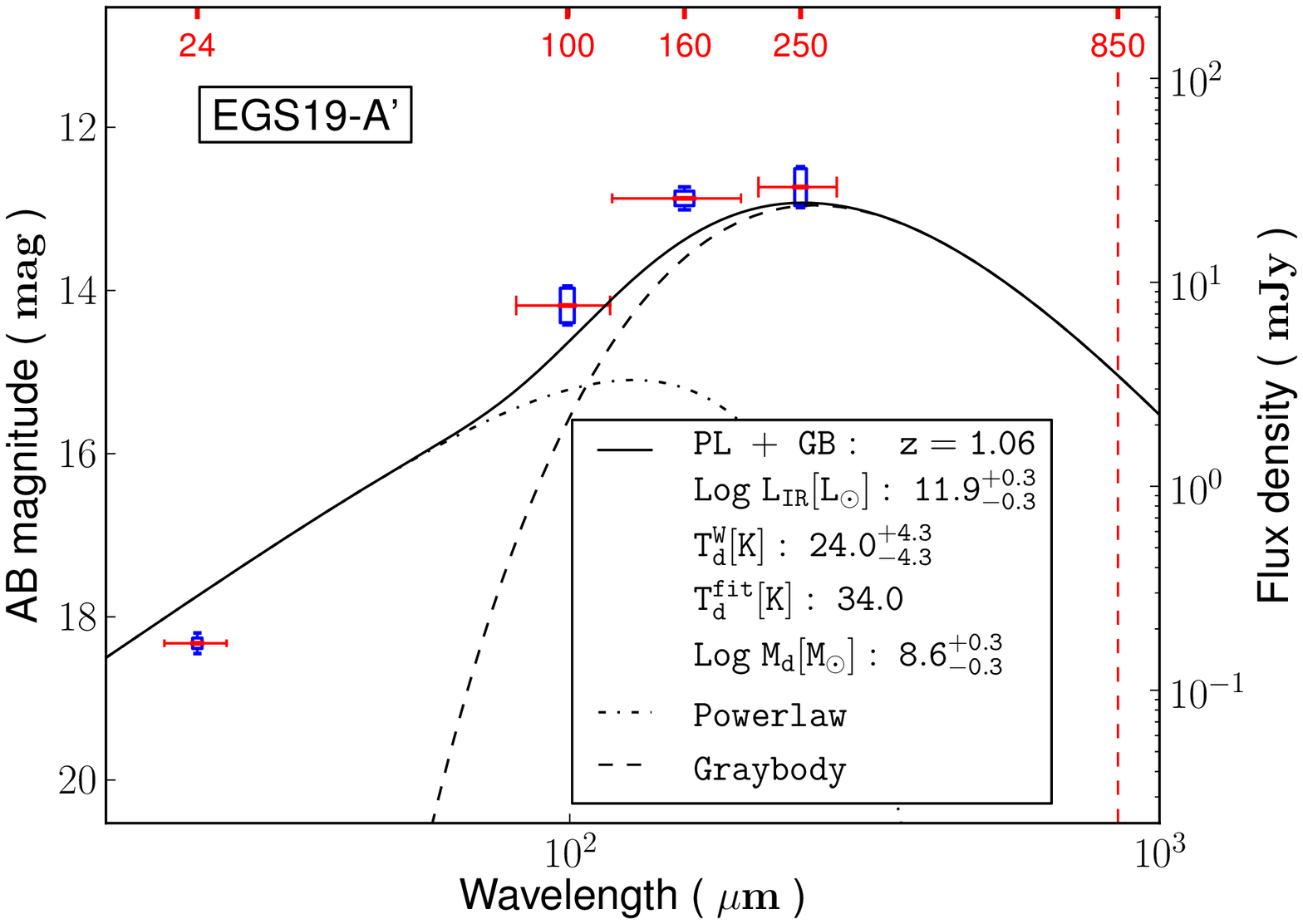}
}
\caption{Mid-to-FIR SED analysis for EGS19-{\tt A$^\prime$} at $z_{ph}=1.06$.
The left panel shows the fit to the SK07 models, and the right panel the fit 
using the power-law + graybody models. Legends are the same as in Fig. 6.
}
\end{figure*}

\subsubsection{Total IR Emission and Stellar Populations}

  The only X-ray source within the EGS19 area is right on $\mathbb C-1$
(positional offset of only $0\farcs 3$). It has full-band 0.5--10~keV flux of
$1.46^{+0.24}_{-0.22}\times 10^{-16} erg/s/cm^2$, which
corresponds to a total X-ray luminosity in restframe 1.1--21.8~keV of 
$1.16\times 10^{43} erg/s$, implying that $\mathbb C-1$ most likely has an
AGN. The lack of X-ray detections at {\tt A$^\prime$} suggests that the
FIR emission of this object is most likely of stellar origin, as the
sensitivity limit would imply an upper limit of $1.8\times 10^{41}$~$erg/s$
in restframe 1--4~keV at $z=1.06$.

    Fig. 37 summarizes the analysis of the mid-to-far-IR SED of 
{\tt A$^\prime$}. The SK07 models provided a good fit, re-assuring that our
decomposition results for {\tt A$^\prime$} is reasonable. We obtained
$L_{IR}=1.0\times 10^{12} M_\odot$, which means that {\tt A$^\prime$} is an
ULIRG. The analytic fit resulted in $T_d^{fit}=34.0$~K and 
$M_d=4.0\times 10^8 M_\odot$. This implies $M_{gas}=5.6\times 10^{10} M_\odot$.

    The dominant members of {\tt A$^\prime$} are $\mathbb A-1$ and 
$\mathbb A-4$, which have extremely similar stellar populations. They both have
moderate stellar masses (4.1 and 2.9$\times 10^{10} M_\odot$, respectively),
old ages (2.0 and 1.7~Gyr, respectively), and moderately prolonged SFH
($\tau=0.7$ and 0.4~Gyr, respectively). Combining $\mathbb A-1$ and
$\mathbb A-4$, we got
$L_{IR}^{ext}=8.1\times 10^{10}L_\odot$, 
$L_{IR}^{blk}=9.2\times 10^{11}L_\odot$, and $SFR_{IR}^{blk}=92$~$M_\odot/yr$.
Their combined $SFR_{fit}=8$~$M_\odot/yr$, and thus 
$SFR_{tot}=100$~$M_\odot/yr$.
Using their stellar masses, we can get
$SSFR=1.4$~Gyr$^{-1}$, $T_{db}^{tot}=700$~Myr and $T_{db}^{blk}=760$~Myr. The
total amount of gas inferred above could fuel this ULIRG for the next 560~Myr
or its dust-blocked region for the next 608~Myr to add a further $\sim 80$\% of
its existing stellar masses.

     There are two radio sources in this region (Ivison et al. 2010). One of
them coincides with
$\mathbb A-1$, and has $S_{1.4GHz}=0.107\pm 0.012$~mJy. The other one is on top
of $\mathbb B-1$, and has $S_{1.4GHz}=0.119\pm 0.012$~mJy. Here we only discuss
the FIR-radio relation for the former. We obtained $q_{IR}=2.27$, which is 
lower but still consistent with the mean of $2.40\pm 0.24$ in Ivison et al.
(2010).

\section{Discussion}

   Strictly speaking, our sample of SDSS-invisible, bright {\it Herschel}
sources is not flux-limited and could suffer from various types of
incompleteness. Nonetheless, our study can serve as a guide to future 
investigations of similar objects at larger, more complete scales.

\subsection{Decomposition of Bright {\it Herschel} Sources}

   Our decomposition approach provides a promising solution to maximize the 
returns of the large volume of precious {\it Herschel} data.
Our technique is based on the position priors from optical or near-IR
images, and this is different from those using the position priors from the
MIPS 24~$\mu$m images, which suffer from the blending problem themselves.
While it probably cannot decompose {\it all} the contributors to a given 
{\it Herschel} source, our method is capable of at least identifying its major
components and extracting their fluxes. This method should work even when only
using medium-deep optical images for position priors. For example, in the case
of UDS01, our results would largely remain the same should we use the priors 
from the CFHTLS-Wide data instead of the CANDELS WFC3 IR data, because the
input list would essentially be the same. 

    Two interesting conclusions can be drawn from our decomposition. First,
as GOODSN63 shows, the brightest 24~$\mu$m source in the {\it Herschel} beam
could have no contribution in the FIR, and therefore the use of 24~$\mu$m data
should be exercised with caution and not to limit to only the brightest
24~$\mu$m sources for position priors. Second, a bright 
{\it Herschel} source usually is the collective result of multiple
contributors. While the seven sources investigated here all have 
$S_{250}>55$~mJy and hence are $\sim 9.5\times$ above the nominal confusion
limit of 5.8~mJy/beam in HerMES at the SPIRE 250~$\mu$m (Nguyen et al. 2010),
it is not guaranteed that they do not suffer from the blending problem. In
fact, only two (GOODS63 and UDS04) out these seven sources can be safely 
treated as having only one contributor. In
two other cases (UDS01 and EGS14) the source is made of two major, distinct
components that are likely at the same redshifts and associated. In the other
three cases (GOODSN06, EGS07 and EGS19), while there is always a primary
component contributing most of the FIR flux, the other contributors, which are
not physically associated with the primary component, are still non-negligible. 

    Both conclusions above actually echo the results of earlier studies
of SMGs using the SMA (Younger et al. 2009; Wang et al. 2011; 
Barger et al. 2012) and the more recent ones using the ALMA (Hodge et al. 2013;
Karim et al. 2013). Our second point, namely, that in most cases bright
{\it Herschel} sources are made of distinct components, will impact the 
interpretation of the FIR source counts and the construction of the FIR 
luminosity functions, especially at the bright end. This is now 
expected from the theoretical side. Niemi et al. (2012) suggest that source 
blending could be an important cause of the inconsistency between their
semi-analytic model predictions and the actual observations of the bright
{\it Herschel} source number counts. Recently, using cosmological numerical
simulation, Hayward et al. (2013) predict that spatially and physically 
unassociated galaxies contribute significantly to the SMG population. 
Therefore, bright {\it Herschel} sources and their multiplicities warrant
further investigations. As compared to 
addressing this problem through using sub-mm interferometry with the SMA or
the ALMA, our approach offers a much less expensive alternative, albeit at the
price that we are only able to reliably extract the major component(s) of
a given {\it Herschel} source. However, as neither the SMA nor the ALMA is a
survey machine, our method has its value in that in principle it can deal with
a large number of sources. In addition, it is applicable in the {\it Herschel}
fields where there are no MIPS 24~$\mu$m data as ``ladders'' but medium-deep
optical imaging data are available or can be acquired. While in this work we 
still used the 24~$\mu$m image to narrow down the input list for the fit, it is
possible to get rid of this intermediate step once we automate the current
trial-and-error approach (Ma, Stefanon \& Yan, in prep.). The APPENDIX
further demonstrates this point.
Of course, relying on
optical images for priors does have its disadvantage in that a
major contributor to the FIR source, if extremely dusty, could still be missing
from a medium-deep optical survey. In this case, the residual image after the
decomposition will be able to reveal such a source. In fact, in our
trial-and-error approach for a number of sources, the iteration was driven by
the residual left at the locations of the suspected contributors.
   
\subsection{ULIRG Diagnostics}

   Our decomposition directly identified the optical-to-near-IR counterparts of
the major components of the FIR sources. This allows us to investigate the 
nature of the FIR emission and the underlying stellar populations. We have
found the following.

   First of all, the major contributors to our FIR sources are all ULIRGs at
$z>1$. Although this is not surprising given their high FIR flux densities
(suggesting high IR luminosities) and faintness in optical (suggesting being at
high redshifts), our analysis provides solid evidence that this is indeed the
case. While one of these sources (EGS07) could have an embedded AGN, all others
are mainly powered by intense star formation heavily obscured by dust. 

    The exquisite morphological details in the restframe optical from the 
{\it HST} data show that these ULIRG all have complicated structures indicating
either merger or local violent instability. This is generally consistent with
the result of Kartaltepe et al. (2012), who find that the majority of the
PACS-selected ULIRGs at $z\sim 2$ in the GOODS-South field are mergers and
irregular galaxies. While a quantitative morphological modeling is beyond the 
scope of this work, most of the counterparts to the ULIRG in
our sample are not likely to be explained by a single
disc galaxy. This is in contrast to the recent study of Targett et al. (2013),
who find that SMGs, being ULIRGs at $z\approx 2$, are mostly disc
galaxies. This contrast probably should not yet be viewed as a contradiction
for two reasons. First, our objects are mostly at $z\approx 1$. Second, based on
the predicted $S_{850}$ (see Table 3) of our objects, most of them probably
are not SMGs based on the conventional SMG selection criterion
($S_{850}\gtrsim 3$--5~mJy; see also Khan et al. 2009). 
Nonetheless, one of our object, GOODSN63-A, is an
SMG at $z_{ph}=2.28$ and apparently cannot be a disc galaxy. Clearly, the
morphologies of high-z ULIRGs merit further investigation.

   The assertion that our sources are all ULIRGs is based on total IR 
luminosities derived from the direct measurements in the FIR bands that have
been properly treated for the effect of blending. They are more reliable than
the extrapolation from the mid-IR
(e.g., using only the MIPS 24~$\mu$m in the pre-{\it Herschel} era). The 
multiple {\it Herschel} FIR bands sample the peak of dust emission, and thus
are more sensitive in selecting ULIRGs than using a single sub-mm band. In fact,
we find that only two ULIRGs in our sample (GOODSN63-{\tt A} and EGS07-{\tt 1})
would satisfy the nominal SMG selection criterion of $S_{850}>5$~mJy.
The {\it Herschel} bands are also sensitive to a wide temperature range, and
are not biased against ULIRG of high temperatures as the traditional SMG
selection at 850~$\mu$m is (e.g., Chapman et al. 2010; Casey et al. 2012).
Indeed, our sample includes one ULIRG whose dust temperature is higher than 
those of normal SMGs (GOODSN06-{\tt A} with $T_d^{fit}=48.7$~K). As shown in
Fig. 38 (left panel), our small sample already shows a trend that dust 
temperature increases with respect to increasing IR luminosity, and this is 
consistent with the results in the recent literature (e.g., Symeonidis et al.
2013; Swinbank et al. 2013; Magnelli et al. 2014). A much larger sample in the
future, constructed following the decomposition process described in this
work, will be able to enhance this trend (particularly at the bright-end) and
determine whether the wide dispersion currently seen is intrinsic or is due to
the contamination to the FIR fluxes by blending.

    We have also investigated the FIR-radio relation of our sources. We find
that some of our sources follow the relation very well and yet some deviate
from it significantly. Fig. 38 (right panel) compares
the $q_{IR}$ values from our sample, which have the mean of $2.22\pm 0.28$, to
the mean of $2.40\pm 0.24$ of Ivison et al. (2010). It seems that our values are
systematically lower than theirs, however our sample is too small for us to
make any assertion. Nevertheless, we argue that using our method of 
decomposition will result in the most reliable measurement of $S_{IR}$ and
hence reduce the random measurement error in the dispersion, and that future
studies using larger sample will be able to test the FIR-radio relation for
$z>1$ ULIRG in better details.

\begin{figure*}[tbp]
\centering
\includegraphics[width=\textwidth]{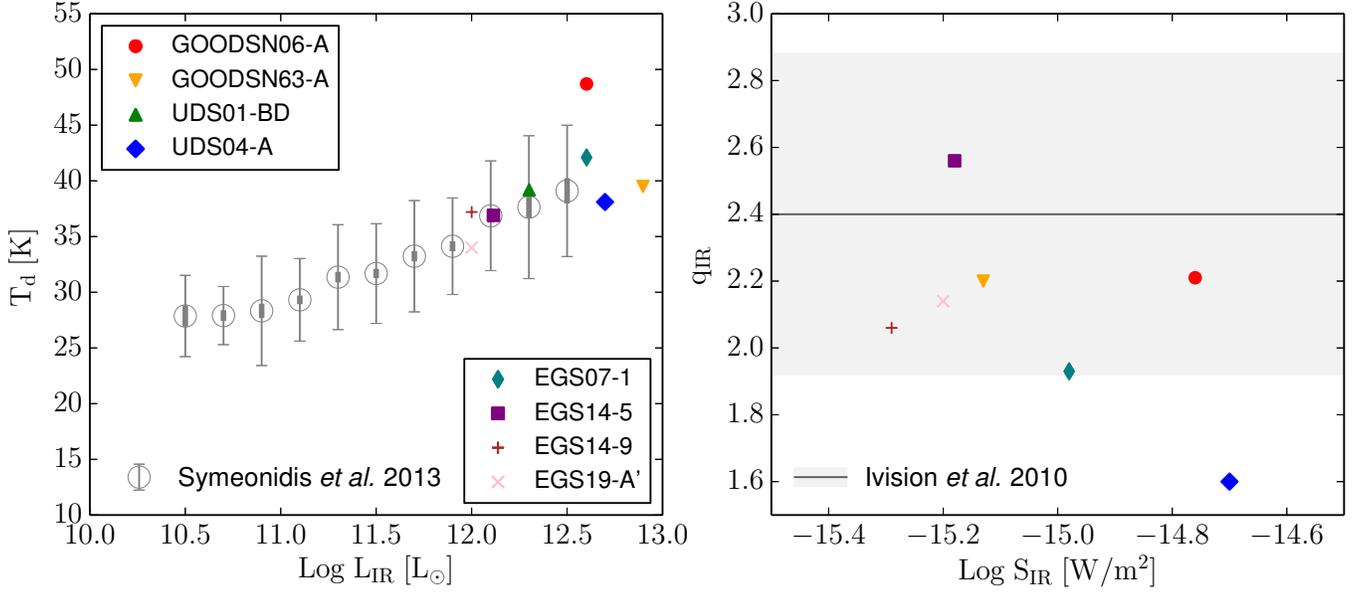}
\caption{(left) Dust temperature of ULIRG as a function of total IR luminosity. 
The circles with error bars are the mean values from Symeonidis et al. (2013)
whose sample is at $0.1<z<2$,
while the results from our sample (as labeled) are plotted as symbols of
various colors. 
(right) Comparison of $q_{IR}$ from our sample, shown as the open squares, to
the result of Ivison et al. (2010). The solid horizontal line indicates their
mean value ($q_{IR}=2.4$) and the gray area represents their 
$\pm 2\sigma_q = \pm 0.48$.
}
\end{figure*}

   It is interesting to relate our ULIRGs to the so-called ``main sequence of 
star formation'' on the SFR (or SSFR) vs. stellar mass plane (e.g., Noeske 
et al. 2007; Elbaz et al. 2007; Daddi et al. 2007; Elbaz et al. 2011; Wuyts
et al. 2011; Rodighiero et al. 2011; Leitner 2012). This is shown in Fig. 39
in terms of SFR (left) and SSFR (right). In the left panel, the solid line 
represents the main sequence of Elbaz et al. (2007) at $z\approx 1$ and the
dashed line $4\times$ above it follows Rodighiero et al. (2011) to indicate the
boundary above which starbursts locate ($\sim 2$~$\sigma$ above the main
sequence). In the right panel, the solid line is from the empirical fit to the
main sequence as a function of redshifts based on Leitner (2012), and is fixed
at the median redshift ($z=1.2$) of our sample. Somewhat
surprisingly, some of our objects are quite close or below this boundary and
would be difficult to qualify as starbursts in this convention. As it turns
out, the existing stellar populations of these objects tend to have an old
age ($T>1$~Gyr). In particular, GOODSN63-{\tt A} and EGS14-{\tt 5}, which are
the closest to the main sequence in SSFR vs. $M^*$, have the oldest ages.

\begin{figure*}[tbp]
\centering
\subfigure{
  \includegraphics[width=0.5\textwidth]{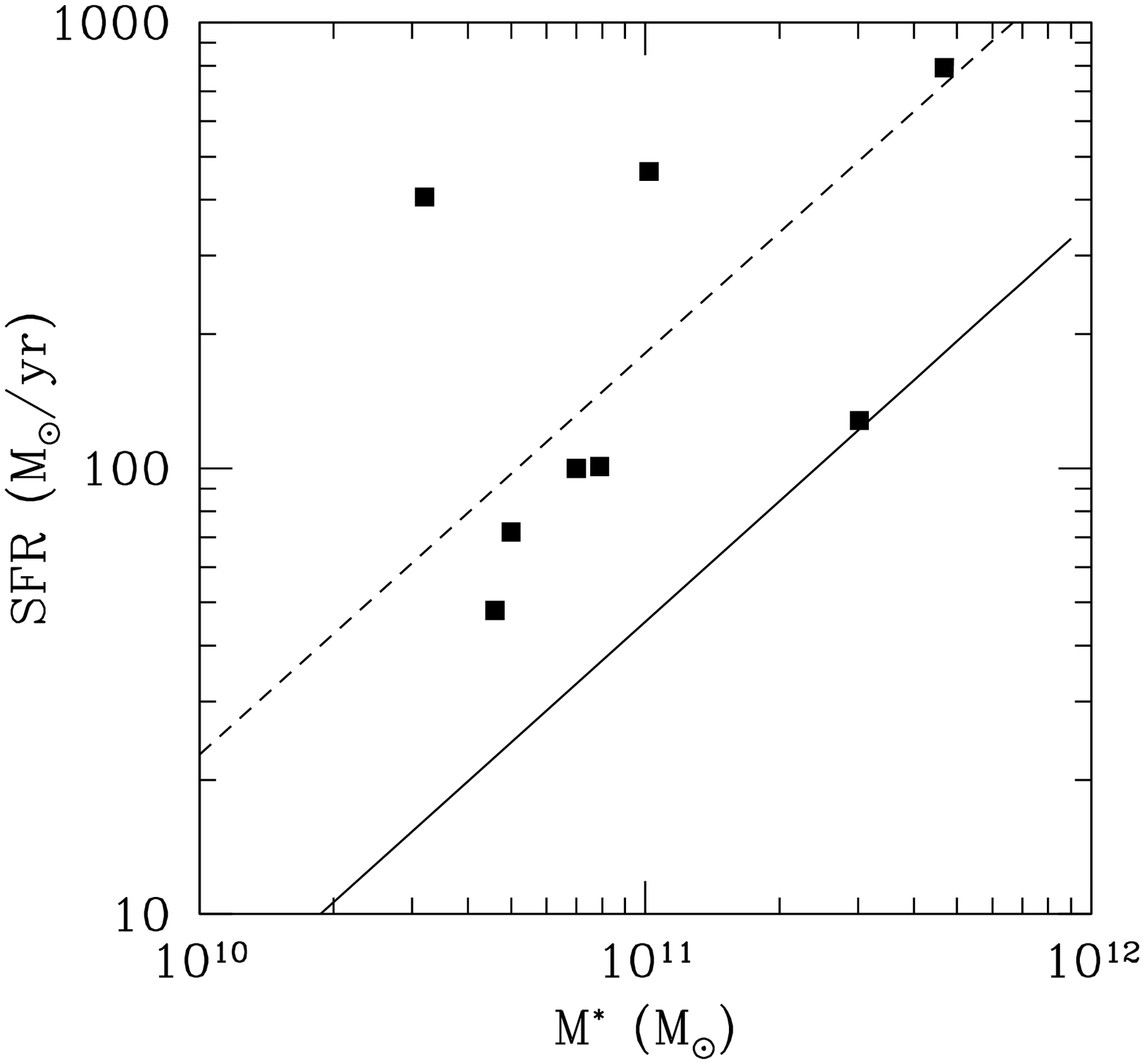}
  \includegraphics[width=0.5\textwidth]{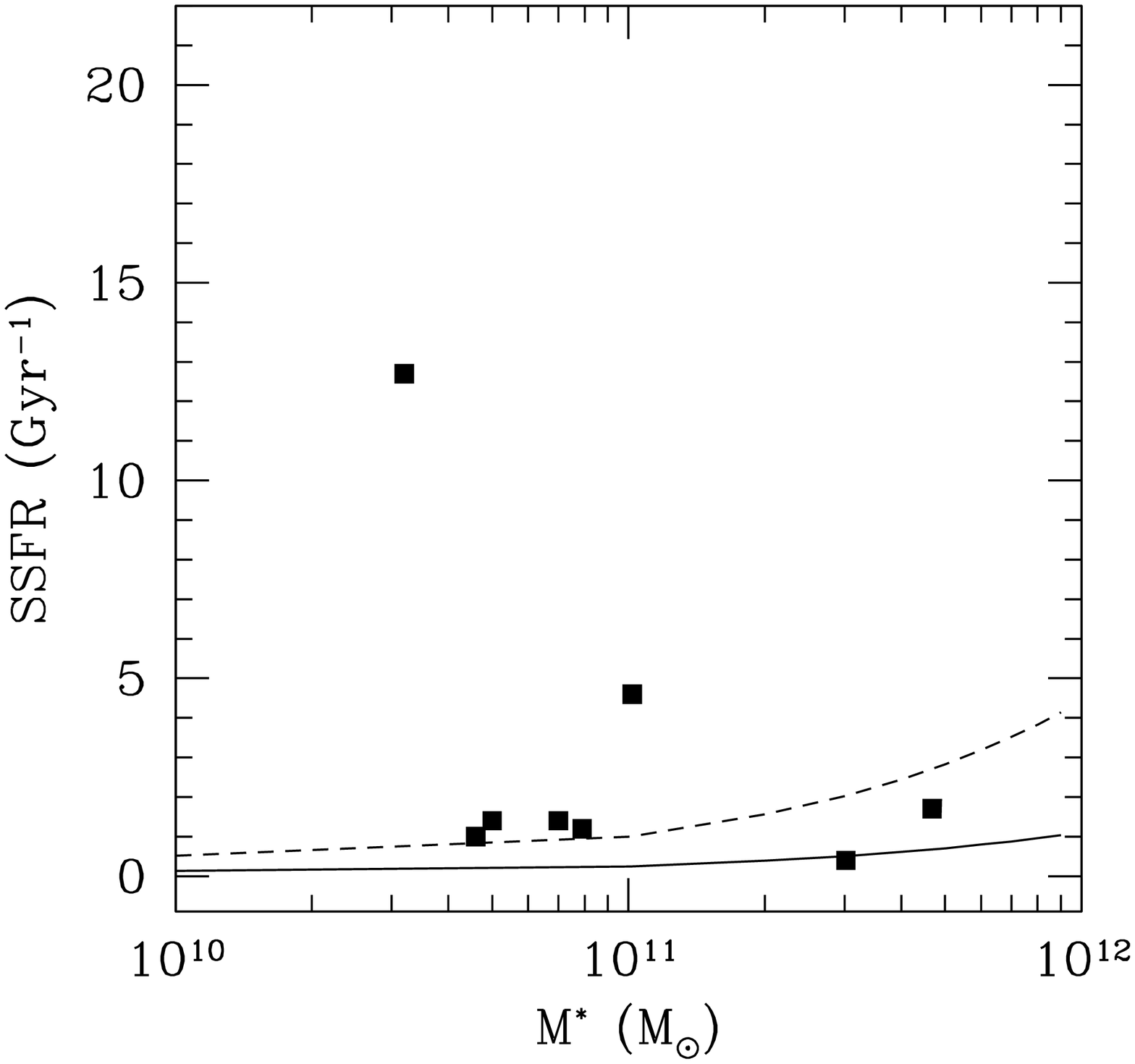}
}
\caption{The locations of the ULIRGs in our sample with respect to the 
``star formation main sequence'' on the SFR versus $M^\ast$ (left) and the
SSFR versus $M^\ast$ (right) planes. The data points are based on $SFR_{tot}$.
In the left panel, the solid line is the
``main sequence'' taken from Elbaz et al. (2007) and has taken into account the
conversion from the Salpeter IMF to the Chabrier IMF. The dashed line follows
that of Rodighiero et al. (2011) and is $4\times$ above the main sequence. In
the right panel, the solid curve is
the main sequence power-law fit taken from Leitner (2012; using the data from
Karim et al. 2011) and fixed at the median redshift ($z=1.2$) of our sample.
The dashed curve is $4\times$ above the main sequence. The objects above the
dashed line/curve can be taken as ``starbursts'' in this context. Some of our
ULIRGs are actually quite close to the main sequence.
}
\end{figure*}

\begin{figure*}[tbp]
\centering
\subfigure{
  \includegraphics[width=0.5\textwidth]{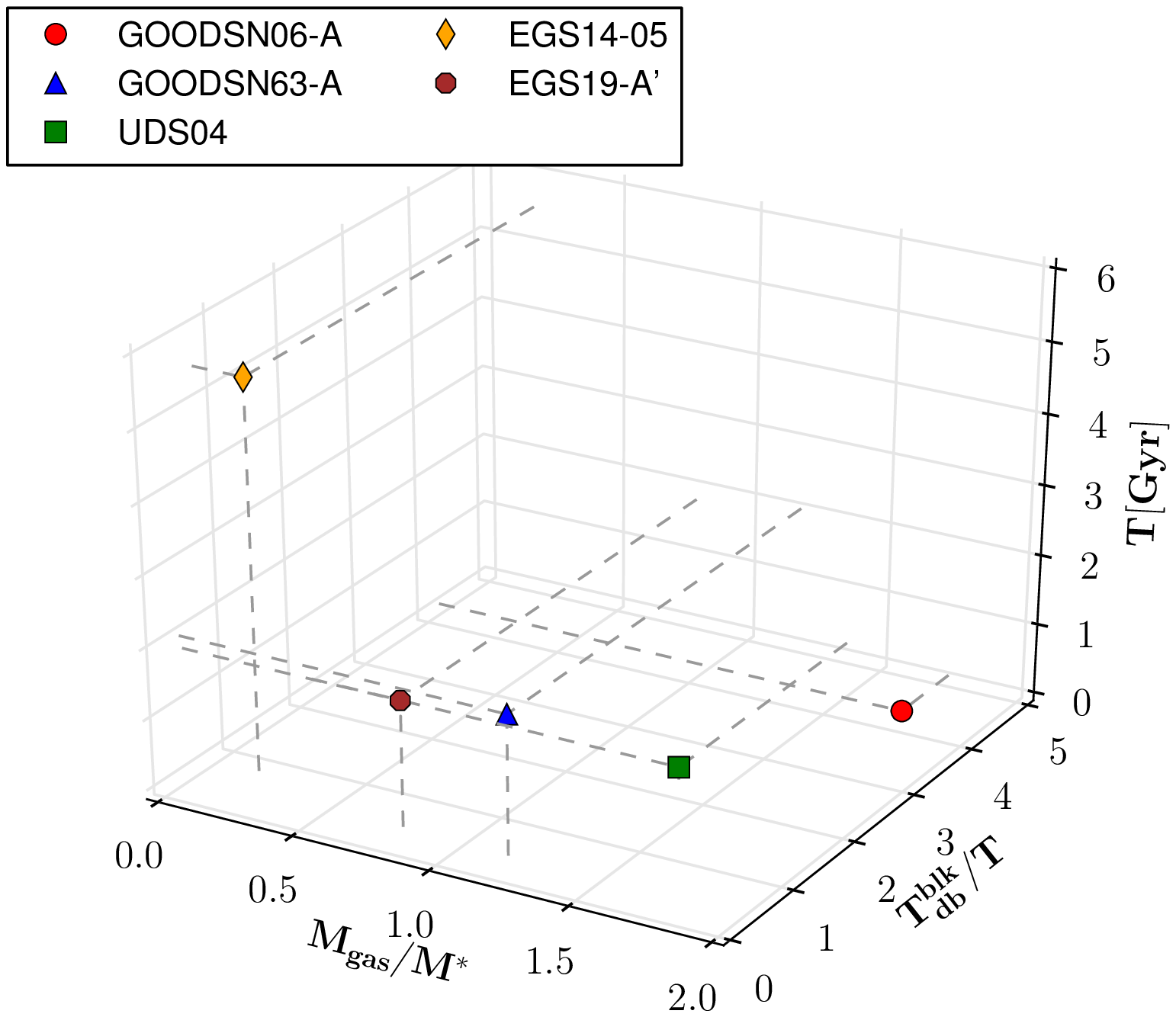}
  \includegraphics[width=0.5\textwidth]{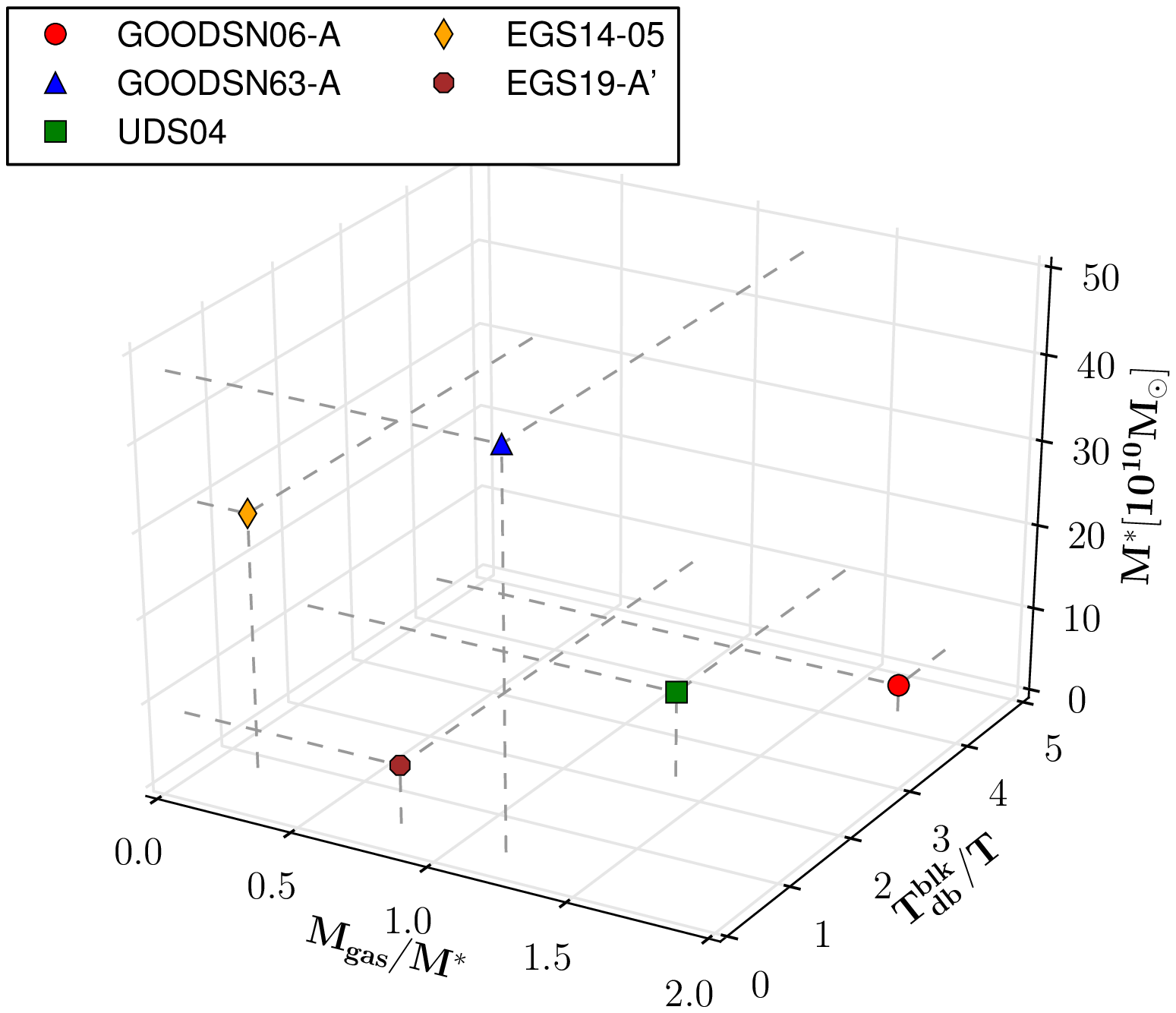}
}
\caption{ULIRG diagnostics involving the gas mass ($M_{gas}$) in the ULIRG
region, the stellar mass ($M^\ast$) and the age ($T$) of the existing 
stellar population in the exposed region, and the stellar mass doubling time
$T_{db}$ due to ULIRG assuming a constant $SFR_{IR}$ into the future.
}
\end{figure*}
 
    The closest analogs to our objects are
SMGs selected at 850~$\mu$m, which are mostly ULIRG at $z\approx 1$--3.
The stellar population studies of SMGs (Borys et al. 2005; 
Dye et al. 2008; Micha{\l}owski et al. 2010; Hainline et al. 2011; 
Micha{\l}owski et al. 2012; Targett et al. 2013) have come to the conclusion
that they have very high stellar masses, with the quoted median ranging from
$7\times 10^{10}M_\odot$ to $2\times 10^{11}M_\odot$. The recent numerical
simulation of Hayward et al. (2011) shows that SMGs should have a minimum
of $M^*\gtrsim 6\times 10^{10}M_\odot$ and that typical masses should be
higher. Micha{\l}owski et al. (2012) and Targett et al. (2013) have found mean
values of $<M^*>=(2.8\pm 0.5) \times 10^{11}$ and
$(2.2\pm 0.2) \times 10^{11} M_\odot$, respectively. The mean of our sample is 
$<M^*>=(1.4\pm 1.5)\times 10^{11} M_\odot$, where the large dispersion
reflects the fact that our estimates range from as low as 
$2.5\times 10^{10} M_\odot$ to as high as 
$4.7\times 10^{11} M_\odot$. In fact, six out of the nine objects are below
$10^{11} M_\odot$ (see Table 2). While the differences in the assumed SFH
could affect the stellar mass estimates systematically (for example, 
Micha{\l}owski et al. (2012) conclude that the estimates using their preferred
two-component models are 2--3$\times$ higher than those using single-component
models), our large dispersion cannot be attributed to this reason. We believe
that this is due to the fact that most of our objects are not SMGs. In other
words, ULIRGs selected by {\it Herschel} bands are more heterogeneous
in their existing stellar populations than SMGs. This probably is 
understandable given that starbursting galaxies at high redshifts are found
over a wide range of stellar masses (e.g., Rodighiero et al. 2011). However, a
starburst (defined as being objects far away from the ``main sequence'') is
not necessarily an ULIRG, and therefore the spread in ULIRG stellar masses
should not be taken for granted. It thus will be important to further
investigate this question with a much increased, properly deblended sample.

    While the stellar mass obtained through SED fitting is robust,
it is known that the age inferred from the same process can be more uncertain
(see e.g. Wuyts et al. (2011) for a recent reference). Presenting a 
comprehensive examination to address the age estimate issue is beyond the 
scope of this paper. Nevertheless, we argue that the age estimate in our case
could still be useful.
Dust extinction and metallicity are the two most severe sources of uncertainty.
In the SED fitting for the existing (exposed) stellar populations, we allowed
$A_V$ to vary from 0 to 4~mag, and assumed the solar metallicity.
As our objects are rather dusty, there is a chance that we might still
underestimate the extinction. On the other hand, these objects could have lower
metallicity because they are at $z>1$. Generally speaking, a higher extinction
would tend to result in a younger age in the fit because it would make the 
templates redder, and a lower metallicity would tend to result in a older
age because it would make the templates bluer. Therefore, the two possible 
biases in our treatment tend to cancel out. For this reason, we believe that
the ages that we derived still provide useful hints to understanding the
progenitor of these ULIRGs. 

    The derived ages of our objects have a wide spread, 
ranging from as young as $\sim 20$~Myr to as old as almost the age of the
universe at the observed redshift, and do not have any obvious trend with 
respect to the existing stellar mass. This implies that the progenitor of an
ULIRG could be formed at any redshift, and that the ULIRG phase can be turned
on at any point during its lifetime. Furthermore, it can be turned on at any
mass range within $10^{10-11} M_\odot$. This is in contrast to the SMG 
population, where the observed ULIRG phase is believed to only happen at the
late stage of their stellar mass assembly and probably will not add much to
the existing mass (see e.g., Micha{\l}owski et al. 2010). 

   In light of this, we propose new diagnostics of ULIRGs, which is shown as
3-D plots in Fig. 39 for six ULIRGs from Table 3. The ratio $M_{gas}/M^\ast$ is
a measure of the gas content in the ULIRG region with respect to the existing
stellar mass in the exposed region. If $M_{gas}/M^\ast>1$, in principle the
ULIRG will be capable of doubling the mass. The ratio $T_{db}/T$ is a measure 
of how quickly the ULIRG can double the mass as compared to the age of the
existing stellar population
\footnote{Numerically, $T_{db}/T$ is equivalent to $1/b$, where $b$ is the
``birthrate'' parameter defined as $b=SFR/<SFR>$ and $<SFR>=M^\ast/T$
(Scalo 1986)}.
The age $T$ is a measure of how closely the
current ULIRG is related to the past SFH, i.e., it can be understood as at
what stage the ULIRG happens since the birth of the galaxy. Finally, $M^\ast$,
together with $T$, indicates how mature the existing stellar population is.

   It is intriguing to see how these six ULIRGs spread in the
$M_{gas}/M^{\ast} - T_{db}^{blk}/T - T - M^\ast$ space. Two of them, namely,
GOODSN06-{\tt A} and UDS04-{\tt A}, are the most gas rich and the most
efficient in their mass assembly. As their hosts are all quite young, these
ULIRGs are likely very close to, or even being an extension of the past episode
of active star formation in the hosts.  The next in line is GOODSN63-{\tt A},
which is capable of doubling its existing mass in a period even shorter than
the age of the host.  Considering that it has a very high stellar mass already
(the highest among all of the six), this is rather extreme. However, its
current ULIRG phase should have little to do with the bulk of the assembly of
the existing mass given the old age of the host. The gas reservoirs of
EGS14-{\tt 9} and EGS19-{\tt A$^\prime$} are somewhat insufficient for them to
double their existing masses, and hence their current ULIRG phase is less 
important in terms of stellar mass assembly. In this regard, the current ULIRG
phase of EGS14-{\tt 5} is the least important, as it does not have sufficient
amount of gas to get close to double the existing mass. However, it is also
very unusual that a very high mass, extremely old galaxy like EGS14-{\tt 5}
still has not yet shut off its star formation processes at $z\approx 1$. 

   All this suggests that the high-z ULIRGs make a diverse population. The 
diagnostics such as shown in Fig. 40 have the potential of revealing the role of
ULIRG in assembling high-mass galaxies. For this purpose, a much larger sample
will be necessary.

\section{Summary}

   We studied a sample of seven very bright {\it Herschel} 
sources ($S_{250}>55$~mJy) from the HerMES program that are not visible in the
SDSS. Such sources have a surface density of roughly 10/deg$^2$, and comprise a
non-negligible fraction of the brightest FIR sources on the sky. In order to 
understand their nature, we selected these seven sources in particular because
they are in the CANDELS fields where a rich set of multi-wavelength data sets
are available for our study.

    Due to the large beam sizes of the {\it Herschel} instruments, the deep
optical and near-IR images in these fields readily reveal many possible 
counterparts within the footprints of these FIR sources. To combat this
problem, we took a new approach to decompose the heavily blended, potential
counterparts, using their centroids at the high-resolution near-IR or optical
images as the position priors. Such an elaborated counterpart identification is
superior to the simple treatment of using a mid-IR proxy such as the MIPS
24~$\mu$m image and claiming the brightest mid-IR source within the FIR beam as
the counterpart. In fact, we show that in at least one case (GOODSN63) the 
brightest 24~$\mu$m source within the 250~$\mu$m beam actually does not 
contribute to the FIR emission, which agrees with the SMA interferometry. Our
method is also an improvement to using the mid-IR data for position priors, as
such images are often already confused in the first place. In this regard, our
approach has the advantage of unambiguously identifying the counterparts and
extracting their fluxes at the same time. While in this work we still use the
MIPS 24~$\mu$m data to narrow down the number of input objects to the fitting
procedure, it is possible to eliminate this step when we fully automate the
entire process in the near future (see APPENDIX).
Once implemented, the automatic routine 
will have a wide application in the {\it Herschel} fields where the 24~$\mu$m
data are not available. 

   Our result shows that in most cases multiple objects contribute to the
ostensibly single FIR
source. While some of them are multiples at the same redshifts, others are
superposition by chance and are physically unrelated. If left untreated, the
contamination to the FIR flux due to the latter case could be as high as 40~\%.
In either case, the decomposition could possibly reconcile the FIR bright-end
source count discrepancy between the observations and the model predictions,
and will be necessary in deriving the FIR luminosity functions, particularly
in the bright-end. Our approach provides a much less expensive alternative to
doing sub-mm interferometry, and has the capability of dealing with a large
number of objects upon full automation.

   The properly extracted, multi-band FIR fluxes allow us to determine the
total IR luminosities with high accuracy, and also enable us to derive other
physical parameters in the dusty region, such as the dust temperature, the dust
mass and the gas mass. We find that all these seven objects are ULIRG
($L_{IR}\geq 10^{12}L_\odot$) at $z\geq 1$. The deep {\it HST}
images show that they all have very disturbed morphologies, indicating either
merger or violent instability. Using the radio data in these fields, we 
investigate the FIR-radio relation, and find that it generally holds but still
breaks down in a number of cases. However, a proper interpretation of the 
break-down will have to wait for a much larger sample.

    The {\it Herschel} FIR bands sample the peak of dust emission over a wide
range of temperatures and redshifts, and hence the selection of ULIRG at $z>1$
using the {\it Herschel} bands will be more comprehensive and less prone to 
selection biases. Our limited sample already shows that this is indeed the
case. The closest analog to our objects are SMG, which are known to be biased
against high dust temperature. Our small sample includes one ULIRG that has a
high dust temperature, which was barely selected by the previous SCUBA
survey. The majority of our objects actually are of similar dust temperatures
as normal SMG, however they all fall below the nominal SMG flux selection
limit, presumably due to their lower stellar masses that the SMGs.

    The detailed SED analysis of the objects in our sample shows that the
host galaxies of these ULIRGs at $z\geq 1$ are heterogeneous. Within our small
sample, the hosts span a wide range in their stellar masses and star formation
histories. This suggests that an ULIRG phase, if happens, can occur at any stage
during the evolution of high-mass galaxies. We provide a new diagnostics of 
high-z ULIRGs, which utilizes the stellar mass and the age of the host, the 
ratio of the gas mass to the existing stellar mass, and the ratio of the 
stellar mass doubling time to the age of the host. This can be used
to reveal how efficient and effective the current ULIRG phase is producing
stars, and its relation to the existing stellar population. With a much
increased sample in the future, it is promising to improve such diagnostics to
shed new light on the role of ULIRGs in the global stellar mass assembly.

\acknowledgements

We thank the anonymous referee for the critical reading and useful comments,
which helped improve the quality of the paper.
We thank M. Bolzonella for providing the latest update of the Hyperz code, 
C. Peng for the useful discussion of the GALFIT code, R. Siebenmorgen for
explaining a number of details of the SK07 models, R. Ivison for allowing
the use of the non-public radio map in the EGS field, and E. Laird and K.
Nandra for the AEGIS Chandra X-ray catalog that is not yet public.
We also thank M. Dickinson, H. Ferguson, J. Newman, M. Salvato for their 
useful comments. 
H.Y., M.S. and Z.M. acknowledge the support of Spitzer RSA 1445905.
This work is based on observations made by Herschel, an ESA space
observatory with science instruments provided by European-led Principal
Investigator consortia and with important participation from NASA.
Support for Program number HST-GO-12060 was provided by NASA through a grant
from the Space Telescope Science Institute, which is operated by the
Association of Universities for Research in Astronomy, Incorporated, under
NASA contract NAS5-26555. 
This work is also based in part on observations made with the Spitzer Space 
Telescope, which is operated by the Jet Propulsion Laboratory, California
Institute of Technology under a contract with NASA. Support for this work was
provided by NASA through an award issued by JPL/Caltech.
This work also makes use of the SDSS. Funding for the creation and distribution
of the SDSS Archive has been provided by the Alfred P. Sloan Foundation, the 
Participating Institutions, the National Aeronautics and Space Administration,
the National Science Foundation, the Department of Energy, the Japanese 
Monbukagakusho, and the Max Planck Society. The SDSS Web site is
http://www.sdss.org. The SDSS is managed by the Astrophysical Research 
Consortium for the Participating Institutions. The Participating Institutions
are the University of Chicago, Fermilab, the Institute for Advanced Study, the
Japan Participation Group, The Johns Hopkins University, Los Alamos National 
Laboratory, the Max Planck Institute for Astronomy, the Max Planck Institute
for Astrophysics, New Mexico State University, the University of Pittsburgh,
Princeton University, the US Naval Observatory, and the University of
Washington.
This paper has used data obtained by the SHARDS project, funded by the Spanish
MINECO grant AYA2012-31277, and based on observations made with the Gran 
Telescopio Canarias (GTC).

\begin{deluxetable}{llcccc}
\tablewidth{0pt}
\tablecolumns{6}
\tabletypesize{\scriptsize}
\tablecaption{Sample Summary}
\tablehead{
\colhead{HerMES ID} &
\colhead{Short ID} &
\colhead{RA \& DEC (J2000.0)} &
\colhead{$S_{250}$} &
\colhead{$S_{350}$} &
\colhead{$S_{500}$}
}
\startdata

1HERMES S250 SF J123634.3+621241 & GOODSN06 & 12:36:34.3 +62:12:41 & $72.1\pm 5.4$ & $44.4\pm 3.2$ & $17.3\pm 3.7$ \\
1HERMES S250 SF J123730.9+621259 & GOODSN63 & 12:37:30.9 +62:12:59 & $55.7\pm 5.4$ & $59.9\pm 3.1$ & $46.3\pm 3.5$ \\
1HERMES S250 SF J021806.0-051247 & UDS01    & 02:18:06.0 -05:12:47 & $55.2\pm 9.7$ & $40.0\pm 7.6$ & $15.3\pm 6.1$ \\
1HERMES S250 SF J021731.1-050711 & UDS04    & 02:17:31.1 -05:07:11 & $67.6\pm 9.7$ & $49.0\pm 7.7$ & $12.2\pm 6.3$ \\
1HERMES S250 SF J141900.3+524948 & EGS07    & 14:19:00.3 +52:49:48 & $81.2\pm 6.0$ & $71.7\pm 4.9$ & $42.9\pm 5.2$ \\
1HERMES S250 SF J142025.9+525935 & EGS14    & 14:20:25.9 +52:59:35 & $57.4\pm 6.0$ & $45.6\pm 4.9$ & $22.3\pm 6.0$ \\
1HERMES S250 SF J141943.4+525857 & EGS19    & 14:19:43.4 +52:58:57 & $57.4\pm 6.0$ & $24.4\pm 5.4$ & $ 9.2\pm 7.1$ \\

\enddata

\tablecomments{The listed flux densities in 250 ($S_{250}$), 350 ($S_{350}$)
   and 500~$\mu$m ($S_{500}$) are in mJy. These values are taken from the HerMES DR1
   catalogs, which are based on the ``xID'' catalogs (Wang et al., in prep.). The errors include the confusion
   noise. ``Short ID'' is a nickname assigned here for simplicity.
  }
\end{deluxetable}

\begin{deluxetable}{lcccccccc}
\tablewidth{0pt}
\tablecolumns{9}
\tabletypesize{\scriptsize}
\tablecaption{Decomposed Mid-to-Far-IR Fluxes of the Major Components}
\tablehead{
\colhead{ID} &
\colhead{RA \& DEC (J2000.0)} &
\colhead{$S_{24}$} &
\colhead{$S_{70}$} &
\colhead{$S_{100}$} &
\colhead{$S_{160}$} &
\colhead{$S_{250}$} &
\colhead{$S_{350}$} &
\colhead{$S_{500}$}
}
\startdata

GOODSN06-{\tt A} & 12:36:34.519 +62:12:40.99 & $0.473\pm 0.023$ & $10.40\pm 0.70$ &  $32.4\pm 1.4$ & $65.6\pm 3.6$ & $59.0\pm 6.5$ &    ---        &  ---      \\
GOODSN63-{\tt A} & 12:37:30.767 +62:12:58.74 & $0.170\pm 0.009$ &    ---          &   $5.0\pm 0.5$ & $22.4\pm 1.1$ & $55.7\pm 5.4$ & $59.9\pm 3.1$ &  $46.3\pm 3.5$ \\
UDS01-{\tt B}    & 02:18:06.114 -05:12:50.11 & $0.319\pm 0.020$ &    ---          &    ---         &   ---         & $32.6\pm 9.5$ &    ---        &  ---      \\
UDS01-{\tt D}    & 02:18:06.159 -05:12:44.93 & $0.465\pm 0.020$ &    ---          &    ---         &   ---         & $22.6\pm 9.5$ &    ---        &  ---      \\
(UDS01-{\tt BD)} &     ---                   & $0.784\pm 0.025$ &    ---          &    ---         &   ---         & $55.2\pm 9.7$ & $40.0\pm 7.6$ &  $15.3\pm 6.1$ \\
UDS04-{\tt A}    & 02:17:31.159 -05:07:09.15 & $0.670\pm 0.020$ &    ---          &    ---         &   ---         & $67.7\pm 9.7$ & $49.0\pm 7.7$ &  $12.2\pm 6.3$ \\
EGS07-{\tt 1}    & 14:19:00.202 +52:49:47.74 & $0.620\pm 0.015$ &  $1.96\pm 0.50$ &  $13.9\pm 1.0$ & $36.8\pm 2.4$ & $64.3\pm 6.5$ &    ---        &  ---      \\
EGS14-{\tt 5}    & 14:20:25.704 +52:59:31.75 & $0.273\pm 0.015$ &     ---         &   $8.0\pm 1.3$ & $28.2\pm 3.2$ & $29.2\pm 6.2$ &    ---        &  ---      \\
EGS14-{\tt 9}    & 14:20:26.424 +52:59:39.33 & $0.243\pm 0.017$ &     ---         &   $7.8\pm 1.6$ & $14.7\pm 3.1$ & $28.2\pm 6.1$ &    ---        &  ---      \\
EGS19-{\tt A-1}  & 14:19:43.436 +52:58:58.19 & $0.170\pm 0.010$ &     ---         &   $7.7\pm 1.5$ & $25.8\pm 2.1$ & $29.3\pm 6.1$ &    ---        &  ---      \\
EGS19-{\tt A-2}  & 14:19:43.583 +52:58:59.02 &                  &                 &                &               &               &               &           \\
EGS19-{\tt A-4}  & 14:19:43.470 +52:58:52.20 &                  &                 &                &               &               &               &           \\

\enddata

\tablecomments{The mid-to-far-IR flux densities (in mJy) of the major components
 of the {\it Herschel} sources in Table 1, based on our decomposition results.
 For completeness, we also include the case of UDS01-{\tt BD} where the
 two components UDS01-{\tt B} and {\tt D} are added together.
 For EGS19, the listed fluxes are for the major component 
 EGS19-{\tt A$^\prime$}, which is a ``composite source'' made of
 EGS19-{\tt A-1}, {\tt A-2} and {\tt A-4}. 
 For GOODSN06, GOODSN63, UDS01, the positions reported here are based
 on the $H_{160}$ images. For UDS04, EGS07, EGS14, and EGS19, the reported
 positions are based on the CFHTLS images. 
}

\end{deluxetable}

\begin{deluxetable}{llccccccccccccll}
\tablewidth{0pt}
\tablecolumns{16}
\tabletypesize{\scriptsize}
\tablecaption{Physical Properties of Revealed ULIRGs}
\tablehead{
  \colhead{ID} &
  \colhead{$z$} &
  \colhead{$L_{IR}$} &
  \colhead{$L_{IR}^{blk}$} &
  \colhead{$S_{850}$} &
  \colhead{$SFR_{IR}^{blk}$} &
  \colhead{$T_d^{fit}$} &
  \colhead{$M_{d}$} &
  \colhead{$M_{gas}$} &
  \colhead{$M^{*}$} &
  \colhead{$SFR_{fit}$} &
  \colhead{$\tau$} &
  \colhead{$T$} &
  \colhead{$SSFR_{tot}$} &
  \colhead{$T_{db}^{tot}$} &
  \colhead{$T_{db}^{blk}$} \\
  &
  &
 \colhead{($10^{12}L_\odot$)} &
 \colhead{($10^{12}L_\odot$)} &
 \colhead{(mJy)} &
 \colhead{($M_\odot$/yr)} &
 \colhead{(K)} &
 \colhead{($10^{8}M_\odot$)} &
 \colhead{($10^{10}M_\odot$)} &
 \colhead{($10^{10}M_\odot$)} &
 \colhead{($M_\odot$/yr)} &
 \colhead{(Gyr)} &
 \colhead{(Gyr)} &
 \colhead{(Gyr$^{-1}$)} &
 \colhead{(Gyr)} &
 \colhead{(Gyr)} 
}
\startdata
  GOODSN06-{\tt A}     & {\bf 1.225} & 4.0 & 1.7  & 2.9 &  170  &   48.7  &  4.0 &  5.6 &  3.2 & 236 &  0.02 & 0.045 & 12.7 & 0.079 & 0.188 \\
  GOODSN63-{\tt A}\tablenotemark{a}
                       &     2.28    & 7.9 & 7.8  & 9.8 &  781  &   39.5  & 39.8 & 55.7 & 46.8 &  11 &   0   & 2.0   &  1.7 & 0.593 & 0.602 \\
  UDS01-{\tt B}        & {\bf 1.042} & 1.0 & 0.48 & 1.5 &   48  &   ---   & ---  & ---  &  4.6 &   0 &   0   & 0.13  &  1.0 & 1.0   & 1.0   \\
  UDS01-{\tt D}        & {\bf 1.042} & 1.0 & 0.62 & 1.5 &   62  &   ---   & ---  & ---  &  7.9 &  39 &  1.0  & 1.4   &  1.2 & 0.882 & 1.3   \\
 (UDS01-{\tt BD})\tablenotemark{b}
                       & {\bf 1.042} & 2.0 & ---  & --- &  ---  &   39.2  &  5.0 &  7.0 & 12.5 & --- &  ---  & ---   &  --- & ---   & ---   \\
  UDS04-{\tt A}        & {\bf 1.267} & 5.0 & 2.5  & 3.7 &  255  &   38.1  & 10.0 & 14.0 & 10.2 & 208 &  0.08 & 0.18  &  4.6 & 0.216 & 0.392 \\
  EGS07-{\tt 1}\tablenotemark{c}
                       & {\bf 1.497} & 4.0 & ---  & 5.4 &  ---  &   42.1  & 12.6 & 17.6 &  6.9 & 137 &  7.0  & 0.72  &  --- & ---   & ---   \\
  EGS14-{\tt 5}        &     1.15    & 1.3 & 0.83 & 2.1 &   83  &   36.9  &  4.0 & 5.6  & 30.2 &  45 &  4.0  & 5.5   &  0.4 & 2.3   & 3.6   \\
  EGS14-{\tt 9}        &     1.15    & 1.0 &   0  & 2.1 &    0  &   37.2  &  3.2 & 4.4  &  5.0 &  72 &  0.04 & 0.1   &  1.4 & 0.694 & ---   \\
EGS19-{\tt A$^\prime$} &     1.06    & 1.0 & 0.92 & 2.1 &   92  &   34.0  &  4.0 & 5.6  &  7.0 &   8 &  0.55 & 1.8   &  1.4 & 0.560 & 0.608 \\
\enddata
\tablecomments{Properties of the major contributors to the seven {\it Herschel} sources.
Under the column for redshift ($z$), the values in bold-face are $z_{spec}$, otherwise they are $z_{ph}$. 
$L_{IR}$ is the total IR (8 to 1000~$\mu$m in restframe) luminosity based on the best-fit SK07 model.
$L_{IR}^{blk}$ is the total IR luminosity in the completely dust-blocked region, obtained by subtracting
the contribution of the dust-reprocessed light in the exposed region from $L_{IR}$. 
$S_{850}$ is the SCUBA-2 850~$\mu$m flux density predicted from the best-fit SK07 model.
$SFR_{IR}^{blk}$ is the SFR in the completely dust-blocked region. $T_d$ and $M_d$ are the
dust temperature and the dust mass, respectively. The gas mass is derived as $M_{gas}=140\times M_d$. 
$M^\ast$, $\tau$ and $T$ are the stellar mass, the exponentially declining SFH time scale and the age
of the stellar population in the exposed region, derived from the best-fit BC03 model. $SFR_{fit}$ is
the on-going SFR pertaining to the best-fit BC03 model. The total SFR of a given system can be
calculated as $SFR_{tot}=SFR_{fit}+SFR_{IR}^{blk}$. The total SSFR is defined as
$SSFR_{tot}=SFR_{tot}/M^\ast$. Finally, two stellar mass doubling time scales are
defined as $T_{db}^{tot}=M^\ast/SFR_{IR}^{tot}$ and $T_{db}^{blk}=M^\ast/SFR_{IR}^{blk}$.
}
\tablenotetext{a.}{For the stellar population of this object, the listed age ($T$) and the characteristic SFH time scale
($\tau$) are for the ``red'' component in its SED. See Fig. 9.}
\tablenotetext{b.}{This is not an independent object but for the case where the two components (UDS01-{\tt B} and {D})
 are combined.}
\tablenotetext{c.}{This object very likely has an AGN, which could be responsible for the IR emission.
}
\end{deluxetable}

\newpage
\appendix
\section{Prospective Automation of the Decomposition Process}

   The decomposition scheme that we developed in this work has the potential
of being automated and being applied to large samples in the future.
Specifically, the ``automatic iterative'' and the ``trial-and-error'' steps can
be integrated, and in principle we do not need a mid-IR ``ladder'' such as an
image in MIPS 24~$\mu$m. There are multiple choices of possible approaches that
involve different algorithms and criteria, and it will require extensive tests
on a large sample before the implement can be finalized. There will be a lot of
degenerated cases, and additional information that we can obtain from
optical/NIR, such as morphology and $z_{ph}$, can be used to further narrow
the possible solutions. While the automation is beyond the scope of this paper,
we show here that it is feasible. We describe one possible approach below for
the decomposition in SPIRE 250~$\mu$m. 

   Basically, this approach is a generalization of the ``trial-and-error''
method. It is reasonable to assume that, generally speaking, the objects that
are closer to the source centroid are more likely the real contributors.
Therefore, we define a core radius from the source centroid, $r_c$, and start
the decomposition from the objects within this radius. The most optimal choice
of this radius will have to wait for extensive tests, and for the illustration
purpose here we choose $r_c=6\arcsec$. We first consider the case that there
is only one real contributor to all the flux, and the decomposition routine
cycles through all the objects within $r_c$ one by one. We then consider the
case where there are two contributors, and run through all the possible pairs
within $r_c$.  If the secondary contributor in a given solution only accounts
for $\leq 5$\% of the total flux, it is deemed insignificant and the solution
falls back to the one-contributor case. The same process repeats for the 
three-contributor case. While we can add more objects and consider the 
four-contributor case, in reality this might not be desirable because most of
the time the added object within $r_c$ will have a separation
$\lesssim 2\arcsec$ to one of the existing object in the group and hence is not
going to add an unique solution. For illustration purpose, here we stop at the
three-contributor case within $r_c$.  We then add an object beyond $r_c$,
cycling through all objects at $r>r_c$ one by one, to the one-, two- and 
three-contributor combinations 
within $r_c$. If a specific addition improves the fit (for example, in terms
of $\chi^2$ as reported by GALFIT), we add a new object to this combination and
repeat the process. Otherwise we terminate the sequence. Finally, we
examine the $\chi^2$ values of all the combinations that have been run through.
The combinations where an object has its flux error larger than the extracted
flux (both as reported by GALFIT) are deemed as ``overstretching'' and are
rejected. We then apply a threshold on $\chi^2$ to define a pool of candidate
solutions. As our goal is to extract the {\it major} component of a
given 250~$\mu$m source, we order these candidate solutions by the major
contributor in the solution. One possible way to reach the final answer is to
deem the object that has the
highest occurrence among all candidate solutions as the major component.
In this stage, the additional
information from the high resolution images can be used to help decide on
the final choice. It is possible that there will still be
degeneracy, and we can offer different interpretations for each should this
happen.

  Here we use UDS01 as an example to illustrate this approach, where no
prior knowledge in 24~$\mu$m is used. Within $18\arcsec$ to the 250~$\mu$m
centroid, there are 33 objects in $H_{160}$ with $S/N\geq 5$ (see \S 5.1.1).
Ten of them are within $r_c=6\arcsec$, which are shown in Fig. 41 (left). 
For the sake of simplicity, we concentrate on the cases within $r_c$ and do 
not go beyond in this illustration. The decomposition was run through the
one-, two- and three-contributor cases as described above. Fig. 41 (right)
also shows the histograms of
the fitting $\chi^2$ (as reported by GALFIT) for these three cases. It is
immediately clear that the two- and three-contributor cases produced
better solutions than the one-contributor case, which was confirmed by
the visual inspection of the residual maps. The inspection also showed that
the residual maps with $\chi^2>1.75$ were all significantly worse than those
that have smaller $\chi^2$ values, and therefore we deemed that the best
solutions were those with $\chi^2\leq 1.75$. 

\begin{figure*}[tbp]
\centering
\subfigure{
  \includegraphics[width=0.5\textwidth]{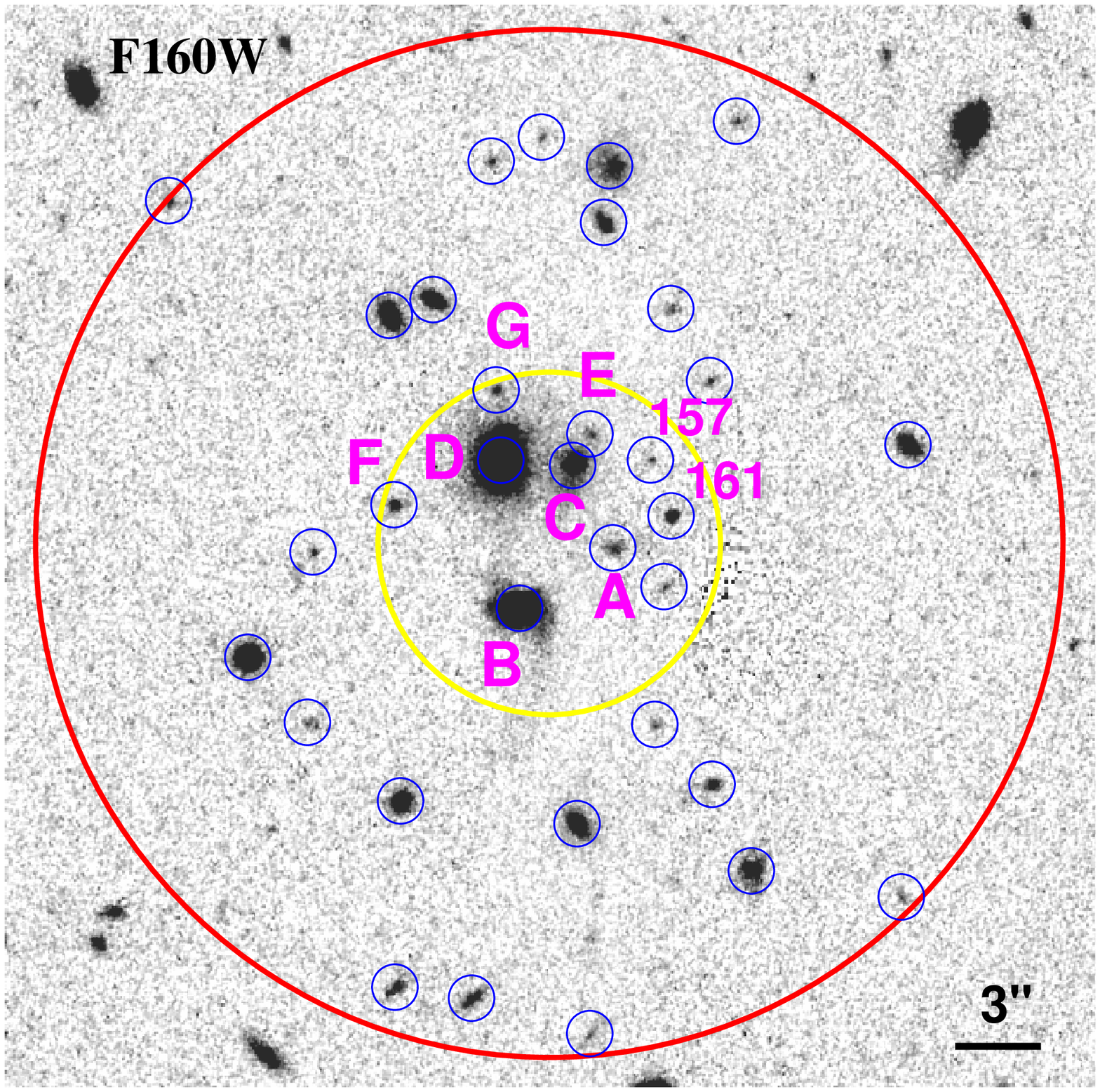} 
  \includegraphics[width=0.5\textwidth]{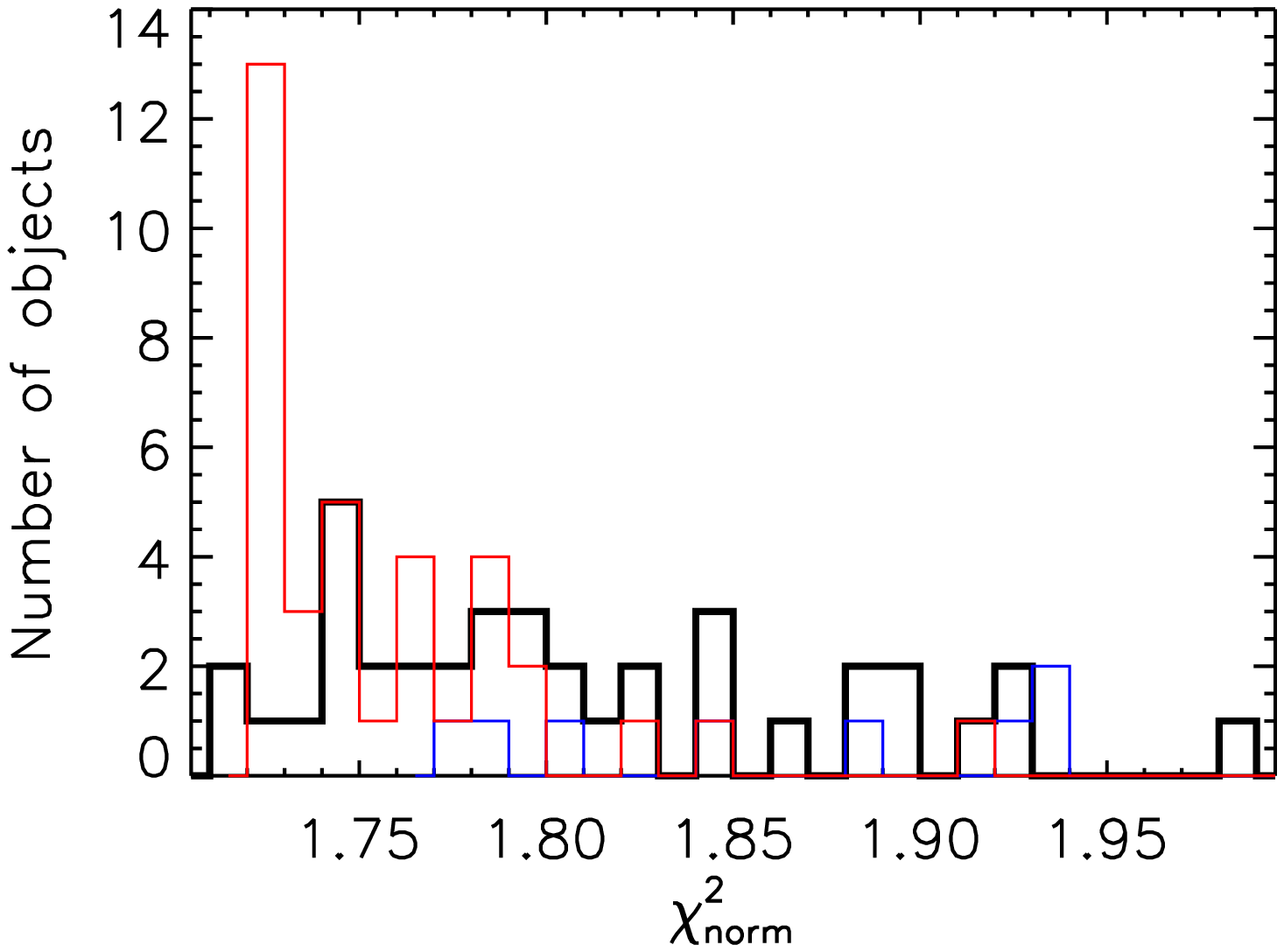}
}
\caption{(left) Zoomed-in view of UDS01 in $H_{160}$. The red and yellow
circles center on the 250~$\mu$m centroid and are $18\arcsec$ and $6\arcsec$ in
radius, respectively. The latter indicates the ``core region''. The objects
detected in $H_{160}$ (with $S/N>5$) are marked by blue circles. The candidate
contributors found by the automatic decomposition routine are labeled in
magenta, among which those the same as in \S 5.1.1 are labeled the same
alphabetically as in Fig. 12 and the two additional ones are labeled by numbers.
(right) Histograms of the fitting $\chi^2$ by the automatic routine as
described in the text. The one-, two- and three-contributor cases are coded in
blue, black and red, respectively.
}
\end{figure*}

   The fitting $\chi^2$ alone would suggest that the 3-component case produced
many more better solutions than the 2-component case does. In particular, there
is a high peak at $\chi^2=1.72$ in the 3-contributor case. However, further 
examination showed that only two of them were unique solutions and the rest
were all ``overstretching''. 
Therefore, the best solutions to choose from were the nine 2-contributor
solutions plus the two surviving 3-contributor solutions, all with 
$\chi^2\leq 1.75$. Fig. 42 shows their residual maps in two panels
in order of their fitting $\chi^2$. The ID (three digit numbers) of
the fitted objects are labeled in descending order of their extracted fluxes.
To compare to the results in \S 5.1 where the potential contributors were 
pre-selected using the MIPS 24~$\mu$m information, we list the correspondence
of the relevant ID's here and those in \S 5.1 as follows:
{\tt 163 = A}, {\tt 166 = B}, {\tt 155 = C}, {\tt 149 = D}, {\tt 151 = E},
{\tt 159 = F}, and {\tt 145 = G}. The label on top of each panel in Fig. 42
reflects the correspondence. Two conclusions are immediately clear. First,
these automated solutions have captured the 24~$\mu$m pre-selected candidates,
and only two objects, {\tt 157} and {\tt 161}, are not among the
24~$\mu$m pre-selected ones. Second, nine out of the eleven solutions involve
{\tt B} ({\tt 166}; labeled on top in either red or blue in Fig. 42), and seven
of these nine have it as the major contributor (labeled on top in red in
Fig. 42). Therefore, our method chose {\tt B} ({\tt 166}) as the major component
of UDS01. The relevant solutions showed that on average {\tt B} ({\tt 166})
accounted for $61\pm 6$\% of the total flux, agreeing with the value quoted in
Table 2 to 2\%. 

\begin{figure*}[tbp]
\centering
\includegraphics[width=\textwidth]{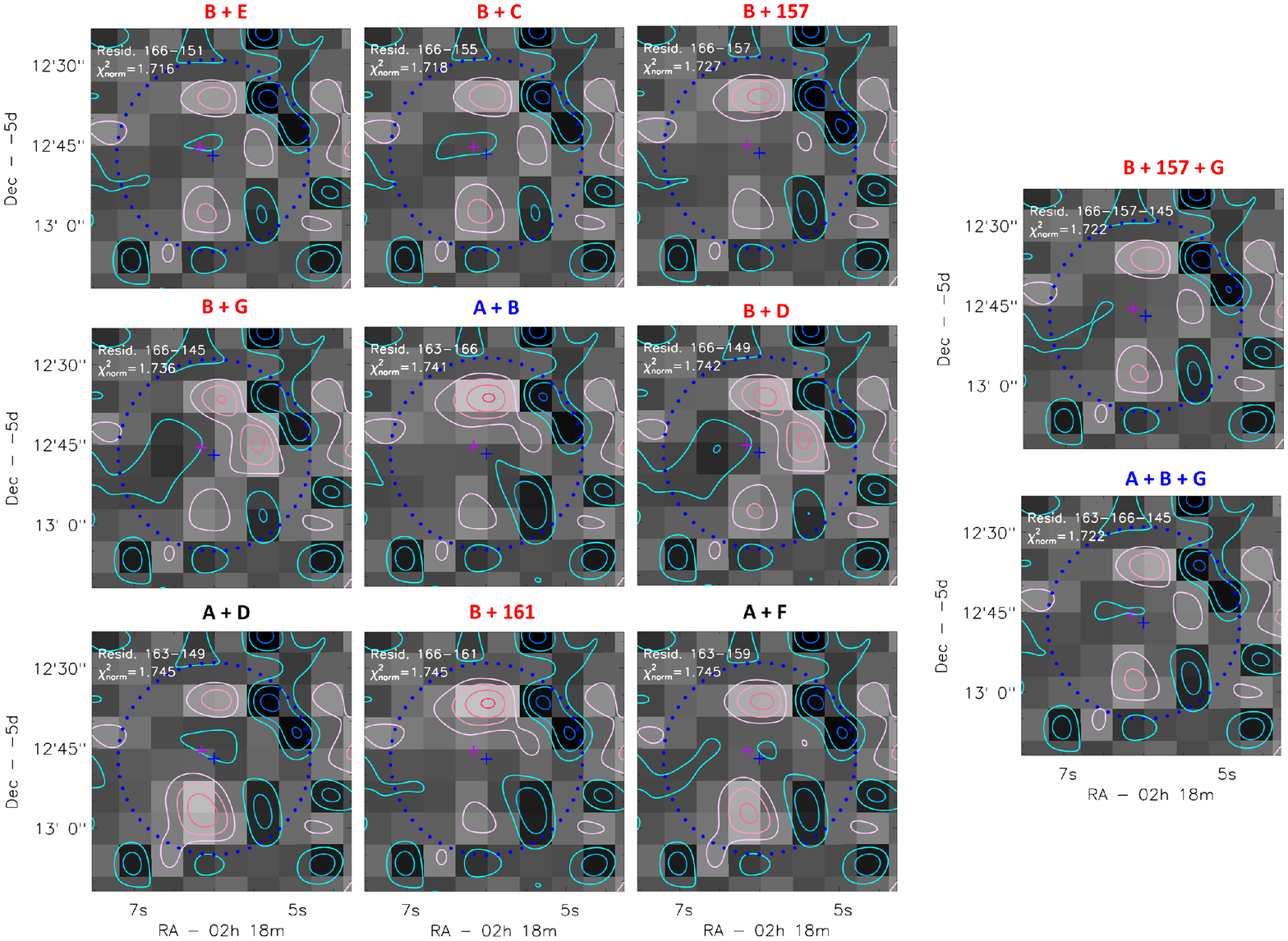}
\caption{Residual maps of the automatic solutions among which the best
solutions are to be decided. Each panel is for one solution as described in the
text. The ID's of the objects involved in each solution are labeled, together
with the fitting $\chi^2$. The labels on top of each panel indicate the
correspondence to the objects in \S 5.1. Other legends are the same as in 
Fig. 15.
}
\end{figure*}

   While more difficult, extracting the secondary component is still possible
if its contribution is significant. For the case of UDS01, $\sim 40$\% of the
total flux should be accounted for by other objects beyond the major component
{\tt B} ({\tt 166}). To be self-consistent, the secondary contributor should be
sought among the solutions that result in the primary contributor, i.e.,
the seven solutions shown in Fig. 42 with red labels on top.
As discussed in \S 5.1.2, {\tt C/E}
({\tt 155/151}) and {\tt D/G} ({\tt 149/145}) should be treated as single
objects. Therefore, these seven solutions reduce to five, namely, {\tt B+C/E},
{\tt B+157}, {\tt B+D/G}, {\tt B+161}, and {\tt B+157+G}. Note that {\tt B+C/E}
and {\tt B+D/G} are also the solutions when using 24~$\mu$m pre-selected
candidates as discussed in \S 5.1.3, where we point out that the formal fitting
preferred {\tt C/E} but we chose {\tt D/G} as the final secondary component
because of its stronger 24~$\mu$m flux. Our inspection of UDS01 in 250~$\mu$m
(see Fig. 12) indeed showed that the light distribution coincides with
{\tt C/E} better than {\tt D/G}, and in fact this is the reason why objects
{\tt 157} and {\tt 161} were among the solutions because they are closer to 
side of {\tt C/E} than to that of {\tt D/G} (see Fig. 41). We could not reject
{\tt 157} or {\tt 161} being the secondary based on the decomposition solutions
alone, and a proper treatment would need to taking into account other
information from optical/NIR images, such as their $z_{ph}$. This will be
deferred to our future paper on the automation. On the other hand, {\tt C/E}
and {\tt D/G} present an interesting case of degeneracy (given that their
$z_{ph}$ are almost the same), and we could provide different interpretation
for each possibility.

\end{document}